\documentclass[floatfix,aps,prc,superscriptaddress,preprint]{revtex4}

\usepackage{amsfonts}                            % Use AMS fonts
\usepackage{amssymb}                             % and symbols
\usepackage{amsmath}                             % and definitions
\usepackage{bm}
\usepackage{graphicx}                            % Graphics package

\usepackage{color}
\usepackage{latexsym}                            % Load additional symbols
\usepackage{dcolumn}                             % Decimal-dot aligned columns
\usepackage[colorlinks=true,linkcolor=blue]{hyperref}
\usepackage{footnote}
\usepackage{slashed}
\usepackage{xfrac}
\usepackage{natbib}
\usepackage{placeins}

\newcommand\be{\begin{equation}}
\newcommand\ee{\end{equation}}
\newcommand\bea{\begin{eqnarray}}
\newcommand\eea{\end{eqnarray}}
\newcommand\bal{\begin{aligned}}
\newcommand\eal{\end{aligned}}
\newcommand\bes{\begin{subequations}}
\newcommand\ees{\end{subequations}}

\newcommand\ct{\cos\theta}
\newcommand\st{\sin\theta}
\newcommand\ctk{\cos\theta_{k_1}}
\newcommand\stk{\sin\theta_{k_1}}
\newcommand\cpk{\cos\phi_{k_1}}
\newcommand\nn{\nonumber}

\newcommand\gep{G_E^p}
\newcommand\gmp{G_M^p}
\newcommand\fonep{F_1'}
\newcommand\ftwop{F_2'}
\newcommand\gap{G_a'}
\newcommand\fonem{\widetilde F_1}
\newcommand\ftwom{\widetilde F_2}
\newcommand\fthreem{\widetilde F_3}

\newcommand\eps{\varepsilon}
\newcommand\ubar{\bar{u}}
\newcommand\ggam{{\gamma\gamma}}
\newcommand\half{\textstyle{\frac{1}{2}}}

\newcommand{\Tr}{\operatorname{Tr}}
\renewcommand{\Re}{\operatorname{Re}}
\renewcommand{\Im}{\operatorname{Im}}

\preprint{JLAB-THY-17-2430}

\begin{document}

\title{A dispersive approach to two-photon exchange \\
	in elastic electron--proton scattering}

\author{P.~G.~Blunden}
\affiliation{\mbox{Department of Physics and Astronomy,
	University of Manitoba}, Winnipeg, Manitoba, Canada R3T 2N2}
\affiliation{Jefferson Lab, Newport News, Virginia 23606, USA}
\author{W.~Melnitchouk}
\affiliation{Jefferson Lab, Newport News, Virginia 23606, USA}

\date{March 10, 2017}

\begin{abstract}
We examine the two-photon exchange corrections to elastic
electron--proton scattering within a dispersive approach, including
contributions from both nucleon and $\Delta$ intermediate states. The
dispersive analysis avoids off-shell uncertainties inherent in
traditional approaches based on direct evaluation of loop diagrams, and
guarantees the correct unitary behavior in the high energy limit. Using
empirical information on the electromagnetic nucleon elastic and $N
\Delta$ transition form factors, we compute the two-photon exchange
corrections both algebraically and numerically.  Results are compared
with recent measurements of $e^+ p$ to $e^- p$ cross section ratios from
the CLAS, VEPP-3 and OLYMPUS experiments, as well as with polarization
transfer observables.
\end{abstract}

\maketitle

\tableofcontents

\newpage
%%%%%%%%%%%%%%%%%%%%%%%%%%%%%%%%%%%%%%%%%%%%%%%%%%%%%%%%%%%%%%%%%%%%%%%%
%%%%%%%%%%%%%%%%%%%%%%%%%%%%%%%%%%%%%%%%%%%%%%%%%%%%%%%%%%%%%%%%%%%%%%%%
\section{Introduction}
\label{sec:Intro}

The nucleon's electroweak form factors are some of the cornerstone
observables that characterize its extended spatial structure. Since the
original observation~\cite{Hofstadter:1955ae} some 60 years ago that
elastic scattering from the proton deviates from point-like behavior at
large scattering angles, considerable information has been accumulated
on the detailed structure of the proton's and neutron's electric and
magnetic responses.  Almost universally the underlying scattering
reaction has been assumed to proceed through the exchange of a single
photon between the lepton (typically electron) beam and nucleon target.

A major paradigm shift occurred around the turn of the last century with
the observation of a significant discrepancy between the ratio of
electric to magnetic form factors of the proton measured using the
relatively new polarization transfer technique~\cite{Jones:1999rz,
Gayou:2001qd} and previous extractions of the same quantity from cross
section measurements via Rosenbluth separation. It was soon
realized~\cite{Blunden:2003sp, Guichon:2003qm} that a large part of the
discrepancy could be understood in terms of additional, hadron
structure-dependent two-photon exchange (TPE) contributions, which had
not been included in the standard treatments of electromagnetic
radiative \mbox{corrections~\cite{Tsai:1961zz, Mo:1968cg}}.

A number of approaches have been adopted to computing the TPE
corrections to elastic scattering, including direct calculation of the
loop contributions in terms of hadronic degrees of
freedom~\cite{Blunden:2003sp, Kondratyuk:2005kk, Blunden:2005ew,
Kondratyuk:2007hc, Nagata:2008uv, Graczyk:2013pca, Lorenz:2014yda,
Zhou:2014xka, Zhou:2016psq}, modeling the high energy behavior of box
diagrams at the quark level through generalized parton
distributions~\cite{Chen:2004tw, Afanasev:2005mp}, or more recently
dispersion \mbox{relations~\cite{Gorchtein:2006mq, Borisyuk:2008es,
Borisyuk:2015xma, Tomalak:2014sva}}. Each of these methods has its own
advantages as well as limitations (for reviews, see
Refs.~\cite{Carlson:2007sp, Arrington:2011dn, Afanasev:2017}), and to
date no single approach has been able to provide a universal description
valid at all kinematics.

Most of the attention on the TPE corrections in recent experiments has
been focussed on the region of small and intermediate values of the
four-momentum transfer squared, $Q^2 \lesssim$~few~GeV$^2$, where the
expectation is that hadrons retain their identity sufficiently well that
calculations in terms of physical degrees of freedom give reliable
estimates.  Traditionally, this approach has required direct evaluation
of the real parts of the two-photon box and crossed-box diagrams, with
nucleons or other excited state hadrons in the intermediate state
parametrized through half off-shell form factors (with one nucleon
on-shell and one off-shell).  Because the off-shell dependence of these
form factors is not known, usually one approximates the half off-shell
form factors by their on-shell limits.

For nucleon intermediate states, the off-shell uncertainties are not
expected to be severe.  On the other hand, for transitions to excited
state baryons described by effective interactions involving derivative
couplings, such as for the $\Delta$ resonance, the off-shell dependence
leads to divergences in the forward angle (or high energy) limit, and
signals a violation of unitarity.  Furthermore, from a more technical
perspective, in order to evaluate the TPE corrections analytically in
terms of Passarino-Veltman functions, the loop integration method
requires the transition form factors to be parametrized as sums or
products of monopole functions.  This can prove cumbersome in some
applications, since such parametrizations are usually only valid in a
limited region of spacelike $Q^2$, and may be prone to roundoff errors
in numerical evaluation.  It would naturally be highly desirable to be
able to compute the loop integrations with a more robust numerical
method that is valid for form factor parametrizations based on more
general classes of functional forms.

The limitations of the previous loop calculations are especially
problematic in view of new measurements of ratios of $e^+ p$ to $e^- p$
elastic scattering cross sections \cite{Rimal:2016toz, Rachek:2014fam,
Nikolenko:2014uda, Henderson:2016dea}, which have provided high
precision data that are directly sensitive to TPE effects. Some of these
data are in the small-angle region, where the off-shell ambiguities in
the loop calculations make the calculations unreliable. To enable
meaningful comparison between the data and TPE calculations over the
full range of kinematics currently accessible, clearly a different
approach to the problem is needed.

In this paper, we revisit the calculation of TPE corrections within the
hadronic approach, but using dispersion relations to construct the real
part of the TPE amplitude from its imaginary part.  The dispersive
method involves the exclusive use of on-shell transition form factors,
thereby avoiding the problem of unphysical violation of unitarity in the
high energy limit.  The dispersive approach to TPE was developed at
forward angles by Gorchtein~\cite{Gorchtein:2006mq}, and at non-forward
angles by Borisyuk and Kobushkin \cite{Borisyuk:2008es, Borisyuk:2012he,
Borisyuk:2013hja, Borisyuk:2015xma}, and more recently by Tomalak and
Vanderhaeghen \cite{Tomalak:2014sva}.  A feature of the latter two
analyses has been the use of monopole form factor parametrizations,
which allowed the computations to be performed semi-analytically. In
this work we extend the dispersion relation approach to allow for more
general classes of transition form factors.

In Sec.~\ref{sec:formalism} of this paper we review the formalism for
elastic electron--nucleon scattering for both one-photon and two-photon
exchange processes, and introduce the main elements of the dispersive
approach.  We describe analytical calculations of the imaginary part of
the TPE corrections using the more restrictive monopole form factors,
for which one can obtain analytic expressions in terms of elementary
logarithms.  We also describe the more general numerical method that
allows standalone calculation of the imaginary part using a general
class of transition form factors.

The results of the calculations are presented in Sec.~\ref{sec.results},
where we critically examine the differences between the new dispersive
method and the previous loop calculations with off-shell intermediate
states.  While the differences are relatively small for the nucleon
elastic contributions, the effects for $\Delta$ intermediate states are
dramatic at high energies and forward scattering angles. We also compare
in Sec.~\ref{sec.results} the new results with the recent data on
$e^+ p$ to $e^- p$ cross section ratios from the
CLAS \cite{Rimal:2016toz},
VEPP-3 \cite{Rachek:2014fam, Nikolenko:2014uda} and
OLYMPUS \cite{Henderson:2016dea} experiments,
as well as with polarization data sensitive to TPE contributions
\cite{Meziane:2010xc}.
Finally, in Sec.~\ref{sec.outlook} we summarize our results, and discuss
possible future developments in theory and experiment. For completeness,
in the appendices we give the full expressions for the generalized form
factors in Appendix~\ref{app:genFF}, and analytic expressions for the
imaginary parts of Passarino-Veltman functions in Appendix~\ref{app:PV}.
 We also provide convenient reparametrizations of the nucleon and
$\Delta$ vertex form factors in Appendix~\ref{app:param} that can be
used in the analytic calculations.

%%%%%%%%%%%%%%%%%%%%%%%%%%%%%%%%%%%%%%%%%%%%%%%%%%%%%%%%%%%%%%%%%%%%%%%%
%%%%%%%%%%%%%%%%%%%%%%%%%%%%%%%%%%%%%%%%%%%%%%%%%%%%%%%%%%%%%%%%%%%%%%%%
\section{Formalism}
\label{sec:formalism}

In this section we present the formalism on which the electron--nucleon
scattering analysis in this paper will be based.  After summarizing the
kinematics and main formulas for the elastic scattering amplitudes and
cross sections at the Born and TPE level, we proceed to describe the new
elements of the analysis that make use of dispersive methods, including
both analytic and numerical evaluation of integrals.

%%%%%%%%%%%%%%%%%%%%%%%%%%%%%%%%%%%%%%%%%%%%%%%%%%%%%%%%%%%%%%%%%%%%%%%%
\subsection{Elastic $ep$ scattering}
\label{ssec:kinematics}

\begin{figure}[t]
\centering
\includegraphics[width=0.75\linewidth]{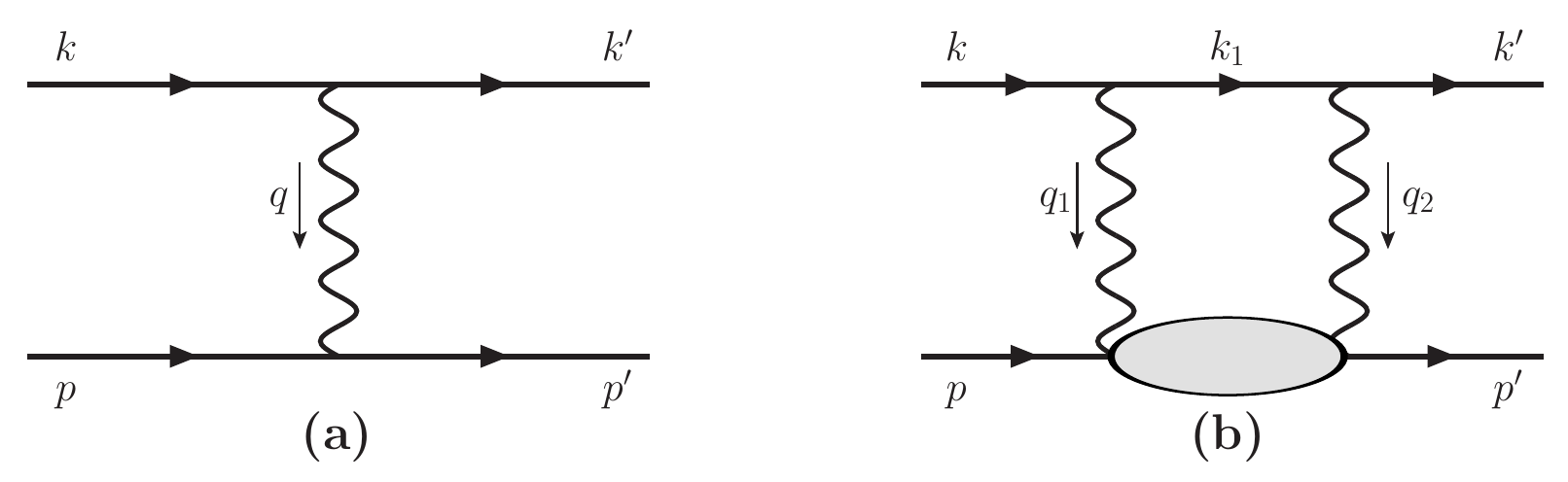}
\caption{Contributions to elastic electron--nucleon scattering from
	(a) one-photon exchange, and (b) two-photon exchange amplitudes,
	with particle momenta as indicated.  For TPE we have $q_1+q_2=q$.
	Only the $s$-channel ``box'' diagram is drawn.
	The ``crossed-box'' contribution, which can be obtained by
	applying crossing symmetry $s\to u$, is implied.}
\label{fig:OPETPE}
\end{figure}

For the elastic scattering process $e p \to e p$ the four-momenta of
the initial and final electrons (taken to be massless) are labeled
by $k$ and $k'$, and of the initial and final protons (mass $M$) by
$p$ and $p'$, respectively, as depicted in Fig.~\ref{fig:OPETPE}.
The four-momentum transfer from the electron to the proton is given
by $q = p'-p = k-k'$, with $Q^2 \equiv -q^2 > 0$.  One can express
the elastic cross section in terms of any two of the Mandelstam
variables $s$ (total electron--proton invariant mass squared),
$t$, and $u$, where
\be
\bal
s&=(k+p)^2=(k'+p')^2\,,\\
t&=(k-k')^2=q^2\,,\\
u&=(p-k')^2=(p'-k)^2\,,
\label{eq:Mandelstam}
\eal
\ee
with the constraint $s + t + u = 2 M^2$.

The elastic scattering cross section can be defined in terms of
any two of the dimensionless quantities
\be
\bal
\eps &= \frac{\nu^2 - \tau (1+\tau)}{\nu^2 + \tau (1+\tau)}
      =\frac{2 \left(M^4-s u\right)}{s^2+u^2-2 M^4}\,,\\
\tau &= \frac{Q^2}{4M^2}\, ,\qquad
\nu = \frac{k\cdot p}{M^2} -\tau\, .
\eal
\label{eq:kin1}
\ee
The inverse relationships are also useful,
\be
\bal
\nu &= \frac{s-u}{4 M^2} = \sqrt\frac{\tau (1+\tau)(1+\eps)}{1-\eps}\, ,\\
  s &= M^2(1+2\tau+2\nu)\, .
\eal
\label{eq:kin2}
\ee
In the target rest frame the variables are given by
\be
\bal
\eps &= \left(1 + 2 (1+\tau) \tan^2{\frac{\theta_e}{2}}\right)^{-1}\, ,\\
\tau &= \frac{E-E'}{2M}\, ,\quad
\nu   = \frac{E+E'}{2M}\, ,\quad
E   = M(\tau + \nu)\, ,
\eal
\label{eq:kin3}
\ee
where $E$ $(E')$ is the energy of the incident (scattered) electron,
$\theta_e$ is the electron scattering angle, and $\eps$ ($0< \eps< 1$)
is identified with the relative flux of longitudinal virtual photons.

% ....................................................................... 
\subsubsection{One-photon exchange}

In the Born (OPE) approximation the electron--nucleon scattering
invariant amplitude can be written as
\be
{\cal M}_\gamma = -\frac{e^2}{q^2}\, j_\mu^\gamma\, J_\gamma^\mu\, ,
\label{eq:Mg}
\ee
where $e$ is the electric charge, and the matrix elements of the
electromagnetic leptonic and hadronic currents are given in terms
of the lepton ($u_e$) and nucleon ($u_N$) spinors by
\be
\bal
j_\mu^\gamma &= \ubar_e(k')\, \gamma_\mu\, u_e(k)\,, \\ 
J_\gamma^\mu &= \ubar_N(p')\, \Gamma^\mu(q)\, u_N(p)\,.
\eal
\ee
The electromagnetic hadron current operator $\Gamma^\mu$ is
parametrized by the Dirac ($F_1$) and Pauli ($F_2$) form factors as
\be
\Gamma^\mu(q) = F_1(Q^2)\, \gamma^\mu\ 
+\ F_2(Q^2)\,\frac{i \sigma^{\mu\nu} q_\nu}{2 M}\, ,
\label{eq:Jg}
\ee
where the Born form factors are functions of a single variable, $Q^2$.
In our convention, the reduced Born cross section $\sigma_R^{\rm Born}$
is given by
\be
\sigma_R^{\rm Born} = \eps\, G_E^2(Q^2)\ + \tau\, G_M^2(Q^2)\, ,
\label{eq:sigmaR}
\ee
where the Sachs electric and magnetic form factors $G_{E,M}(Q^2)$
are defined in terms of the Dirac and Pauli form factors as
\be
\bal
\label{eq:GEMdef}
G_E(Q^2) &= F_1(Q^2) - \tau F_2(Q^2)\, ,\\
G_M(Q^2) &= F_1(Q^2) + F_2(Q^2)\, .
\eal
\ee

% ....................................................................... 
\subsubsection{Two-photon exchange}

Using the kinematics illustrated in the box diagram in
Fig.~\ref{fig:OPETPE}(b), the contribution to the TPE box amplitude
from an intermediate hadronic state $R$ of invariant mass $M_R$ can
be written in the general form~\cite{Blunden:2003sp, Kondratyuk:2005kk}
\be
{\cal M}_{\ggam}^{\rm box}
= -ie^4 \int \frac{d^4 q_1}{(2\pi)^4}\
    \frac{L_{\mu\nu} H_R^{\mu\nu}}
	 {(q_1^2-\lambda^2)(q_2^2-\lambda^2)}\,,
\label{eq:Mggbox}
\ee
with $q_2=q-q_1$, and an infinitesimal photon mass $\lambda$ is
introduced to regulate any infrared divergences.  (In general the mass
$M_R$ can have a distribution $W$ which can be integrated over, but here
we specialize to the case of a narrow state $R$.) The leptonic and
hadronic tensors here are given by
\bea
L_{\mu\nu}
&=& \ubar_e(k')\, \gamma_\mu\, S_F(k_1,m_e)\, \gamma_\nu\, u_e(k)\,,
\label{eq:Lgg}\\
H_R^{\mu\nu}
&=& \ubar_N(p')\, \Gamma_{R\to\gamma N}^{\mu\alpha}(p_R,-q_2)\,
    S_{\alpha\beta}(p_R,M_R)\,
    \Gamma_{\gamma N\to R}^{\beta\nu}(p_R,q_1)\, u_N(p)\,,
\label{eq:Hgg}
\eea
with $k_1=k-q_1$, $p_R=p+q_1$, and the electron propagator is
\be
S_F(k_1,m_e)
= \frac{(\slashed{k}_1 + m_e)}
       {(k_1^2 - m_e^2 + i 0^+)}\, .
\label{eq:sfe}
\ee
The hadronic transition current operator $\gamma N\to R$ is written
in a general form $\Gamma_{\gamma N\to R}^{\alpha\mu}(p_R,q)$ that
allows for a possible dependence on the {\em incoming} momentum $q$
of the photon and the {\em outgoing} momentum $p_R$ of the hadron,
while $\mu$ and $\alpha$ are Lorentz indices.

The hadronic state propagator $S_{\alpha\beta}(p_R,M_R)$ in this work
will describe the propagation of a baryon with either spin-\sfrac{1}{2}
or spin-\sfrac{3}{2}.  For spin-\sfrac{1}{2} intermediate states,
such as the nucleon, this reduces to
\be
S_{\alpha\beta}(p_R,M_R)\
=\ \delta_{\alpha\beta}\, S_F(p_R,M_R),
\ee
and the transition operator $\Gamma_{\gamma N\to R}$ involves one free
Lorentz index.  For spin-\sfrac{3}{2} intermediate states, such as the
$\Delta$ baryon, the propagator can be written
\be
S_{\alpha\beta}(p_R,M_R)\
=\ -S_F(p_R,M_R)\, {\cal P}^{3/2}_{\alpha\beta}(p_R),
\ee
where the projection operator
\be 
{\cal P}^{3/2}_{\alpha\beta}(p_R)\
=\ g_{\alpha\beta}\
-\ {1 \over 3} \gamma_\alpha \gamma_\beta\
-\ {1 \over 3 p_R^2}
   \left( \slashed{p}_R \gamma_\alpha (p_R)_\beta
 	+ (p_R)_\alpha \gamma_\beta \slashed{p}_R 
   \right)\, ,
\ee
ensures the presence of only spin-\sfrac{3}{2} components. Unphysical
spin-\sfrac{1}{2} contributions are suppressed by the condition on the
vertex $p_{R \alpha} \Gamma_{\gamma N \to R}^{\alpha\mu}(p_R,q) = 0$.

One can obtain the crossed-box (``xbox'') contribution directly
from the box term \eqref{eq:Mggbox} by applying crossing symmetry.
For example, in the unpolarized case, we have
\be
{\cal M}_{\ggam}^{\rm xbox}(u,t) =
-{\cal M}_{\ggam}^{\rm box}(s,t)
\Big|_{s\to u}\, .
\label{eq:Mggxbox}
\ee
In general, ${\cal M}_{\ggam}^{\rm box}(s,t)$ has both real and
imaginary parts, whereas ${\cal M}_{\ggam}^{\rm xbox}(u,t)$ is
purely real.
The total squared amplitude for the sum of the one- and two-photon
exchange processes shown in Fig.~\ref{fig:OPETPE} is then
\be
\bal
\left| {\cal M}_\gamma + {\cal M}_{\ggam} \right|^2
&\approx \left| {\cal M}_\gamma \right|^2
	+ 2 \Re \left( {\cal M}_\gamma^\dagger {\cal M}_{\ggam} \right) \\
&\equiv \left| {\cal M}_\gamma \right|^2
	\left( 1 + \delta_{\rm TPE} \right),
\eal
\ee
where the relative correction to the cross section due to the
interference of the one- and two-photon exchange amplitudes is defined as
\be
\delta_{\rm TPE}
= \frac{2 \Re \left( {\cal M}_\gamma^\dagger {\cal M}_{\ggam} \right)}
     {\left| {\cal M}_\gamma \right|^2 }\,.
\label{eq:delTPE}
\ee

Within the framework of the simplest hadronic models, analytic
evaluation of $\delta_{\rm TPE}$ is made possible by writing the
transition form factors at the $\gamma$-hadron vertices as a sum and/or
product of monopole form factors~\cite{Blunden:2003sp, Blunden:2005ew},
which are typically fit to empirical transition form factors over a
suitable range in spacelike four-momentum transfer.  Four-dimensional
integrals over the momentum in the one-loop box diagram can then be
expressed in terms of the Passarino-Veltman (PV) scalar functions $A_0$,
$B_0$, $C_0$ and $D_0$~\cite{tHooft:1978jhc, Passarino:1978jh}. This
reduction to scalar integrals is automated by programs such as
FeynCalc~\cite{Mertig:1990an, Shtabovenko:2016sxi}. The PV functions can
then be evaluated numerically using packages such as
LoopTools~\cite{Hahn:1998yk}. In this paper we are interested only in
the imaginary parts of these PV functions, which are considerably
simpler than the full expressions.  This will be discussed in detail in
the next section.

Note that the expressions (\ref{eq:Mggbox}) and (\ref{eq:delTPE})
contain infrared (IR) divergences arising from the elastic intermediate
state when the momentum $q_i\ (i=1,2)$ of either photon vanishes.  In
analyzing the TPE corrections for $ep$ scattering, it is convenient to
subtract off these conventional IR-divergent parts, which are
independent of hadronic structure, and which are usually already
included in experimental analyses using a specific prescription ({\em
e.g.} Mo~\&~Tsai~\cite{Mo:1968cg}, Grammer \&
Yennie~\cite{Grammer:1973db}, or Maximon \& Tjon~\cite{Maximon:2000hm}).

In general, the TPE amplitude at the IR poles ($q_1\to 0$ or $q_2\to 0$)
has the form
\be
{\cal M}_\ggam \longrightarrow {\cal M}_\gamma \Delta_{\rm IR}\, ,
\label{eq:MggIR}
\ee
where $\Delta_{\rm IR}$ is a function containing all the IR divergences
that is independent of hadronic structure. Its form depends on the
particular IR prescription being used. This is discussed extensively in
the TPE review by Arrington, Blunden, and
Melnitchouk~\cite{Arrington:2011dn}, and we defer to that paper for
details.  The hard-TPE correction of interest is then
\be
\delta_{\ggam} \equiv \delta_{\rm TPE} - 2 \Re \Delta_{\rm IR}\, .
\label{eq:delggIR}
\ee

In this paper we follow the prescription used by Maximon and
Tjon~\cite{Maximon:2000hm}, which is to evaluate the contribution to the
numerator of Eq.~(\ref{eq:Mggbox}) arising from the poles $q_i \to 0$,
while keeping the propagators in the denominator intact. In this
prescription,
\be
\Delta_{\rm IR}({\rm MTj})
= - \frac{\alpha}{\pi} \log\left(\frac{M^2-s}{M^2-u}\right)
			\log\left(\frac{Q^2}{\lambda^2}\right)\, .
\label{eq:delMTj}
\ee
This expression has both real and imaginary parts.  In our convention,
$\log(-x) = \log{x} - i\pi$ for $x>0$, so explicitly the real and
imaginary parts are
\bes
\bea
\Re \Delta_{\rm IR}({\rm MTj})
&=& - \frac{\alpha}{\pi}\log\left(\frac{s-M^2}{M^2-u}\right)
\log\left(\frac{Q^2}{\lambda^2}\right)\, ,	\label{eq:delMTjRe}\\
\Im \Delta_{\rm IR}({\rm MTj})&=&\alpha 
\log\left(\frac{Q^2}{\lambda^2}\right)\, .
\label{eq:delMTjIm}
\eea
\ees

After accounting for conventional radiative corrections, the measured
reduced cross section $\sigma_R$ is related to the Born cross section
by
\be
\sigma_R = \sigma_R^{\rm Born} \left( 1 + \delta_{\ggam} \right)\, .
\label{eq:sigmaRmeas}
\ee
In practice, most experimental cross section analyses use the
IR-divergent expression of Mo and Tsai~\cite{Mo:1968cg}, so that if
one uses the Maximon and Tjon prescription~\cite{Maximon:2000hm}
(as we do in this paper) then the difference should be accounted
for when comparing to experimental data
(see Ref.~\cite{Arrington:2011dn} for further discussion).

The total TPE amplitude ${\cal M}_{\ggam}$ can be rewritten in terms
of ``generalized form factors'', generalizing the expressions of
Eqs.~(\ref{eq:Mg})--(\ref{eq:Jg}), as described by Guichon and
Vanderhaeghen~\cite{Guichon:2003qm}.  Although the decomposition
is not unique, and different generalized form factor conventions
have been used in the literature, in this paper we use the basis of
form factors denoted by $\fonep$, $\ftwop$ and $\gap$, defined via
\bea
{\cal M}_{\gamma\gamma}
&=& - {e^2 \over q^2}\,
    \bar{u}_e(k') \gamma_\mu u_e(k)\ \bar{u}_N(p') 
    \left[ F_1'(Q^2,\nu)\,\gamma^\mu\,
       +\, F_2'(Q^2,\nu)\, {i \sigma^{\mu\nu} q_\nu \over 2 M}
    \right] u_N(p)						\nn\\
& & - {e^2 \over q^2}\, \bar{u}_e(k')\gamma_\mu\gamma_5 u_e(k)\,
    \bar{u}_N(p')\, G_a'(Q^2,\nu)\,\gamma^\mu\gamma_5 u_N(p)\, ,
\label{eq:Mgen}
\eea
where the vector $\fonep$ and $\ftwop$ generalized form factors
are the TPE analogs of the Dirac and Pauli form factors, while the
axial vector $\gap$ generalized form factor has no Born level analog.

Rather than construct $\delta_{\rm TPE}$ and subtract the IR-divergent
terms, as in Eq.~(\ref{eq:delggIR}), it is convenient to incorporate the
IR subtractions directly into the generalized form factors $\fonep$ and
$\ftwop$ ($\gap$ is not IR-divergent),
\begin{subequations}
\label{eq:IRsubtraction}
\bea
\fonep &\equiv& F_{1,{\rm TPE}}' - F_1(Q^2)\, \Delta_{\rm IR}\, ,\\
\ftwop &\equiv& F_{2,{\rm TPE}}' - F_2(Q^2)\, \Delta_{\rm IR}\, ,
\eea
\end{subequations}
where $F_{i,{\rm TPE}}'$ refer to the unregulated expressions. In terms
of these regulated generalized form factors, the relative TPE correction
is given by
\be
\delta_\ggam
= 2\Re \frac{\eps G_E (\fonep - \tau \ftwop)
	     + \tau G_M (\fonep + \ftwop)
	     + \nu (1-\eps) G_M \gap}
	    {\eps G_E^2 + \tau G_M^2}\, .
\label{eq:delgg}
\ee

%%%%%%%%%%%%%%%%%%%%%%%%%%%%%%%%%%%%%%%%%%%%%%%%%%%%%%%%%%%%%%%%%%%%%%%%
\subsection{Dispersive approach}
\label{ssec:dispersive}

As noted earlier, the TPE amplitude ${\cal M}_{\gamma\gamma}$ has both
real and imaginary parts.  The real and imaginary parts can be related
through dispersion relations~\cite{Gorchtein:2006mq, Borisyuk:2008es},
which forms the basis of the dispersive method discussed in this
section.  Our discussion in this section follows the formalism of
Tomalak and Vanderhaeghen~\cite{Tomalak:2014dja, Tomalak:2014sva}. An
alternative treatment by Borisyuk and Kobushkin~\cite{Borisyuk:2008es}
starts from the annihilation channel, $e^-+e^+\to p+\bar{p}$.

Using the parametrization of the TPE amplitude ${\cal M}_{\gamma\gamma}$
in terms of the generalized form factors $\fonep$, $\ftwop$ and $\gap$,
we note that these TPE amplitudes have the symmetry properties
\cite{Gorchtein:2006mq,Borisyuk:2008es}
\bes
\label{eq:genFFsymm}
\bea
F_{1,2}'(Q^2,-\nu) &=& -F_{1,2}'(Q^2,\nu)\, ,\\
G_a'(Q^2,-\nu) &=& +G_a'(Q^2,\nu)\, ,
\eea
\ees
and satisfy the fixed-$t$ dispersion relations
\bes
\label{eq:genFFdisp}
\bea
\Re F_1'(Q^2,\nu)
&=& \frac{2}{\pi} {\cal P} \int_{\nu_{\rm th}}^\infty d\nu'\ 
    {\nu\over \nu'^2-\nu^2}\, \Im F_1'(Q^2,\nu')\, ,\\
\Re F_2'(Q^2,\nu)
&=& \frac{2}{\pi} {\cal P} \int_{\nu_{\rm th}}^\infty d\nu'\ 
    {\nu\over \nu'^2-\nu^2}\, \Im F_2'(Q^2,\nu')\, ,\\
\Re G_a'(Q^2,\nu)
&=& \frac{2}{\pi} {\cal P} \int_{\nu_{\rm th}}^\infty d\nu'\ 
    {\nu'\over \nu'^2-\nu^2}\, \Im G_a'(Q^2,\nu')\, .
\eea
\ees
Here ${\cal P}$ denotes the Cauchy principal value integral, and
$\nu_{\rm th} = -\tau$ is the threshold for the elastic cut,
corresponding to an electron of energy $E = 0$. The physical threshold
for electron scattering is at $\eps = 0$ (or $\cos\theta_e = -1$), which
requires $E \ge M(\tau+\nu_{\rm ph})$, with $\nu_{\rm ph} \equiv
\sqrt{\tau(1+\tau)}$. This integral therefore extends into an unphysical
region of parameter space, which requires knowledge of the transition
form factors in the timelike region of four-momentum transfer. The
crossed-box terms in the real part of the TPE amplitudes are generated
by incorporating the symmetry properties into the dispersive integrals,
which is equivalent to the use of Eq.~(\ref{eq:Mggxbox}) in the loop
calculation.

For the interaction of point particles, such as in elastic $e \mu$
scattering, the real parts generated in this way agree completely with
those obtained directly from the four-dimensional loop integrals of
Eq.~\eqref{eq:Mggbox}~\cite{Tomalak:2014dja}. In general, however, there
may be momentum dependence in the $\gamma$-hadron interaction, such as
for the $\gamma N \Delta$ vertex (see Sec.~\ref{ss.Delta} below).
In fact, the momentum dependence in a transition vertex function allows
one to construct different parametrizations of that vertex function, for
example, by using the Dirac equation, that are equivalent on-shell but
differ off-shell.  The additional momentum-dependence associated with
this freedom will affect one-loop integrals because the intermediate
hadronic states are not on-shell.  This ambiguity is not present in the
dispersive method, for which all the intermediate states are on-shell.
In the context of TPE, this means that for any momentum-dependent
interactions one should not expect agreement between the real parts of
the generalized form factors calculated from Eqs.~\eqref{eq:genFFdisp}
and those calculated using the loop integration method. We will quantify
these differences for the cases of the nucleon and $\Delta$ intermediate
states in Sec.~\ref{sec.results}.

%%%%%%%%%%%%%%%%%%%%%%%%%%%%%%%%%%%%%%%%%%%%%%%%%%%%%%%%%%%%%%%%%%%%%%%%
\subsubsection{Analytic method}
\label{ssec:analytic}

The analytic approach used in previous work~\cite{Blunden:2003sp,
Kondratyuk:2005kk, Blunden:2005ew, Kondratyuk:2007hc, Tjon:2007wx,
Nagata:2008uv, Tjon:2009hf, Graczyk:2013pca, Lorenz:2014yda,
Zhou:2014xka} relies on a parameterization of the transition form
factors as a sum and/or product of monopole form factors. The most basic
relation is
\be
  \frac{1}{q_i^2} \left(\frac{\Lambda_i^2}{\Lambda_i^2-q_i^2}\right)
= \frac{1}{q_i^2} - \frac{1}{q_i^2-\Lambda_i^2}\, ,
\label{eq:monopole}
\ee
which is to be applied at each photon--hadron vertex ($i=1,2$). More
complicated constructions are straightforward to generate by repeatedly
applying the feature that the product of any two monopoles is
proportional to their difference.  The general expression for an
amplitude with form factors will thus involve a sum of ``primitive''
integrals with different photon mass parameters $\Lambda_{1}$ and
$\Lambda_{2}$ for each of the two photon propagators, modified according
to Eq.~(\ref{eq:monopole}).  The primitive integrals may yield spurious
ultraviolet or infrared divergences, but these divergences will cancel
when taking the sum. We give details on these constructions in
Appendix~\ref{app:PV}.

By means of the PV reduction scheme, a one-loop integral for the
box-diagram amplitude can be written in terms of a set of scalar PV
functions $A_0$, $B_0$, $C_0$ and $D_0$, corresponding to one-, two-,
three-, and four-point functions.  This can be visualized as a
``pinching'' of the four various propagators in the box diagram due to
cancellations of the terms in the numerator with the propagator terms in
the denominator. The PV functions can be evaluated numerically, and
there are various computer programs to do this~\cite{Hahn:1998yk,
vanHameren:2010cp, Carrazza:2016gav, Denner:2016kdg}.

\begin{figure}[t]
\centering
\includegraphics[width=0.6\linewidth]{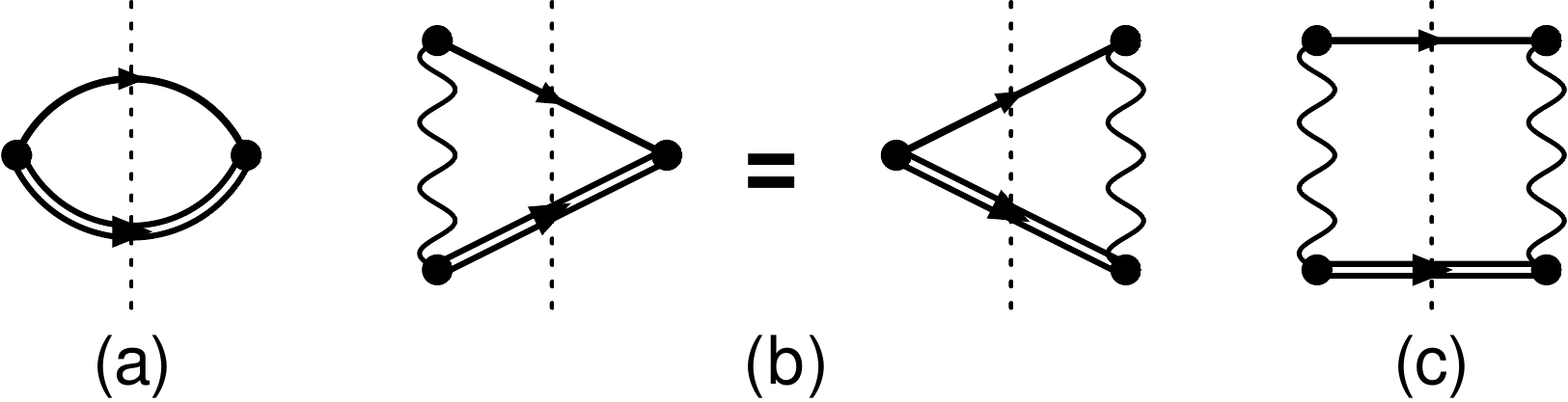}
\caption{Visualization of the Passarino-Veltman functions from
	the TPE amplitudes that have an imaginary part, for
	(a) $b_0(s)$;
	(b) $c_0(s;\Lambda)$;
	(c) $d_0(s;\Lambda_1,\Lambda_2)$,
	where $s$ is the total invariant mass of the system.	
	Here external legs have been amputated.
	The double line indicates a hadronic state of invariant
	mass $W$, and the dotted line indicates that the electron
	and hadronic states are on-shell.
	For elastic scattering, $c_0(s;\Lambda)$ is the same
	whichever photon propagator is pinched, and
	$d_0(s;\Lambda_2,\Lambda_1) = d_0(s;\Lambda_1,\Lambda_2)$.}
\label{fig:B0C0D0}
\end{figure}

In general the scalar PV functions are complex-valued. The imaginary
parts of the TPE amplitudes are contained entirely in these functions. 
For the box (and crossed-box) diagrams in elastic $ep$ scattering there
are only four of the PV functions that have imaginary parts.  These four
functions are the ones that arise in the $s$-channel box diagram with
the electron and intermediate hadronic states on-shell.  This is
illustrated in Fig.~\ref{fig:B0C0D0}.

Recall that an amplitude becomes imaginary when the intermediate state
particles become real, or on their mass shells. This is formalized by
the well-known Cutkosky cutting rules~\cite{Cutkosky:1958zz}. Namely, as
a consequence of unitarity, the imaginary part of a scattering amplitude
can be obtained by summing all possible cuttings of the corresponding
Feynman diagram, where a cut is across any two internal propagators
separating the external states from the rest of the diagram.  Cut
propagators are then put on-shell according to the rule
$1/(p^2-m^2+i0^+)\ \to\ -2\pi i\, \theta(p_0)\, \delta(p^2-m^2)$.

For elastic scattering, the two functions $C_0(s)$ arising when either
photon propagator is pinched are identical, so there are only three
distinct PV functions.  Other PV functions where the electron or
hadronic intermediate state (or both) are pinched have no imaginary
parts for $ep$ scattering, and the $u$-channel crossed-box diagram also
has no imaginary part. (Recall that the crossed-box amplitude can be
obtained by replacing $s \to u$, with an appropriate overall changed in
sign given in Eq.~(\ref{eq:genFFsymm}).) We will denote these three
functions as $B_0(s)$, $C_0(s;\Lambda_1)$, and
$D_0(s;\Lambda_1,\Lambda_2)$. The full expression for these functions is
\bea
\lefteqn{
\left\{ B_0(s),
	C_0(s;\Lambda_1^2),
	D_0(s;\Lambda_1^2,\Lambda_2^2)
\right\}
\equiv \frac{1}{i\pi^2}\,\int d^4q_1}	\nn\\
& & \hspace*{4cm}\times
\left\{ 1,
	\frac{1}{\left(q_1^2-\Lambda_1^2\right)},
	\frac{1}{\left(q_1^2-\Lambda_1^2\right)
		 \left(q_2^2-\Lambda_2^2\right)}
\right\}					\\
& & \hspace*{4cm}\times
\frac{1}{\left[ (k-q_1)^2-m_e^2+i 0^+ \right]
	 \left[ (p+q_1)^2-W^2+i 0^+   \right]}\, .\nn
\eea
In addition to the explicit dependence on $s$ and $\Lambda_i^2$,
there is also an implied dependence on $M$, $W$, and $Q^2$ that
is suppressed for clarity of notation (see Appendix~\ref{app:PV}
for details).

We define the imaginary parts of the PV functions by
$\{ b_0(s), c_0(s), d_0(s)\} \equiv
\{\Im B_0(s), \Im C_0(s), \Im D_0(s)\}$.
According to the Cutkosky rules, the imaginary parts correspond to
putting the electron and intermediate hadronic states on-shell,
$k_1^2 \equiv (k-q_1)^2 = m_e^2$ and $(p+q_1)^2 = W^2$.  Working
in the center-of-mass (CM) frame, we define the electron variables
\bea
k   &=& E_k (1; 0, 0, 1),		\nn\\
k'  &=& E_k (1; \sin\theta,0,\cos\theta),\\
k_1 &=& E_{k_1} (1; \sin\theta_{k_1}\cos\phi_{k_1},
		    \sin\theta_{k_1}\sin\phi_{k_1},
		    \cos\theta_{k_1}).	\nn
\eea
In this frame we have
\be
E_k     = \frac{s_M}{2\sqrt{s}}\, ,\quad
E_{k_1} = \frac{s_W}{2\sqrt{s}}\, ,\quad 
\ct     = 1 - \frac{Q^2}{2 E_k^2}\, ,
\label{eq:costheta}
\ee
with the shorthand notation
\be
s_M\equiv (s-M^2)\, ,\qquad s_W\equiv (s-W^2)\, .
\label{eq.sMsW}
\ee
In the physical region, the CM scattering angle $\theta$ satisfies
the constraint $-1\le \cos\theta\le 1$, requiring $s_M^2\ge s Q^2$.
However, the dispersive integral of Eq.~(\ref{eq:genFFdisp}) only
requires $s_M>0$, meaning there is an unphysical region of parameter
space where $\cos\theta<-1$, and $\sin\theta$ is purely imaginary.
Therefore, in the dispersive approach expressions for the imaginary
parts of the TPE amplitudes need to be analytically continued into
this unphysical region.

Recall that in terms of the electron energy in the laboratory
frame, $E$, the $ep$-invariant mass squared is $s = M^2 + 2 M E$.
After changing the integration variable from $q_1$ to $k_1$,
and using the on-shell conditions, we find after some algebra
the expressions
\bea
\lefteqn{
\left\{ b_0(s), c_0(s;\Lambda_1^2), d_0(s;\Lambda_1^2,\Lambda_2^2)
\right\}
\equiv \frac{s_W}{4 s}\theta(s_W)}		\nn\\
& &\times
\int d\Omega_{k_1}
\left\{ 1,
	\frac{-1}{\left(Q_1^2+\Lambda_1^2\right)},
	\frac{1}{\left(Q_1^2+\Lambda_1^2\right)
		 \left(Q_2^2+\Lambda_2^2\right)}
\right\}\, ,
\label{eq:bcddef}
\eea
where $Q_i^2=-q_i^2$ are the squared four-momenta of the virtual
photons ($i=1,2$), with
\be
\bal
Q_1^2 &= Q_0^2 \left(1-\ctk\right),\\
Q_2^2 &= Q_0^2 \left(1-\ct \ctk - \st\stk\cpk\right),
\eal
\label{eq:Q12defs}
\ee
and $Q_0^2=2 E_k E_{k_1} = s_M s_W/(2 s)$.

The $b_0$ integral in Eq.~(\ref{eq:bcddef}) is trivial. Through the use
of Eq.~(\ref{eq:Q12defs}), the other integrals can be brought into the
form
\be
J = \int d\Omega_{k_1}\,
    \frac{1}{(a_1-b_1\ctk)(a_2-b_2\ctk - c_2 \stk\cpk)}\, .
\label{eq:Jdef}
\ee
The integrand here has poles when $|b_2|>|a_2|$, which can arise in the
unphysical region when $Q_2^2$ becomes timelike. A simpler version of
this integral was considered by Mandelstam~\cite{Mandelstam:1958xc} for
the case where the target and scattering particles have equal masses.
The general expression has been given by Beenakker and
Denner~\cite{Beenakker:1988jr},
\bea
J &=& \frac{2\pi}{X}\,
      \log\left(\frac{a_1 a_2 - b_1 b_2 + X}{a_1 a_2 - b_1 b_2 - X}
          \right)\, ,				\label{eq:Jexpr}\\
{\rm with}\quad
X^2 &=& (a_1 a_2-b_1 b_2)^2 - (a_1^2-b_1^2)(a_2^2-b_2^2-c_2^2)\, .\nn
\eea

For the $c_0(s)$ function, we set
$a_1=1+\Lambda_1^2/Q_0^2$,
$b_1=1$, $a_2=1$, $b_2=0$, and $c_2=0$.
For $d_0(s)$, we set
$a_i=1+\Lambda_i^2/Q_0^2$, $b_1=1$, $b_2=\cos\theta$,
and $c_2=\sin\theta=\sqrt{1-b_2^2}$.
In the unphysical region, $\cos\theta < -1$, so that $\sin\theta$
is purely imaginary.  However, we note that the combination
$b_2^2 + c_2^2 = 1$, and therefore $0 \le X \le (a_1 a_2 - b_1 b_2)$,
independent of the value of $\cos\theta$. Thus Eq.~(\ref{eq:Jexpr}) for
$J$ is the proper analytic continuation of the integral for $d_0(s)$
into the unphysical region. Explicit expressions for $b_0(s)$,
$c_0(s;\Lambda)$ and $d_0(s;\Lambda_1,\Lambda_2)$, including the IR
limits $\Lambda \to \lambda$, are given in Appendix~\ref{app:PV}.

In previous work~\cite{Blunden:2003sp, Blunden:2005ew} the TPE
amplitudes were obtained by numerical evaluation of the PV functions
using the program LoopTools~\cite{Hahn:1998yk}.  The real parts were
used directly, and the imaginary parts were not needed.  Here, we have
constructed analytic expressions for the imaginary parts in terms of
elementary logarithms, thus allowing a completely analytic evaluation of
the imaginary parts of the TPE amplitudes.  The real parts are then
constructed from a numerical evaluation of the dispersion integrals of
Eq.~(\ref{eq:genFFdisp}).  The imaginary parts obtained here are, of
course, identical with those obtained numerically in the earlier
\mbox{work~\cite{Blunden:2003sp, Blunden:2005ew}}.  The real parts are
numerically identical for the elastic $F_1'$ and $F_2'$ TPE amplitudes,
while the $G_a'$ amplitude differs, but in a numerically insignificant
way.  For the inelastic $\Delta$ states there are significant
differences, especially as $\eps \to 1$.  These differences will be
discussed further in Sec.~\ref{sec.results}.

As an alternative to using the PV reduction method implemented in
FeynCalc~\cite{Mertig:1990an, Shtabovenko:2016sxi}, one can work
entirely with on-shell quantities.  Using the on-shell conditions, we
find that the TPE amplitudes are sums of integrals of the general form
\be
I = \frac{s_W}{4 s} \int\, d\Omega_{k_1}\ 
\frac{f\left(Q_1^2,Q_2^2\right)}
     {\left(Q_1^2+\lambda^2\right)\left(Q_2^2+\lambda^2\right)}
\frac{\Lambda_1^2}{Q_1^2+\Lambda_1^2}
\frac{\Lambda_2^2}{Q_2^2+\Lambda_2^2}\, ,
\label{eq:Iintegral}
\ee
where $f(Q_1^2,Q_2^2)$ is a polynomial function of combined degree $N$
in $Q_1^2$ and $Q_2^2$,
\be
f(Q_1^2,Q_2^2)
= \sum_{i=0}^N\, \sum_{j=0}^{N-i}\ f_{ij}\,Q_1^{2i}\,Q_2^{2j}\, .
\label{eq:fij}
\ee
The coefficients $f_{ij}$ are functions of $s$, $W$, and $Q^2$, and
satisfy $f_{ij}=f_{ji}$ for elastic scattering due to the symmetry under
$Q_1^2 \leftrightarrow Q_2^2$. Thus we can write
\be
I = \sum_{i=0}^N\, \sum_{j=0}^{N-i}\ f_{ij} I_{ij}\, ,
\ee
with the ``primitive'' integrals $I_{ij}$ defined as
\be
I_{ij} = \frac{s_W}{4 s} \int\, d\Omega_{k_1}\ 
\frac{Q_1^{2i} Q_2^{2j}}
     {\left(Q_1^2+\lambda^2\right)\left(Q_2^2+\lambda^2\right)}
\frac{\Lambda_1^2}{Q_1^2+\Lambda_1^2}
\frac{\Lambda_2^2}{Q_2^2+\Lambda_2^2}\, .
\label{eq:Iij}
\ee
For nucleon intermediate states we find $N=2$, indicating that monopole
form factors are sufficient to eliminate the UV divergences (there is
one power of $q_i$ at each photon--nucleon vertex in the $F_2$ term of
$\Gamma^\mu(q_i)$). For $\Delta$ intermediate states, however, we find
$N=3$, which implies that dipole form factors (or a product of
monopoles) are needed to eliminate the UV divergences, as there are up
to two powers of $q_i$ at each vertex in $\Gamma_{\gamma N \to
\Delta}^{\alpha\mu}$ (see Secs.~\ref{ss.N} and \ref{ss.Delta} below).
The integrals of Eq.~(\ref{eq:Iij}) up to $N=2$ are given in
Table~\ref{tab:Iij} of Appendix~\ref{app:PV}, and can easily be extended
to more complicated form factor constructions.

%%%%%%%%%%%%%%%%%%%%%%%%%%%%%%%%%%%%%%%%%%%%%%%%%%%%%%%%%%%%%%%%%%%%%%%%
\subsubsection{Numerical method}
\label{ssec:numerical}

In analogy with Eq.~(\ref{eq:Iintegral}), the TPE amplitudes of
interest have the general form
\be
\frac{s_W}{4 s} \int\, d\Omega_{k_1}\ 
\frac{f\left(Q_1^2,Q_2^2\right) G_1(Q_1^2)\, G_2(Q_2^2)}
     {\left(Q_1^2+\lambda^2\right)
\left(Q_2^2+\lambda^2\right)}\, ,
\label{eq:ampgen}
\ee
where $f(Q_1^2,Q_2^2)$ is a polynomial function of combined degree~2~(3)
in $Q_{1,2}^2$ for $N~(\Delta)$ intermediate states, and $G_i(Q_i^2)$
are form factors that are real-valued and finite for all spacelike
values of $Q_i^2$.  For elastic scattering, the total integral is
symmetric under the interchange $Q_1^2 \leftrightarrow Q_2^2$. For the
nucleon intermediate state, it is convenient to bring the IR
subtractions of Eq.~(\ref{eq:IRsubtraction}) into this integral. This is
consistent with the Maximon and Tjon IR regularization scheme whereby
the numerator of Eq.~(\ref{eq:ampgen}) vanishes whenever $Q_i^2\to 0$.
It also vanishes for excited states under these conditions. Therefore we
could actually set $\lambda\to 0$ without encountering any singularities
in the integrals.  This is unlike the analytic expressions of the
previous section, where only the sum of individual IR-divergent
expressions is independent of $\lambda$.

In the physical region there are no singularities in the integrand of
Eq.~(\ref{eq:ampgen}), so evaluation of the integral is a
straightforward 2-dimensional numerical quadrature over the domain $-1
\leq \cos\theta_{k_1} \leq 1$ and $0 \leq \phi_{k_1} \leq 2\pi$,
following Eq.~(\ref{eq:Q12defs}). However, this approach fails in the
unphysical region. To get around this, Tomalak and Vanderhaeghen used a
contour integration method~\cite{Tomalak:2014sva}, and applied it to the
calculation of TPE amplitudes with monopole form factors. By summing the
residue at the poles enclosed by the contour they were able to obtain
algebraic expressions for the TPE amplitudes. These expressions are
equivalent to the ones we obtained in the previous section using
algebraic expressions for the PV functions. In this section we will
follow this method, with modifications, to implement a numerical contour
integration of Eq.~(\ref{eq:ampgen}) that allows for a more general
parametrization of the transition form factors than a sum and/or product
of monopoles.

In the complex $Q_i^2$ plane, we define a timelike half with $\Re Q_i^2
< 0$, and spacelike half with $\Re Q_i^2 > 0$. In general, the allowed
form factors can have poles in $Q_i^2$ anywhere in the timelike half of
the complex plane. With certain restrictions, which we will state
explicitly, the form factors can have poles in the spacelike half as
well.

Without providing a rigorous mathematical proof, we can nonetheless
specify certain restrictions on the type of allowed form factors.
Namely, they should have a simple functional form in $Q_i^2$, such as
exponentials, polynomials, or inverse polynomials, that can be
analytically continued to the complex plane.  There should be no branch
cuts, and any poles should either lie along the negative, real axis (at
timelike $Q_i^2$), or occur in complex conjugate pairs. This is the case
for a commonly used form factor parametrization [see
Eq.~(\ref{eq:Gform})] in terms of a ratio of
polynomials~\cite{Kelly:2004hm,Arrington:2007ux,Venkat:2010by}.

The area of integration in Eq.~(\ref{eq:ampgen}) can be visualized as an
integral over the photon virtual momenta $Q_1^2$ and $Q_2^2$ of
Eq.~(\ref{eq:Q12defs}), which form a symmetric ellipse in $Q_1^2$
vs.~$Q_2^2$, centered at $\{Q_0^2,Q_0^2\}$~\cite{Gorchtein:2008dy,
Tomalak:2014sva}.  The boundary of the ellipse is defined by
$\cos\phi_{k_1}=1$.  Following Ref.~\cite{Tomalak:2014sva}, we make a
change of variables to elliptic coordinates,
\be
\int d\Omega_{k_1} \to 2\int_0^1 d\alpha \int_0^{2\pi} d\theta_{k_1}\, .
\ee
The contours of constant $\alpha$ represent concentric ellipses with
radial parameter $r=\sqrt{1-\alpha^2}$.  From Eq.~(\ref{eq:Q12defs}),
in elliptic coordinates we have
\be
\bal
Q_1^2 &= Q_0^2 \left(1-r\ctk\right)\, , \\
Q_2^2 &= Q_0^2 \left(1-r\cos\theta \ctk - r \sin\theta\stk\right)\, .
\label{eq:Q12ell}
\eal
\ee
In the physical region the integral over $\theta_{k_1}$ can be rewritten
as a contour integral on the unit circle $z = \exp(i\theta_{k_1})$, with
$Q_i^2(z)$ regarded as functions of $z$ (and $r$) using
\be
\cos\theta_{k_1}=\frac{1}{2}\left(z+\frac{1}{z}\right)\, ,\quad 
\sin\theta_{k_1}=\frac{1}{2 i}\left(z-\frac{1}{z}\right)\, ,
\ee
and
\be
\int_0^{2\pi} d\theta_{k_1} \to \oint \frac{dz}{i z}\, .
\ee
In anticipation of extending this formalism to the unphysical region,
Eq.~(\ref{eq:Q12ell}) can be simplified into the form
\be
\bal
Q_1^2(z)
&= Q_0^2 \left[1-\frac{r}{2}\left(z+\frac{1}{z}\right)
	 \right]\, , 		\\
Q_2^2(z)
&= Q_0^2 \left[1-\frac{r}{2}\left(\frac{z}{\beta}+\frac{\beta}{z}\right)
	 \right]\, ,
\eal
\label{eq:Q12z}
\ee
with
\be
\beta \equiv
   \left\{ \begin{array}{l c l}
      e^{i \theta} & \mbox{ for} & -1\le \cos\theta \le 1 \, ,\\
      \cos\theta-\sqrt{\cos^2\theta-1} & \mbox{ for} & \cos\theta<-1\, .
    \end{array}\right.
\ee
Recall that $\cos\theta$ is given in Eq.~(\ref{eq:costheta}) in terms of
$s$ and $Q^2$. Expressed in this way, it is clear that $Q_1^2(z) =
Q_2^2(\beta z)$. Because the integrands of the full TPE amplitudes are
symmetric under the interchange $Q_1^2 \leftrightarrow Q_2^2$, this
means that for every value $z_1$ associated with the poles in $Q_1^2$ in
the complex $z$ plane, there is a corresponding pole at $z_2=\beta z_1$.
In addition, for every pole in $Q_1^2$, both $z_{\rm 1i}$ and $z_{\rm
1o}=1/z_{\rm 1i}$ are poles in the complex $z$ plane, where $|z_{\rm
1i}|\le 1$ lies inside the unit circle, and $|z_{\rm 1o}|\ge 1$ lies
outside the unit circle.
The values $z_{\rm 1i}=z_{\rm 1o}=1$ are the IR-divergent poles, where
$Q_1^2=0$ for $r=1$.  These could be regulated by introducing the
$\lambda$ photon mass parameter, but as the integrand vanishes in our
regularization scheme when $Q_1^2=0$, there is no IR divergence, and one
can set $\lambda=0$.  Analogously, for $Q_2^2$, both $z_{\rm 2i}$ and
$z_{\rm 2o}=\beta^2/z_{\rm 2i}$ are poles inside and outside a circle of
radius $|\beta|$, respectively. The values $z_{\rm 2i}=z_{\rm 2o}=\beta$
are the IR-divergent poles, where $Q_2^2=0$ for $r=1$. These results are
valid in either the physical region, where $\beta=\exp(i\theta)$, or the
unphysical region, where $\beta < -1$.

\begin{figure}[tb]
\centering
\begin{minipage}{0.5\textwidth}
\centering
\includegraphics[width=\linewidth]{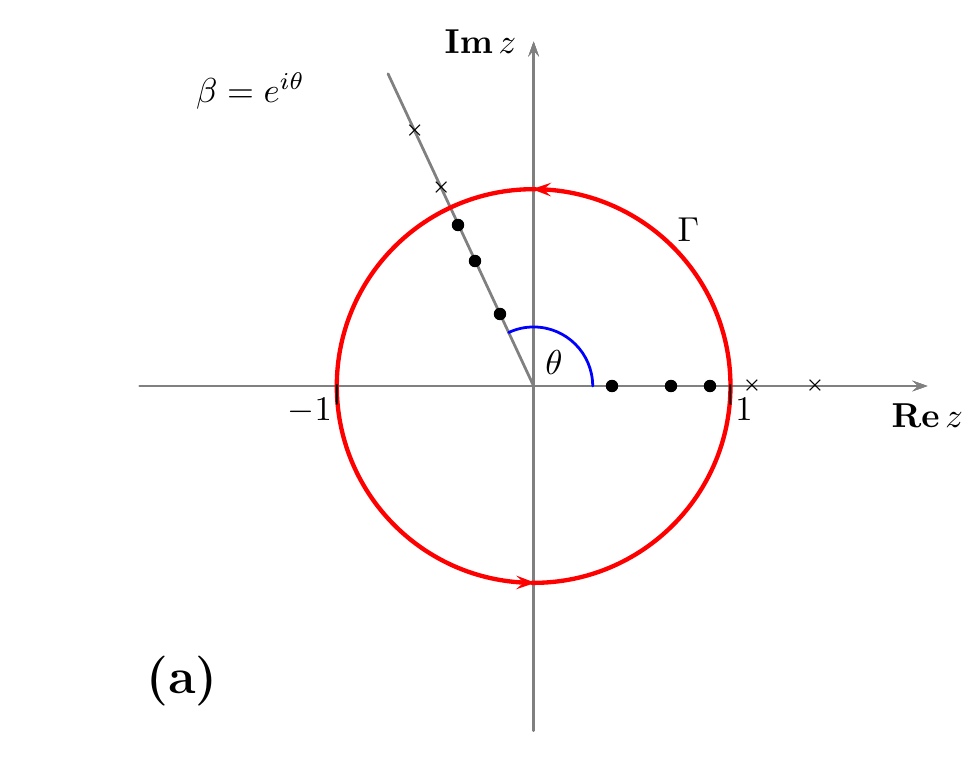}
\end{minipage}%
\begin{minipage}{0.5\textwidth}
\centering
\vspace*{-1ex}
\includegraphics[width=\linewidth]{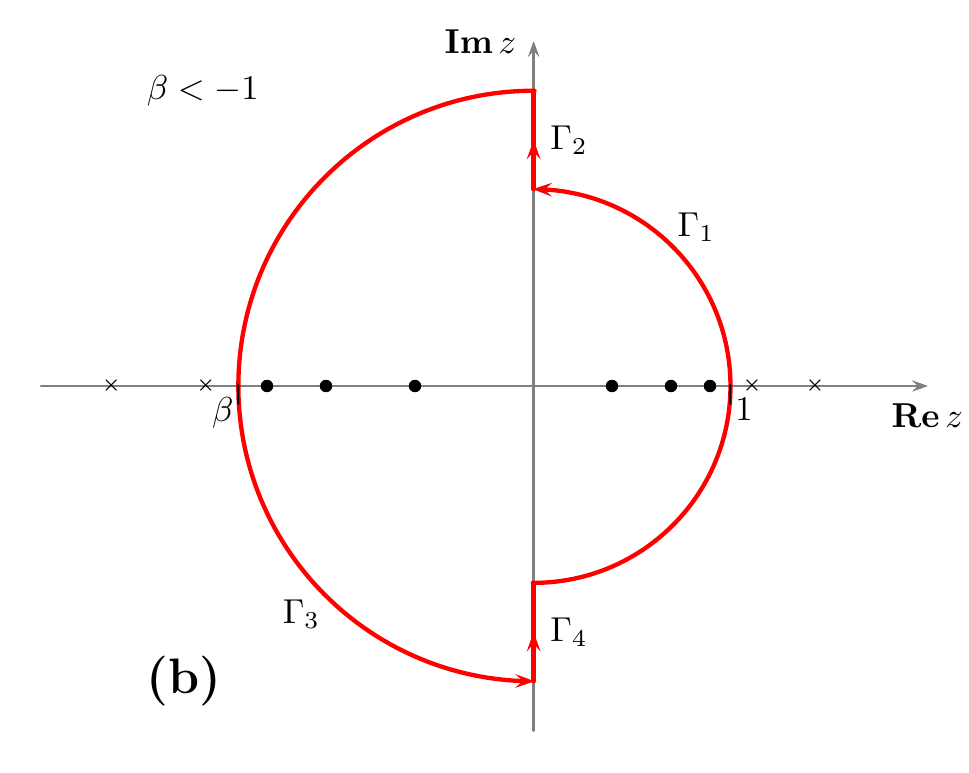}
\end{minipage}
\caption{(a):
	The contour around the unit circle in the physical region,
	$-1\le\cos\theta\le 1$. For every pole $z_{\rm i}$ inside the
	contour (denoted by a dot), there is a corresponding pole at
	$z_{\rm o}=1/z_{\rm i}$ lying outside the contour (denoted
	by a cross).  For monopole form factors, interior poles in
	$Q_1^2$ appear along the real $z$-axis between 0 and 1,
	and corresponding poles in $Q_2^2$ along the line at angle
	$\theta$.
	(b):
	One possible choice of contour in the unphysical region,
	$\cos\theta<1$, which we denote as $\Gamma_{\rm CD}$.
	For monopole form factors, interior poles
	in $Q_1^2$ lie along the real $z$-axis between 0 and 1, and
	corresponding poles in $Q_2^2$ lie between $\beta$ and 0.}
\label{fig:contours1}
\end{figure}

As an illustration of these points, consider the monopole form factors
as given in Eq.~(\ref{eq:Iintegral}), for which there is a pole in
$Q_1^2$ along the negative real axis at $-\Lambda_1^2$.  This yields
poles $z_{\rm 1i}$ and $z_{\rm 1o}$ along the positive real axis in the
complex $z$ plane, with the interior pole $z_{\rm 1i}$ lying between 0
and 1. For the corresponding poles associated with $Q_2^2$, in the
physical region, with $\beta=\exp(i\theta)$, these lie along a line at
angle~$\theta$.  In the unphysical region, with $\beta < -1$, they lie
along the negative real axis, with $z_{\rm 2i}$ lying between $\beta$
and 0. A graphical representation of these results in the complex $z$
plane is shown in Fig.~\ref{fig:contours1}, with the dots representing
the ``inside'' points $z_{\rm 1i}$ and $z_{\rm 2i}$, and the crosses
representing the ``outside'' points $z_{\rm 1o}$ and $z_{\rm 2o}$.

For more general form factors, we restrict ourselves in the first
instance to those functions with poles in the timelike half of the
complex $Q_i^2$ plane.  In this case it can be shown that all poles in
$z_{\rm 1i}$ will be clustered around the positive real axis, with
$|z_{\rm 1i}| \le 1$ and $\Re z_{\rm 1i} > 0$.  Recall that for every
pole $z_1$ there is a corresponding pole $z_2 = \beta z_1$. In the
physical region, all poles in $z_{\rm 2i}$ will therefore be clustered
around the line at angle $\theta$, with $|z_{\rm 2i}| \leq 1$. In the
unphysical region, the $z_{\rm 2i}$ poles must therefore satisfy
$|z_{\rm 2i}| < |\beta|$ and $\Re z_{\rm 2i} < 0$.

By Cauchy's theorem, a closed contour integral is equivalent to summing
the residue at the poles of the integrand enclosed by the contour. In
continuing the integral from the physical to the unphysical region, the
contour $\Gamma$ must be deformed in such a way that no poles cross the
boundary defined by the contour.  There are many possible choices of an
appropriate contour in the unphysical region.  However, based on the
symmetry of the poles in $z_1$ and $z_2$, as discussed above, the closed
contour shown in Fig.~\ref{fig:contours1}(b) merits special
consideration.  It consists of two semicircles connected by two straight
line segments, which we denote as~$\Gamma_{\rm CD}$. It clearly
satisfies the criterion that no poles $z_{\rm 2i}$ or $z_{\rm 2o}$ cross
the boundary of the contour, provided that the form factors only have
poles in the timelike region of the $Q_i^2$ plane. Moreover, because of
the symmetry of the TPE amplitudes under the interchange of $Q_1^2$ and
$Q_2^2$, the line integrals along the contours $\Gamma_1$ and $\Gamma_3$
are equal, as are the line integrals along the contours $\Gamma_2$ and
$\Gamma_4$.
Furthermore, the real part of the integrand is symmetric between the
upper and lower half planes, while the imaginary part is antisymmetric
(hence the imaginary part integrates to 0 in any closed contour). Thus
one only needs to compute the real part of the line integral along the
contour $\Gamma_1$ in the upper half plane, plus the contribution
from~$\Gamma_2$,
\be
\oint_\Gamma \frac{dz}{i z} f(z)\ \longmapsto\
   4 \int_0^{\pi/2} dt\, \Re\Big\{f(z)\big|_{z=e^{i t}}\Big\}
 + 2 \int_1^{-\beta} dt\, \Re\left\{\frac{f(z)}{z}\bigg|_{z=i t}\right\}\, .
\label{eq:contour1}
\ee

Form factor parametrizations $G(Q_i^2)$ are driven almost exclusively
by fits to data at spacelike values of $Q_i^2$.  Aside from requiring
no poles along the positive real axis, typically there is little or no
consideration given to the location of poles in the complex $Q_i^2$
plane.  Therefore, the appearance of poles in the spacelike half of
$Q_i^2$ should be regarded as a nuisance rather than a reflection of
any underlying physics.  Nevertheless, there are form factor
parametrizations in the literature that do have such poles.
For example, the ratio of polynomials,
\be
G(Q_i^2) = \frac{\sum_{j=0}^N a_j Q_i^{2 j}}
		{\sum_{k=0}^{N+2} b_k Q_i^{2k}}\, ,
\label{eq:Gform}
\ee
is commonly used~\cite{Kelly:2004hm,Arrington:2007ux, Venkat:2010by}.
Requiring all $b_k \ge 0$ is sufficient to eliminate zeros in the
denominator for positive real $Q_i^2$, but zeros can still occur in
complex conjugate pairs with $\Re Q_i^2>0$.

\begin{figure}[tb]
\centering
\begin{minipage}{0.5\textwidth}
\centering
\includegraphics[width=1.0\linewidth]{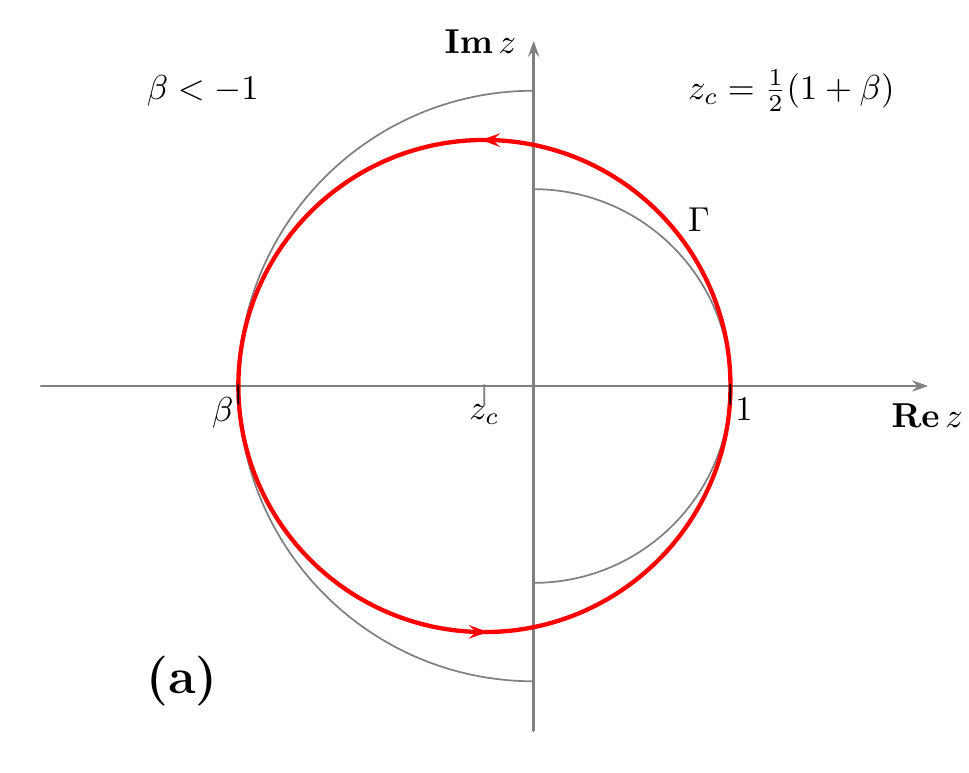}
\end{minipage}%
\begin{minipage}{0.5\textwidth}
\centering
\vspace*{-1ex}
\includegraphics[width=1.0\linewidth]{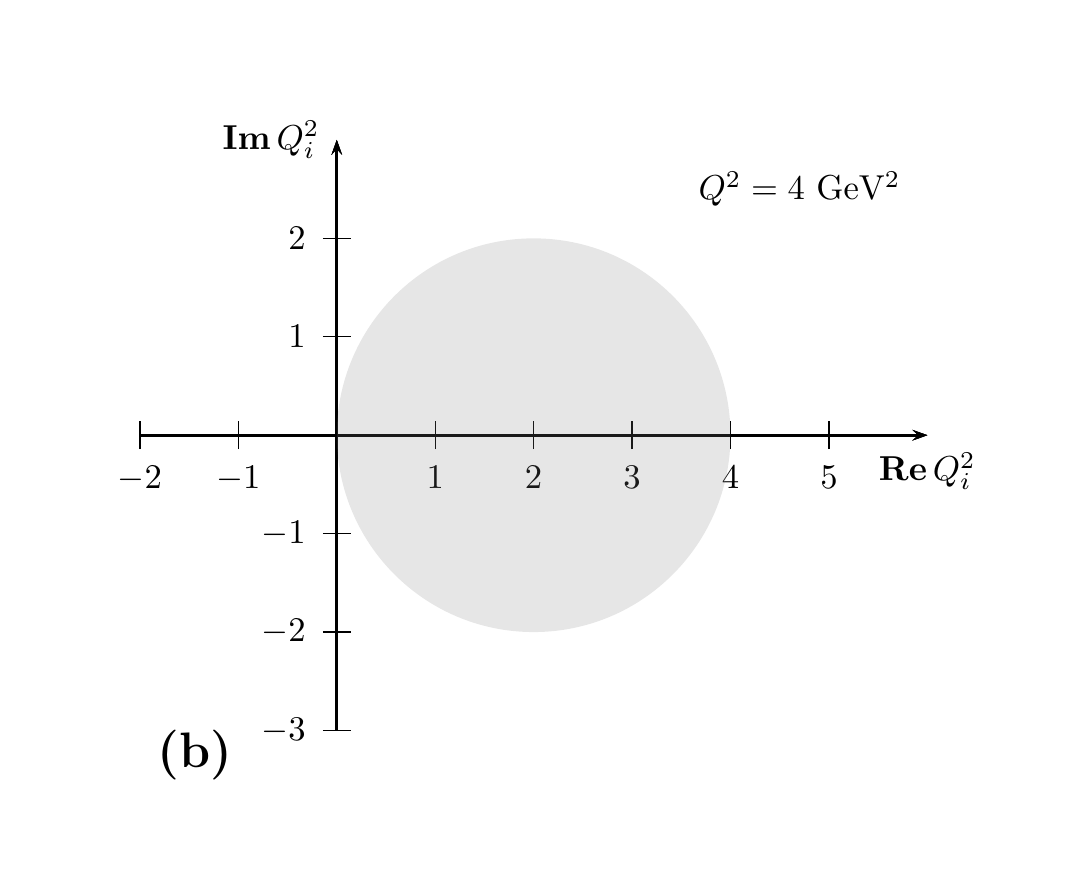}
\end{minipage}
\caption{(a):
	The circular contour $\Gamma$ (shown in red) in the
	unphysical region is valid for form factors having
	poles $Q_p^2$ satisfying the condition $R > \cos\eta$,
	where $Q_p^2/Q^2 = R e^{i\eta}$.  For comparison,
	the light gray line is the contour $\Gamma_{\rm CD}$
	of Fig.~\ref{fig:contours1}(b).
	(b):
	Visualization of the condition on the poles of the form
	factors in the complex $Q_i^2$ plane, using $Q^2=4$~GeV$^2$
	as an example.  Any form factors with poles in the shaded
	region of a circle centered at $\{\half Q^2,0\}$ of radius
	$\half Q^2$ are not allowed.  This will always occur for a
	large enough value of $Q^2$, since the limit $Q^2 \to \infty$
	is the whole spacelike region.}
\label{fig:contours2}
\end{figure}

To handle such cases we modify the contour of integration in the
unphysical region, $\Gamma_{\rm CD}$. We choose instead a circular
contour centered at $z_c=\half(1+\beta)$ of radius $\half(1-\beta)$, as
illustrated in Fig.~\ref{fig:contours2}(a).  The new contour is shown in
red, while the $\Gamma_{\rm CD}$ contour is in light gray. As we require
no poles to cross the boundary defined by the new contour, there are two
possible failures.
Firstly, the poles $z_{\rm 1i}$ are no longer restricted to the positive
half plane, and can lie anywhere inside a circle of unit radius.  This
gives the possibility of a pole $z_{\rm 1o}$ crossing into the interior
of the new contour, since the only restriction is $|z_{\rm 1o}| > 1$.
Secondly, the corresponding poles $z_{\rm 2i}$ are no longer restricted
to the negative half plane, and can lie outside the new contour, since
the only restriction here is $|z_{\rm 2i}| < |\beta|$. By careful
analysis of the location of the poles for arbitrary values of $\beta$
and $Q^2$, we have derived the following condition:
\begin{quote}
{\em The validity of the contour defined by a circle of radius
$\half(1-\beta)$, centered at $z_c=\half(1+\beta)$,
is that the poles $Q_p^2$ in the form factor $G(Q_i^2)$
must satisfy the condition}
\end{quote}
\be
\left| 2\frac{Q_p^2}{Q^2} - 1\right| > 1\, .
\label{eq:Qp2cond1}
\ee
The condition (\ref{eq:Qp2cond1}) is equivalent to
\be
R > \cos\eta,\quad {\rm with }\quad \frac{Q_p^2}{Q^2} = R\, e^{i\eta}\, .
\label{eq:Qp2cond2}
\ee
This condition is satisfied for all $\cos\eta < 0$ (poles in the
timelike half of $Q_i^2$), and for all $R > 1$.  We can visualize the
condition using a circle of radius $\half Q^2$, centered at $\{\half
Q^2,0\}$, in the complex $Q_i^2$ plane, where the poles $Q_p^2$ in the
form factors must lie outside this circle. An example using
$Q^2=4$~GeV$^2$ is shown in Fig.~\ref{fig:contours2}(b).
If a form factor has poles in the spacelike half ($\cos\eta > 0$), then
the maximum $Q^2$ for which we can use the new contour is given by
$Q_{\rm max}^2 = |Q_p|^2/\cos\eta$.

%%%%%%%%%%%%%%%%%%%%%%%%%%%%%%%%%%%%%%%%%%%%%%%%%%%%%%%%%%%%%%%%%%%%%%%%
%%%%%%%%%%%%%%%%%%%%%%%%%%%%%%%%%%%%%%%%%%%%%%%%%%%%%%%%%%%%%%%%%%%%%%%%
\section{Impact of two-photon exchange}
\label{sec.results}

In this section we present the results of the numerical calculations
of the TPE contributions to elastic $ep$ scattering within the
dispersive approach, and discuss the differences with the traditional
loop calculations with off-shell intermediate states.  Including the
nucleon and $\Delta$ intermediate state contributions, we compare the
results with measurements of $e^+ p$ to $e^- p$ cross section ratios
from the recent CLAS~\cite{Rimal:2016toz},
VEPP-3~\cite{Rachek:2014fam, Nikolenko:2014uda} and
OLYMPUS \cite{Henderson:2016dea} experiments, as well as with
polarization observables sensitive to effects beyond the Born
approximation~\cite{Meziane:2010xc}.

% ......................................................................
\subsection{Nucleon intermediate state}
\label{ss.N}

\begin{figure}[t]
\includegraphics[width=0.49\textwidth]{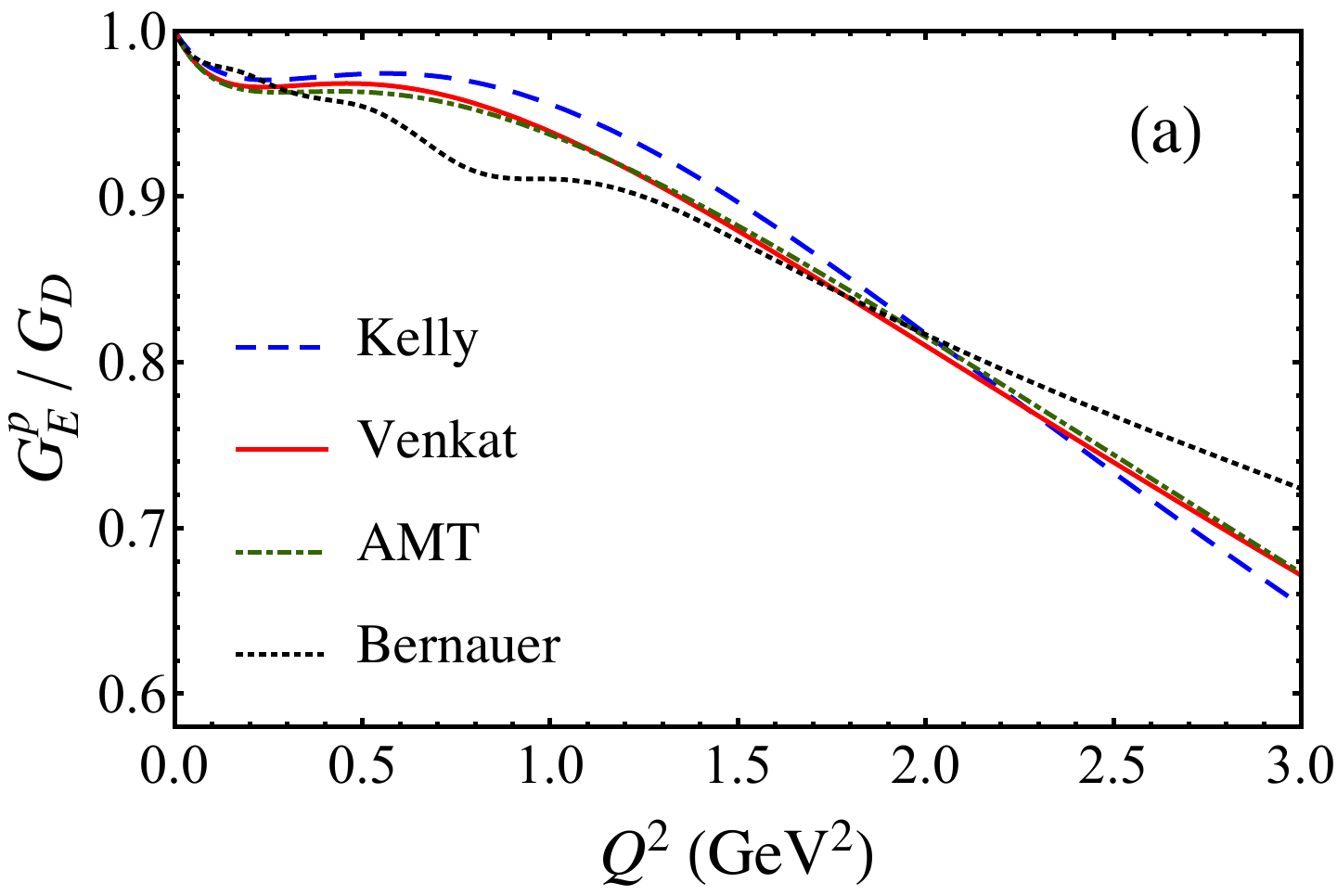}\hspace*{0.2cm}
\includegraphics[width=0.5\textwidth]{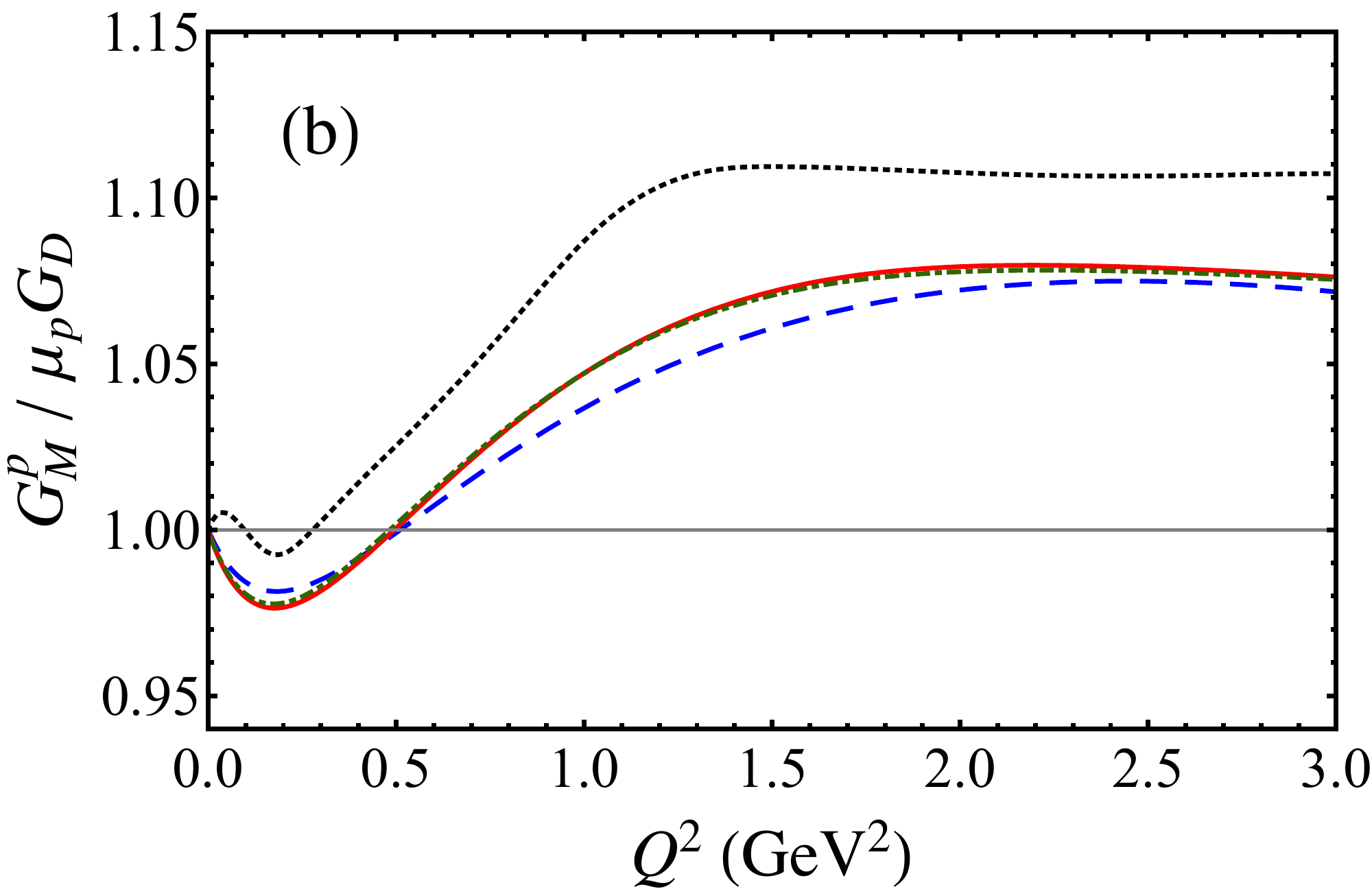}
\caption{$Q^2$ dependence of the proton
	(a) electric and (b) magnetic form factors,
	scaled by the dipole form factor, $G_D(Q^2)$
	[see Eq.~(\ref{eq:Gdipole})], for the
	Kelly~\cite{Kelly:2002if} (dashed blue curves),
	Venkat {\it et al.}~\cite{Venkat:2010by} (solid red curves),
	AMT~\cite{Arrington:2007ux} (dot-dashed green curves), and
	Bernauer {\it et al.} \cite{Bernauer:2010wm} (dotted black
	curves) parametrizations.}
\label{fig:Nff}
\end{figure}

For the contributions to the box diagram from nucleon intermediate
states, the inputs into the calculation are the proton elastic electric
and magnetic form factors, $G_E^p$ and $G_M^p$. In the numerical
calculations in this analysis we use the recent fit by Venkat {\it et
al.}~\cite{Venkat:2010by}, which has the form of Eq.~(\ref{eq:Gform})
with $N=3$ for both $\gep$ and $\gmp$.
As Fig.~\ref{fig:Nff} illustrates, this fit gives similar results to
other parametrizations, such as the ones by Kelly~\cite{Kelly:2004hm}
and Arrington {\it et al.} (AMT)~\cite{Arrington:2007ux}. In contrast,
the recent parametrization by Bernauer {\it et al.}
\cite{Bernauer:2010wm}, which is based on a spline with 8 knots,
displays distinctive wiggles at low $Q^2$ for both the electric and
magnetic form factors.  In Ref.~\cite{Bernauer:2010wm} a number of other
functional forms were considered in fits to the world's elastic
electron--proton scattering data.

In the application of the numerical contour integration method to the
calculation of the box diagram, care must be taken to ensure that all
relevant poles are included inside the contour, following the condition
given in Eq.~(\ref{eq:Qp2cond2}).  For some commonly used fits in the
literature, such as those by Bosted \cite{Bosted:1994tm} or Brash
\cite{Brash:2001qq} (in which the denominators of $\gep$ and $\gmp$ are
fifth-order polynomials in $\sqrt{Q^2}$), or the Kelly parametrization
\cite{Kelly:2002if} (third-order polynomials in $Q^2$), there are no
upper limits on $Q^2$, since all poles occur in the timelike region. On
the other hand, for the AMT \cite{Arrington:2007ux} and Venkat {\it et
al.} \cite{Venkat:2010by} parametrizations, which involve denominators
with 5th-order polynomials in $Q^2$, poles in the spacelike region limit
the range of applicability to $Q^2 < 4.5$~GeV$^2$.  For the purposes of
the data analysis in this paper, all the above parametrizations are
valid; however, for future applications at higher $Q^2$ values care must
be taken to ensure the chosen parametrization has a suitable pole
structure.

\begin{figure}[t]
\includegraphics[width=0.5\textwidth]{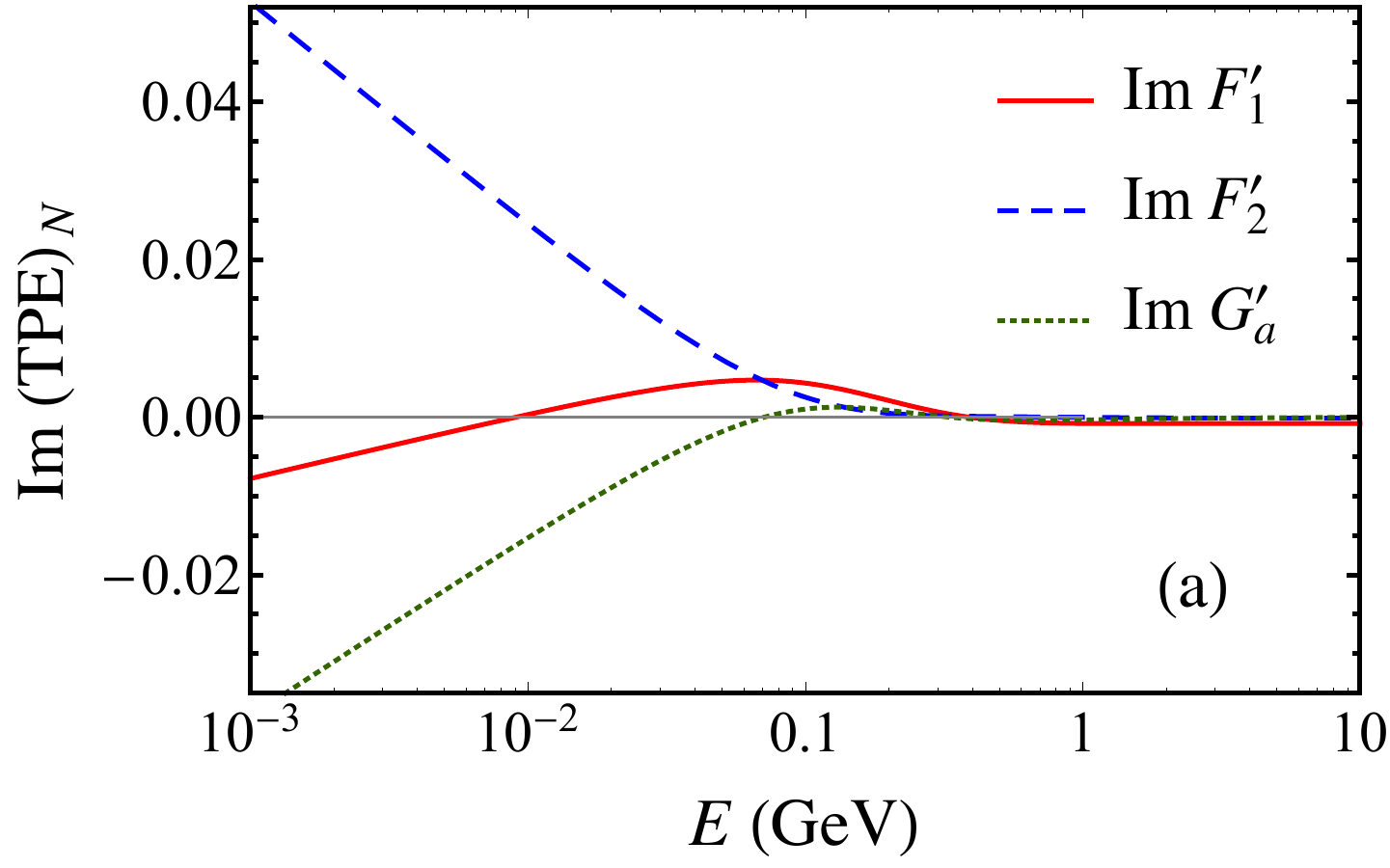}
\hspace*{0.1cm}
\includegraphics[width=0.48\textwidth]{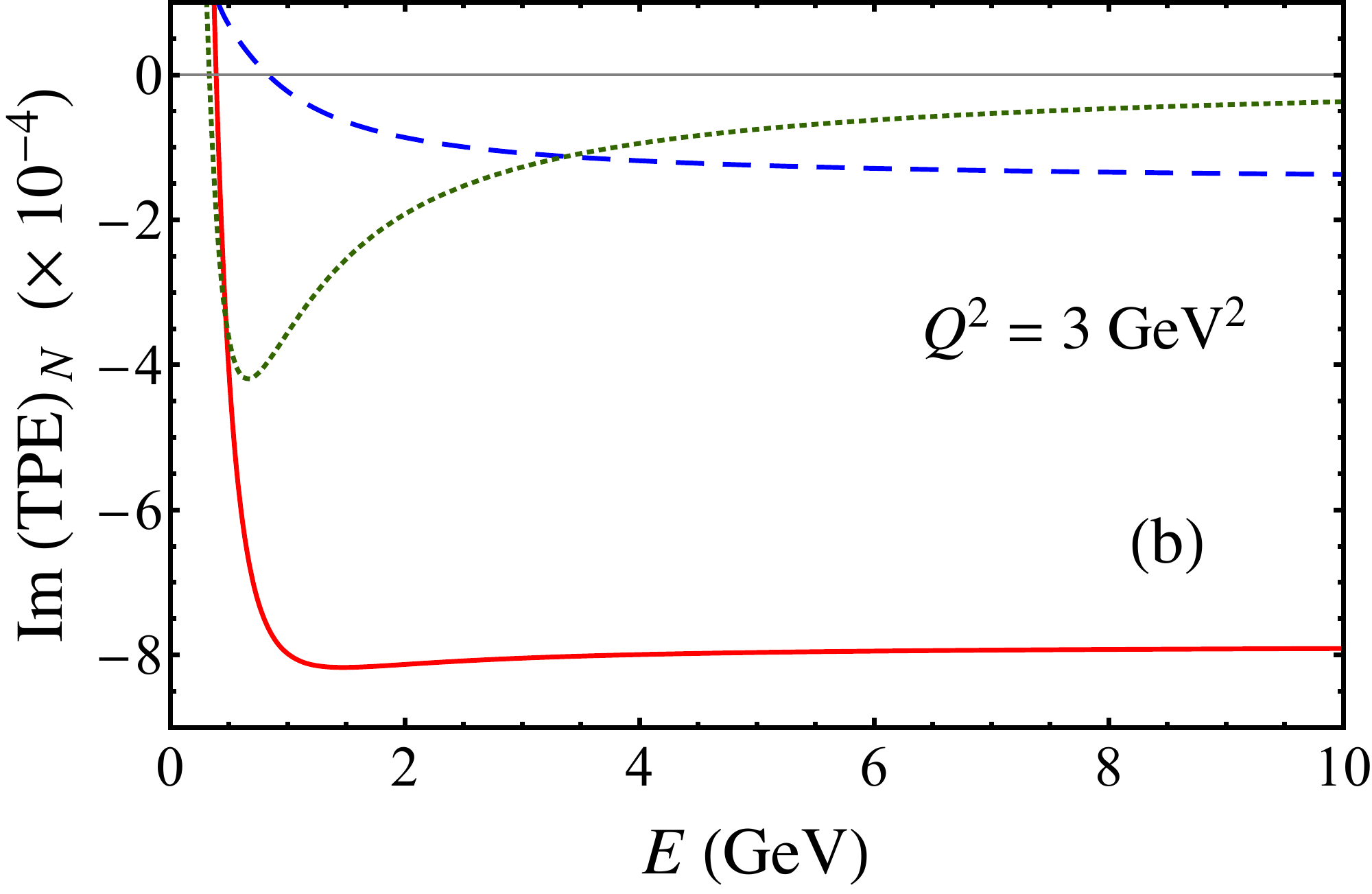}
\caption{Imaginary parts of the TPE contributions from nucleon
	intermediate states to the generalized
	$F_1'$ (solid red curves),
	$F_2'$ (dashed blue curves) and
	$G_a'$ (dotted green curves) form factors as a
	function of energy $E$,	for fixed $Q^2=3$~GeV$^2$:
	(a) illustrating the low energy behavior, $E \to 0$, and
	(b) showing the asymptotic behavior as large $E$.}
\label{fig:NffIm}
\end{figure}

The TPE contributions from nucleon intermediate states to the imaginary
parts of the generalized $F_1'$, $F_2'$ and $G_a'$ form factors are
illustrated in Fig.~\ref{fig:NffIm} as a function of energy $E$ (in the
lab frame), at a representative $Q^2$ value of 3~GeV$^2$. (The results
at other $Q^2$ values are qualitatively similar.) In the low energy
region, $E \lesssim 0.1$~GeV, in Fig.~\ref{fig:NffIm}(a) the TPE
amplitudes display a logarithmic divergence in $E$. Although the
imaginary parts diverge, the dispersive integrals (\ref{eq:genFFdisp})
for the real parts remain finite. To accommodate this in our numerical
analysis, we fit the low-$E$ expressions for $\Im\fonep$, $\Im\ftwop$
and $\Im\gap$ to functions of the form
\be
a + b\log{E}\, ,
\ee
with the parameters $a$ and $b$ determined by a least-squares fit
for $E$ between $0.001$ and $0.01$~GeV.
At high energies, the imaginary parts of the $F_1'$ and $F_2'$
form factors become constant, as is apparent for $E \gtrsim 6$~GeV
from Fig.~\ref{fig:NffIm}(b), where the magnified scale more
clearly illustrates the asymptotic behavior.  The imaginary part of
the axial $G_a'$ form factor falls off as $1/E$ for $E \to \infty$.
This high energy behavior is sufficient to ensure the convergence
of the dispersive integrals of Eq.~(\ref{eq:genFFdisp}).
Note that the corrections to the form factors are relative to
the Maximon-Tjon result for the infrared part of the TPE
\cite{Maximon:2000hm}.

For our numerical calculation, we compute the imaginary part of the TPE
amplitudes on a logarithmic grid of 51 points in $E$, ranging up to
100~GeV. In the unphysical region, the two-dimensional numerical contour
integral is computed using either of the contours discussed in the
previous section, as appropriate to the poles of the form factors. In
the physical region, the numerical contour integral is on the unit
circle (although a direct numerical integration of (\ref{eq:ampgen})
using Eq.~(\ref{eq:Q12defs}) can also be used). We then interpolate
between the grid points with a spline fit to obtain a continuous
function of $E$.  To obtain the real part on a grid of 20 equally spaced
points in $\eps$, the dispersion integral is evaluated using the $\log
E$ fit for $E<0.01$~GeV, the spline fit for $0.01<E<100$~GeV, and an
extrapolation beyond 100~GeV using the known asymptotic behavior.  This
approach can be tested against the analytic results obtained in the
previous section, as well as the known analytic results for $e \mu$
scattering.

\begin{figure}[t]
\includegraphics[width=0.5\textwidth]{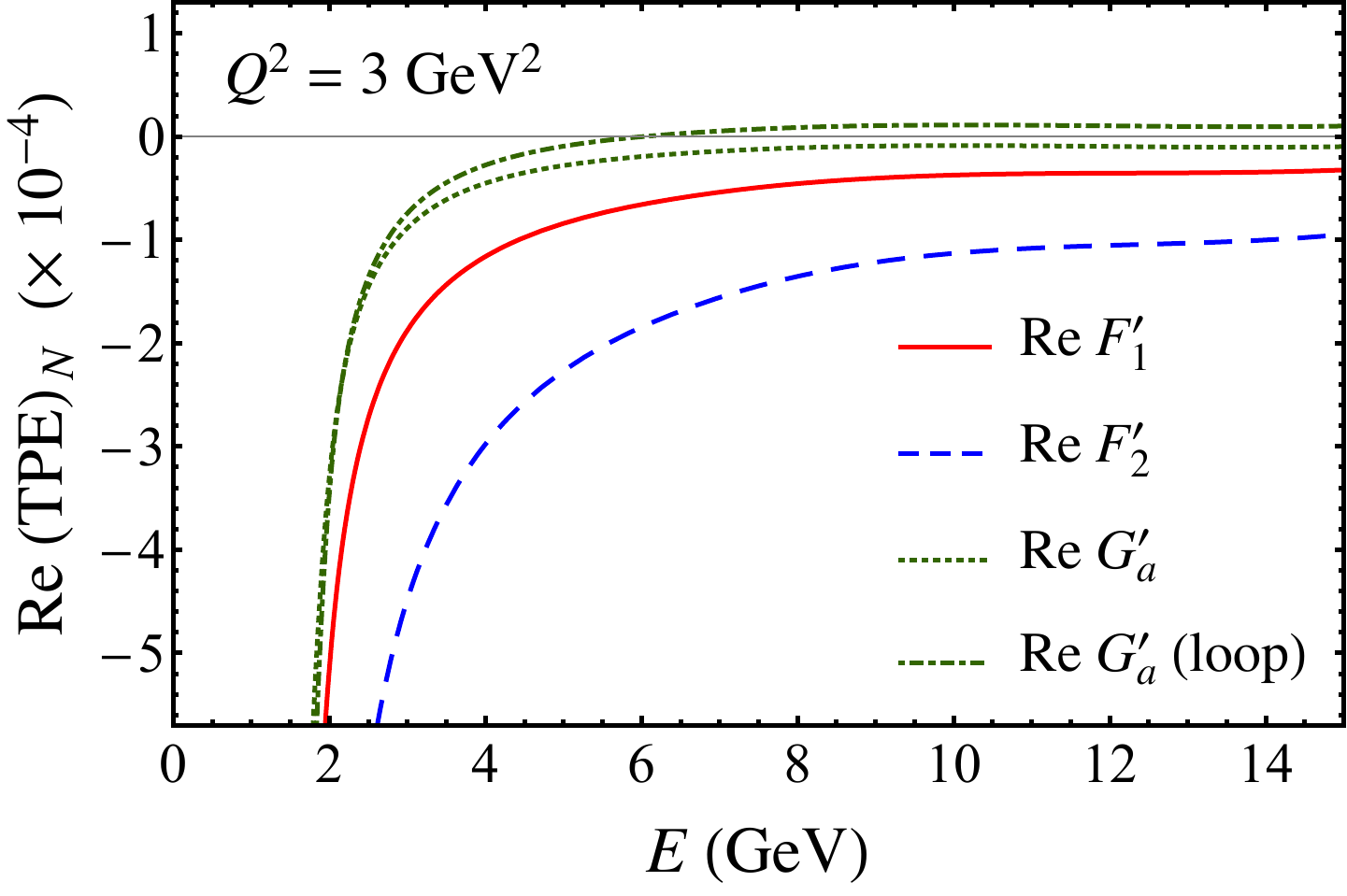}
\caption{Dispersive results for the real parts of the TPE contributions
	from nucleon intermediate states to the generalized
	$F_1'$ (solid red curve),
	$F_2'$ (dashed blue curve) and
	$G_a'$ (dotted green curve)
	form factors as a function of energy $E$,
	for fixed $Q^2=3$~GeV$^2$. The scattering threshold is at $E=1.97$~GeV.
	For $G_a'$ the dispersive results are compared with the
	direct (off-shell) loop calculation (dot-dashed green curve).\\}
\label{fig:NffRe}
\end{figure}

Interestingly, for the real part of the TPE amplitudes the dispersive
integral is dominated by contributions from the unphysical region, $E <
M(\tau+\sqrt{\tau(1+\tau)})$, which for $Q^2 = 3$~GeV$^2$ is $E =
1.97$~GeV. This results in the generally smoothly decaying functions for
$E \gtrsim 2$~GeV observed in Fig.~\ref{fig:NffRe}. The real parts of
each of the form factors are negative in the region illustrated, with
the $F_1'$ form factor having the largest magnitude, and the $G_a'$ form
factor the smallest magnitude.
Compared with the direct loop calculation in terms of the off-shell
nucleon intermediate states, differences arise for the Pauli $F_2$ form
factor term, which translates to a difference for the $G_a'$ form
factor.  The results for the $F_1'$ and $F_2'$ form factors are the same
for both calculations, as was observed previously in
Refs.~\cite{Borisyuk:2008es, Tomalak:2014sva}.  Numerically, the
differences are relatively small, however, as Fig.~\ref{fig:NffIm}
indicates, becoming notable only for $E \gtrsim 3$--4~GeV.

\begin{figure}[t]
\includegraphics[width=0.57\textwidth]{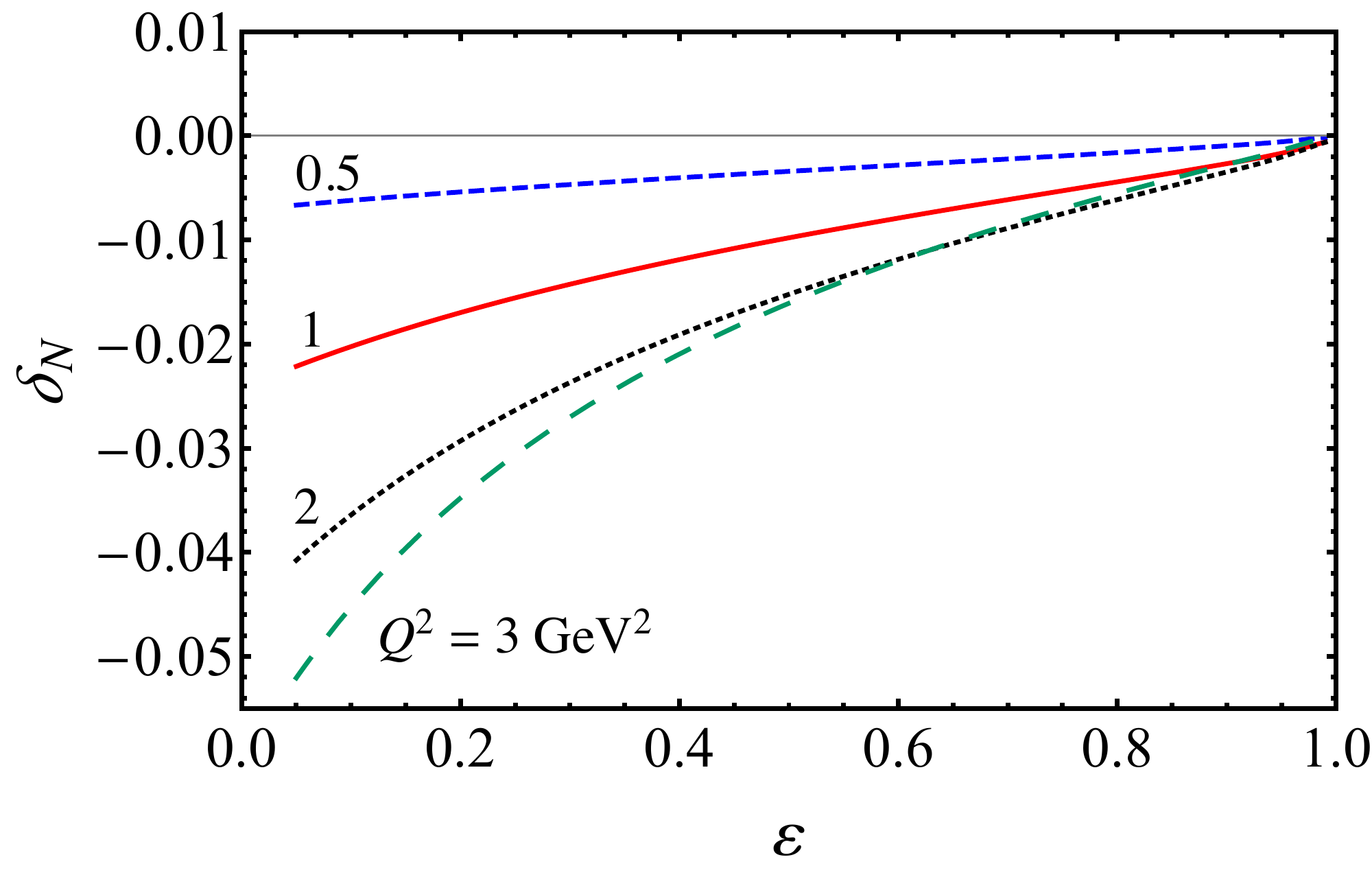}
\caption{Dispersive TPE correction to the cross section, $\delta_N$,
	from nucleon intermediate states as a function of $\eps$
	for fixed values of $Q^2$ (= 0.5, 1, 2 and 3~GeV$^2$).
	The corrections are relative to the Mo-Tsai infrared
	result \cite{Mo:1968cg}.}
\label{fig:deltaN}
\end{figure}

The total TPE correction (\ref{eq:delgg}) to the elastic cross section
from nucleon intermediate states relative to the Mo-Tsai infrared
prescription~\cite{Mo:1968cg}, denoted $\delta_N$, is shown in
Fig.~\ref{fig:deltaN} as a function of $\eps$ for $Q^2$ between
0.5~GeV$^2$ and 3~GeV$^2$. As found in previous loop calculations
\cite{Blunden:2003sp, Blunden:2005ew}, the corrections at these
kinematics are negative, and increase in magnitude with increasing
$Q^2$. For large $Q^2$ values we expect the reliability of the hadronic
calculation to deteriorate, but indications from earlier work
\cite{Blunden:2003sp, Blunden:2005ew} suggest that it remains sizeable.
In practice, since only the $G_a'$ form factor is affected, and its
magnitude is considerably smaller than that of $F_1'$ and $F_2'$, as
illustrated in Fig.~\ref{fig:NffRe}, the off-shell effects play a
relatively minor role in $\delta_N$, with the dispersive and loop
results almost indistinguishable.

% ......................................................................
\subsection{$\Delta$ intermediate state}
\label{ss.Delta}

The contribution to the TPE amplitude from intermediate states involving
the spin-\sfrac{3}{2}, isospin-\sfrac{3}{2} $\Delta$ baryons is computed
from the $\gamma^* N \to \Delta$ electromagnetic transition operator,
$\Gamma_{\gamma N \to \Delta}^{\alpha\mu}$. Usually this is expressed in
terms of three Jones-Scadron transition form factors, $G_M^*(Q^2)$,
$G_E^*(Q^2)$ and $G_C^*(Q^2)$, corresponding to magnetic, electric, and
Coulomb multipole excitations, respectively \cite{Jones:1972ky}.
Although the $\gamma^* N \to \Delta$ cross section is diagonal in these
functions, they are cumbersome to work with in the transition vertex
function, and other parametrizations have also been suggested in the
literature \cite{Pascalutsa:2006up, Kondratyuk:2005kk, Nagata:2008uv,
Pascalutsa:2004pk, Caia:2004pm}.  In this work we follow
Ref.~\cite{Kondratyuk:2005kk} and use the on-shell equivalent
parametrization of the $\gamma^* N \to \Delta$ vertex
\bea
\Gamma_{\gamma N \to \Delta}^{\alpha\mu}(p_\Delta,q)
&=& \frac{1}{2 M_\Delta^2}\sqrt\frac{2}{3}
\Big\{
  g_1(Q^2)
  \left[ g^{\alpha \mu} \slashed{q}\slashed{p}_\Delta
    - \slashed{q} \gamma^\alpha p_\Delta^\mu
    - \gamma^\alpha \gamma^\mu q\cdot p_\Delta
    + \slashed{p}_\Delta\,\gamma^\mu q^\alpha
  \right]                               \nn\\
& &
+\ g_2(Q^2)
  \left[ q^\alpha p_\Delta^\mu - g^{\alpha\mu} q\cdot p_\Delta
  \right]\nn\\
&&+\ \frac{g_3(Q^2)}{M_\Delta}
  \left[ q^2 \left( \gamma^\alpha p_\Delta^\mu
                  - g^{\alpha\mu} \slashed{p}_\Delta
             \right)
       + q^\mu \left( q^\alpha \slashed{p}_\Delta
                    - \gamma^\alpha q\cdot p_\Delta
               \right)
  \right]
\Big\} \gamma_5\,,
\label{eq:gND}
\eea
where $p_\Delta$ and $q$ are the momenta of the {\em outgoing}
$\Delta$ and {\em incoming} photon, respectively.
The $g_i\ (i=1,2,3)$ transition functions are related to the
Jones-Scadron form factors by
\begin{subequations}
\label{eq:g123}
\bea
g_1(Q^2)
&=& \widetilde C \big[ G_M^*(Q^2) - G_E^*(Q^2) \big]\, ,	\\
g_2(Q^2)
&=& g_1(Q^2)
 + \widetilde C \frac{2}{Q_-^2}
   \big[ P G_E^*(Q^2) + Q^2 G_C^*(Q^2) \big]\, ,		\\
g_3(Q^2)
&=& \widetilde C \frac{1}{Q_-^2}
    \big[ P G_C^*(Q^2) - 4 M_\Delta^2 G_E^*(Q^2) \big]\, ,
\eea
\end{subequations}
where
\be
Q_\pm = \sqrt{(M_\Delta\pm M)^2+Q^2}\, ,\qquad
P = M_\Delta^2-M^2-Q^2\, ,\qquad
\widetilde C = \frac{3 M_\Delta^2 (M_\Delta+M)}{M Q_+^2}\, .
\ee
Since in practice the magnetic multipole dominates the $\gamma^* N \to
\Delta$ transition, the $g_1$ function is determined mostly by $G_M^*$. 
The electric form factor $G_E^*$ determines the difference $g_2-g_1$,
while $g_3$ is sensitive to $G_E^*$ and the Coulomb form factor $G_C^*$.

In the present analysis, we take the Jones-Scadron form factors from the
phenomenological parametrization by Aznauryan \cite{Aznauryan:2011qj,
Aznauryan:2016pc},
\begin{subequations}
\label{eq:GMEC}
\bea
\label{eq:GstarM}
G_M^*(Q^2)
&=& 3.0\ G_D(Q^2)\,
    \exp{(-0.21\, Q^2)}\,\frac{Q_+}{M_\Delta+M}\, ,	\\
\label{eq:GstarE}
G_E^*(Q^2)
&=& -R_{\rm EM}\, G_M^*(Q^2)\, ,			\\
\label{eq:GstarC}
G_C^*(Q^2)
&=& -R_{\rm SM}\, G_M^*(Q^2)\,\frac{4 M_\Delta^2}{Q_+ Q_-}\, ,
\eea
\end{subequations}
where
\bea
G_D(Q^2) = \left( \frac{1}{1 + Q^2/0.71} \right)^2\, ,
\label{eq:Gdipole}
\eea
with $Q^2$ in units of GeV$^2$.  Empirical fits to data suggest
that the E1/M1 multipole ratio $R_{\rm EM}$ and the S1/M1 multipole
ratio $R_{\rm SM}$ can be well approximated by
\begin{subequations}
\label{eq:Remsm}
\bea
R_{\rm EM} &=& -0.02\, ,			\\
R_{\rm SM} &=& 0.01\, (1 + 0.0065 \,Q^4)	\nn\\
&&\times (- 6.066 + 5.807\, Q - 8.5639\, Q^2 + 2.37058\, Q^4
	  - 0.75445\, Q^5)\, .
\eea
\end{subequations}
As this parametrization only has poles for timelike $Q^2$,
we can use the contour $\Gamma_{\rm CD}$ of Eq.~(\ref{eq:contour1})
in the unphysical region.

\begin{figure}[t]
\includegraphics[width=0.46\textwidth]{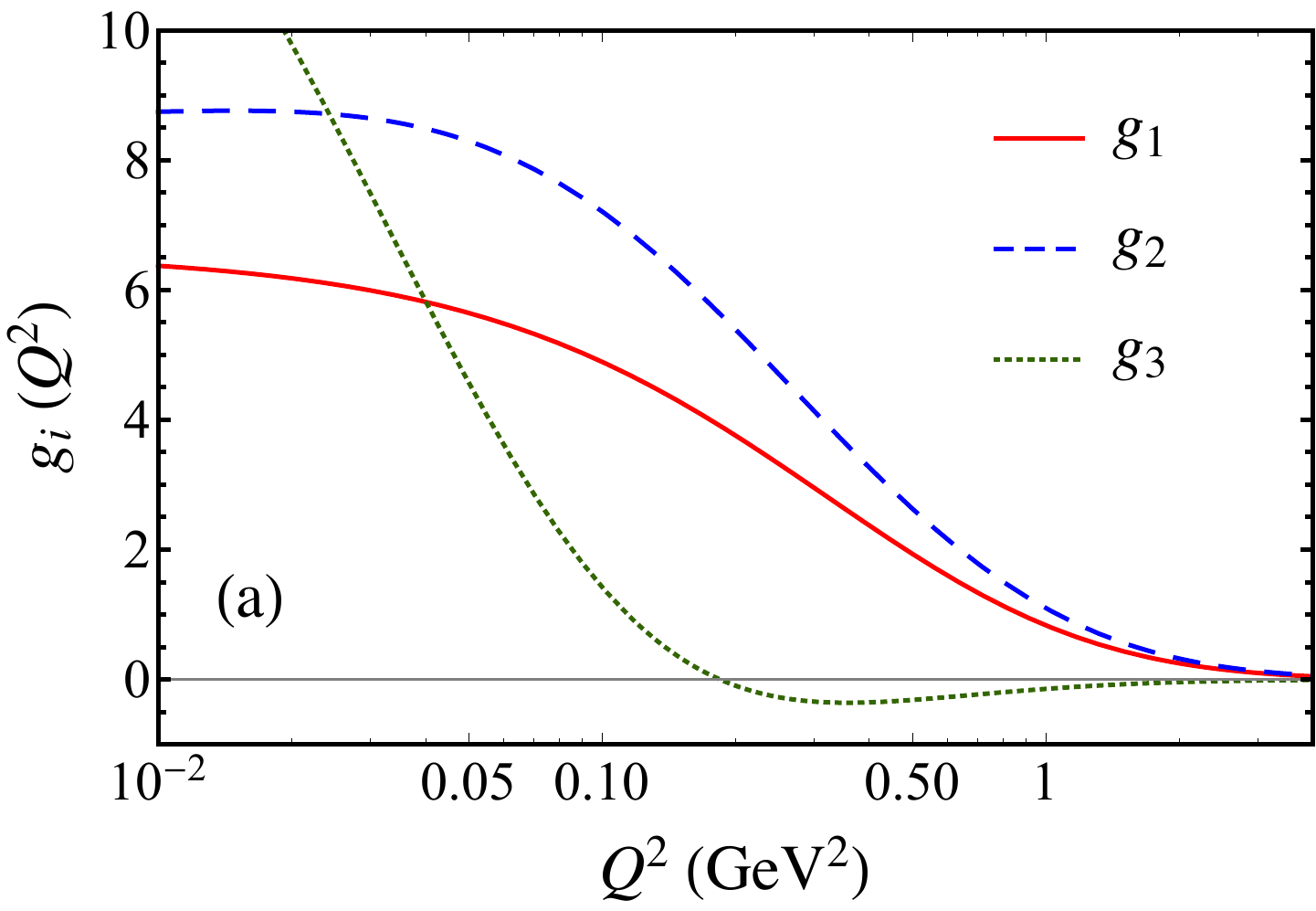}\hspace*{0.4cm}
\includegraphics[width=0.46\textwidth]{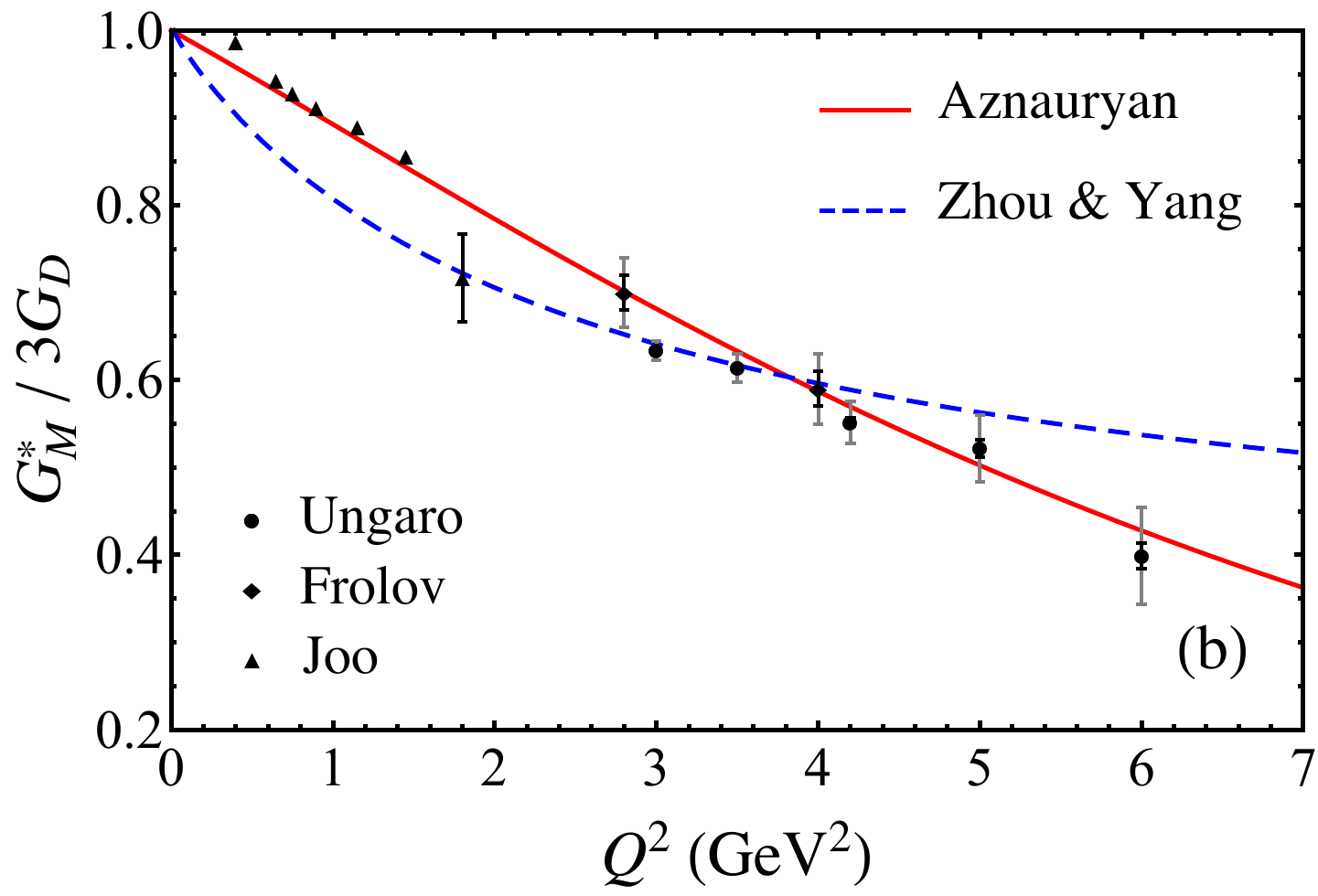} \\
\hspace*{-0.2cm}
\includegraphics[width=0.46\textwidth]{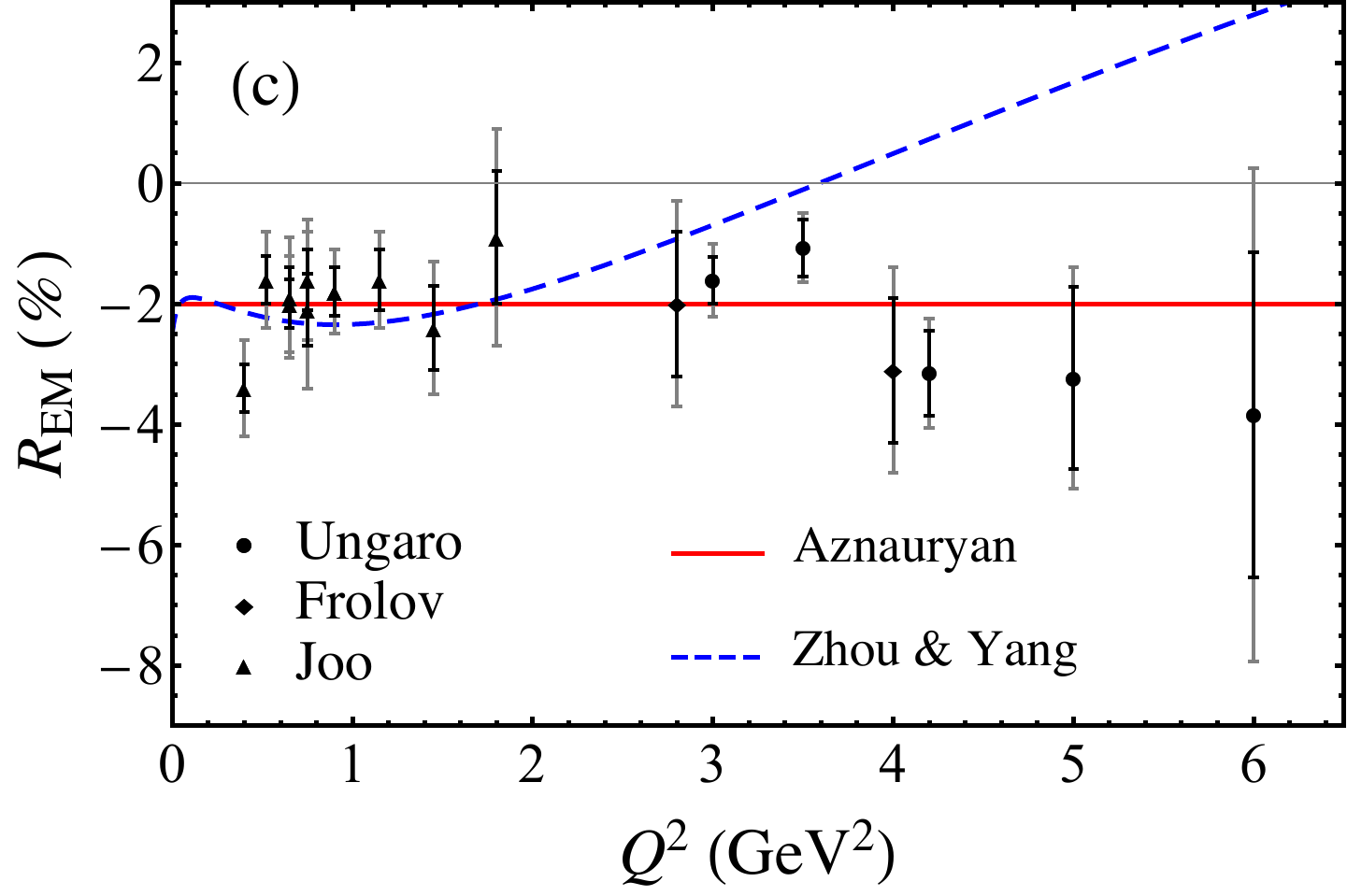}\hspace*{0.4cm}
\includegraphics[width=0.46\textwidth]{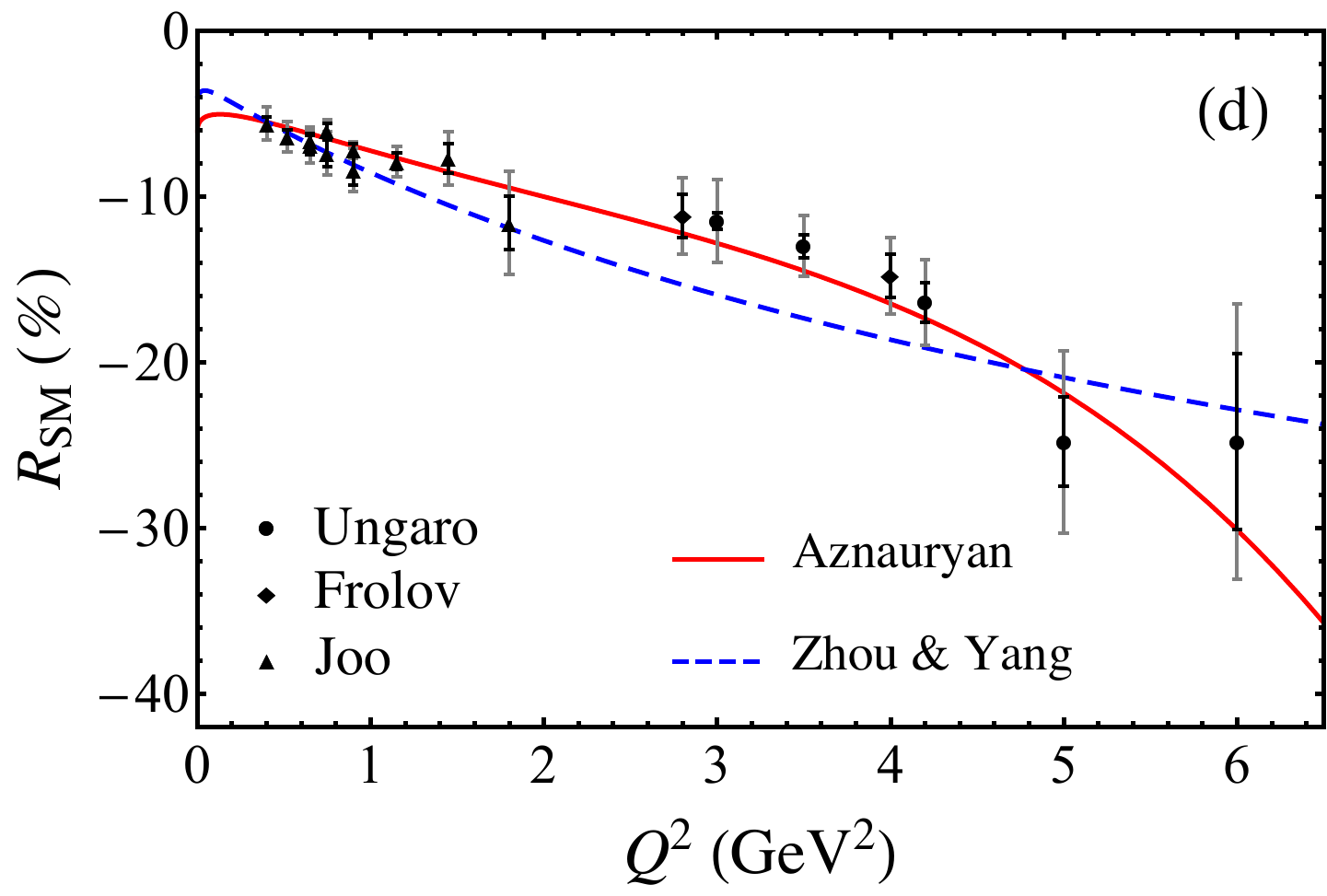}
\caption{(a) $\gamma^* N \Delta$ transition form factors $g_1$,
	$g_2$ and $g_3$ versus $Q^2$, as used in this analysis
	\cite{Aznauryan:2011qj, Aznauryan:2016pc};
	 (b) magnetic $G_M^*$ form factor, scaled by the dipole
	form factor, $3 \times G_D$;
	 (c) electric to magnetic dipole ratio $R_{\rm EM}$ (in percent);
	and
	 (d) Coulomb to magnetic dipole ratio $R_{\rm SM}$ (in percent).
	In (b), (c) and (d) the transition form factors from Aznauryan
	\cite{Aznauryan:2011qj, Aznauryan:2016pc} (solid red curves)
	used in this analysis are compared with data from
	Ungaro {\it et al.} \cite{Ungaro:2006df} (circles),
	Frolov {\it et al.} \cite{Frolov:1998pw} (diamonds), and
	Joo {\it et al.} \cite{Joo:2001tw} (triangles),
	and with the recent parametrization by Zhou and	Yang
	\cite{Zhou:2014xka}.}
\label{fig:NDff}
\end{figure}

The $\gamma^* N \Delta$ transition form factors $g_1$, $g_2$ and $g_3$
from the Eqs.~(\ref{eq:GMEC})--(\ref{eq:Remsm}) are illustrated in
Fig.~\ref{fig:NDff}(a) as a function of $Q^2$. At moderate and large
$Q^2$ values, $Q^2 \gtrsim 0.1$~GeV$^2$, the $g_1$ and $g_2$ form
factors dominate, with the $g_3$ form factor essentially zero.  At very
low $Q^2 \lesssim 0.02$~GeV$^2$, the $g_3$ form factor rises rapidly and
becomes larger than the largest contribution; for the phenomenological
applications relevant to this paper, however, its role is essentially
negligible.

The quality of the fit to the magnetic transition form factor $G_M^*$ is
shown in Fig.~\ref{fig:NDff}(b), compared with data from several
experiments \cite{Ungaro:2006df, Frolov:1998pw, Joo:2001tw} for $Q^2$ up
to $\approx 6$~GeV$^2$. The parametrization in Eq.~(\ref{eq:GstarM}) is
compared with an alternative parametrization from the recent analysis by
Zhou and Yang \cite{Zhou:2014xka}, which agrees with the data in the
intermediate $Q^2$ region, $Q^2 \sim 3$--4~GeV$^2$, but underestimates
(overestimates) the data at lower (higher) $Q^2$ values. Similarly, good
agreement is obtained for the $R_{\rm EM}$ and $R_{\rm SM}$ ratios in
Figs.~\ref{fig:NDff}(c) and \ref{fig:NDff}(d), respectively, for the
parametrizations in Eqs.~(\ref{eq:Remsm}) over the full range of $Q^2$
($\lesssim 6$~GeV$^2$) where data are available.  The parametrization
\cite{Zhou:2014xka} also agrees with the data at low $Q^2$ values, $Q^2
\lesssim 1$~GeV$^2$, but discrepancies appear for larger $Q^2$.

\begin{figure}[t]
\includegraphics[width=0.47\textwidth]{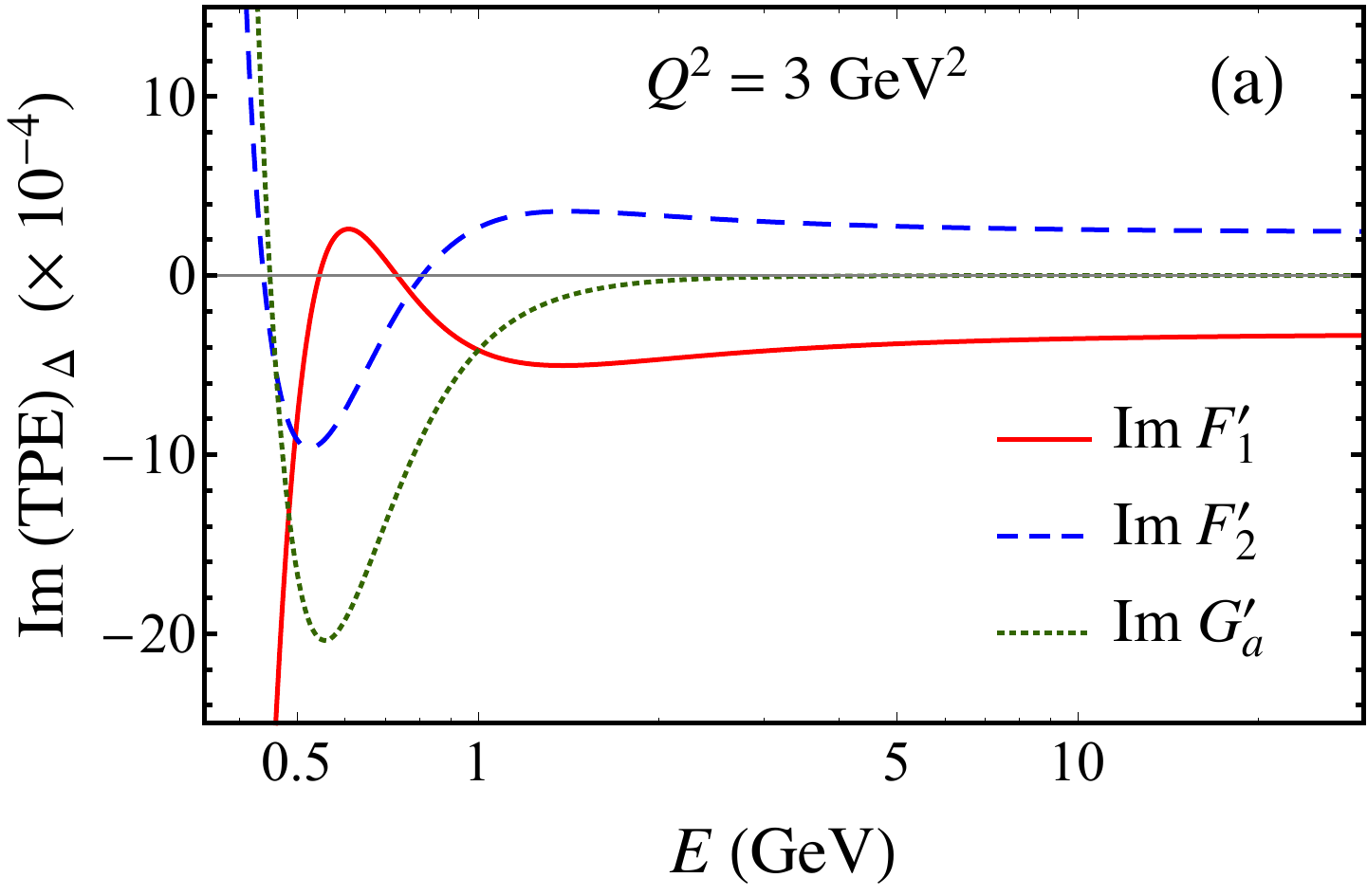}\hspace*{0.3cm}
\includegraphics[width=0.46\textwidth]{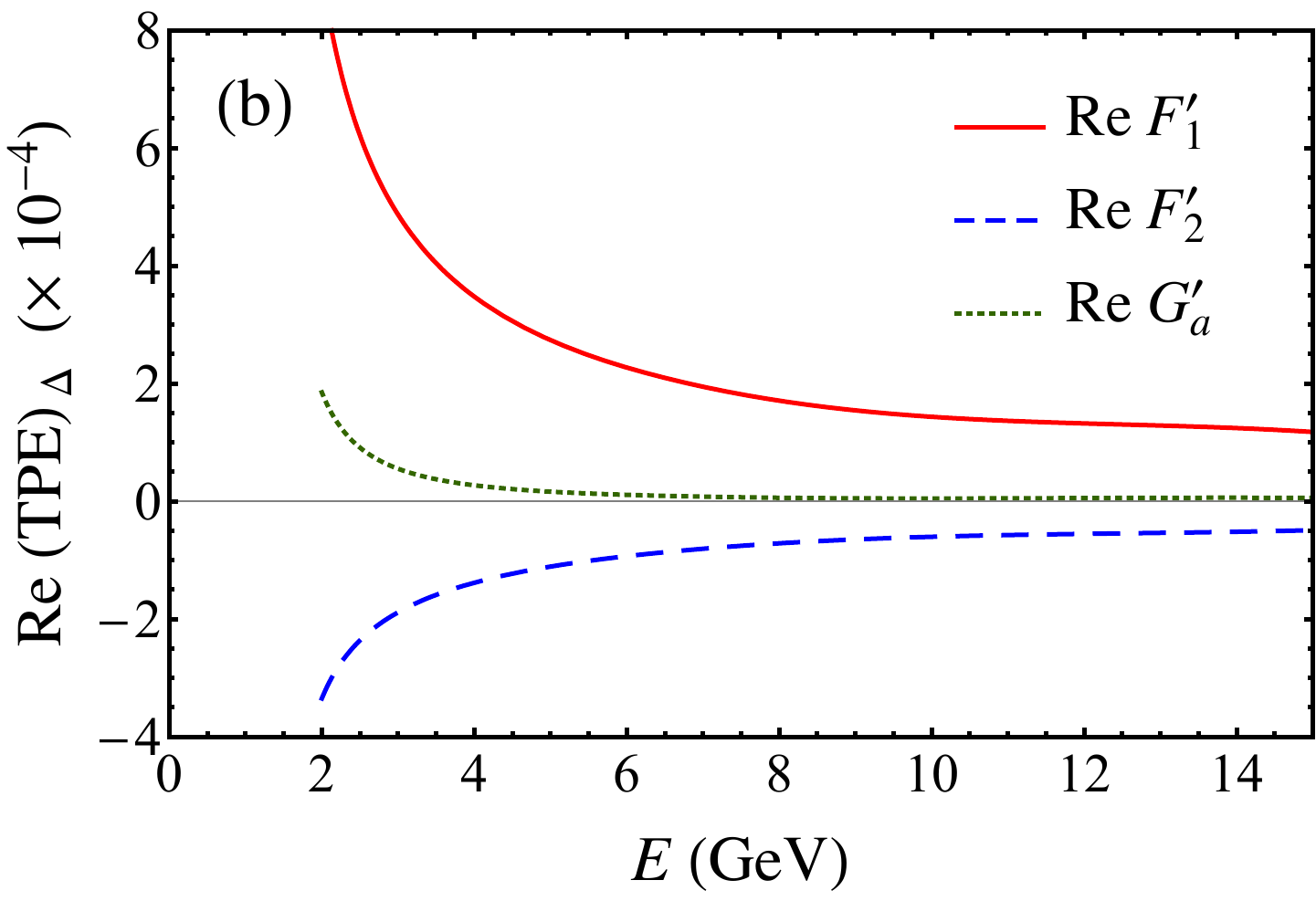}
\caption{TPE contributions from $\Delta$ intermediate states to the
	(a) imaginary part and
	(b) real part of the generalized
	$F_1'$ (solid red curves),
	$F_2'$ (dashed blue curves) and
	$G_a'$ (dotted black curves)
	form factors as a function of energy $E$,
	for fixed $Q^2=3$~GeV$^2$. \\}
\label{fig:Dff}
\end{figure}

Using the Aznauryan parametrization \cite{Aznauryan:2011qj,
Aznauryan:2016pc} of the $\gamma^* N \Delta$ form factors, the TPE
contributions from $\Delta$ intermediate states to the generalized form
factors $F_1'$, $F_2'$ and $G_a'$ are illustrated in Fig.~\ref{fig:Dff}
for both the imaginary and real parts.  For the imaginary parts of the
amplitudes, Fig.~\ref{fig:Dff}(a) shows resonance-like structure
appearing in the unphysical region, $0.34 < E < 2$~GeV.  As for the
nucleon case, the unphysical region accounts for most of the dispersive
integral, giving rise to smoothly decaying real parts of the amplitudes
for $E \gtrsim 3$~GeV, as Fig.~\ref{fig:Dff}(b) illustrates.

\begin{figure}[t]
\includegraphics[width=0.57\textwidth]{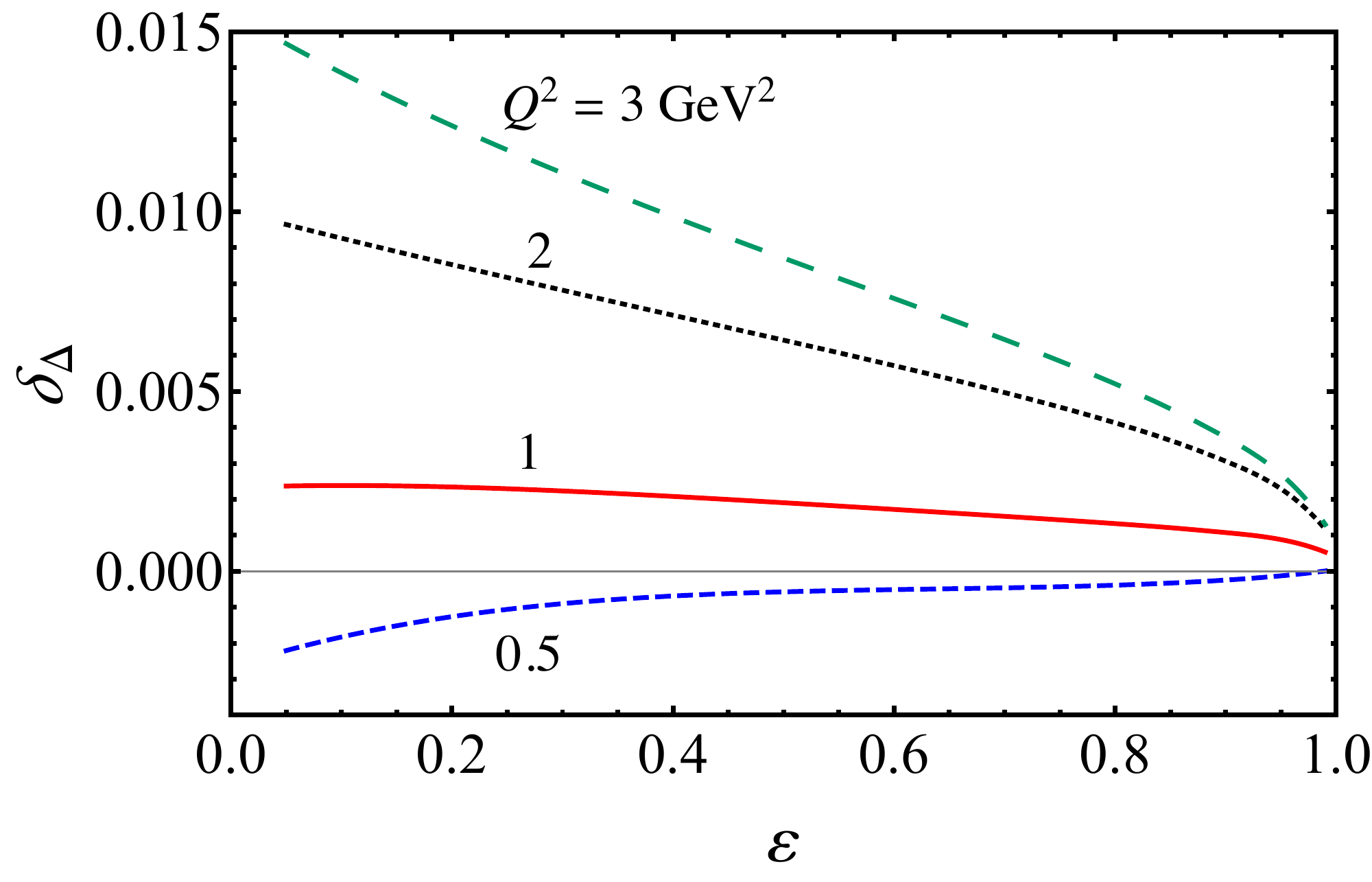}
\caption{Dispersive TPE correction to the cross section,
	$\delta_\Delta$, from $\Delta$ intermediate states
	as a function of $\eps$ for fixed values of
	$Q^2$ (= 0.5, 1, 2 and 3~GeV$^2$). \\}
\label{fig:deltaD}
\end{figure}

\begin{figure}[t]
\includegraphics[width=0.48\textwidth]{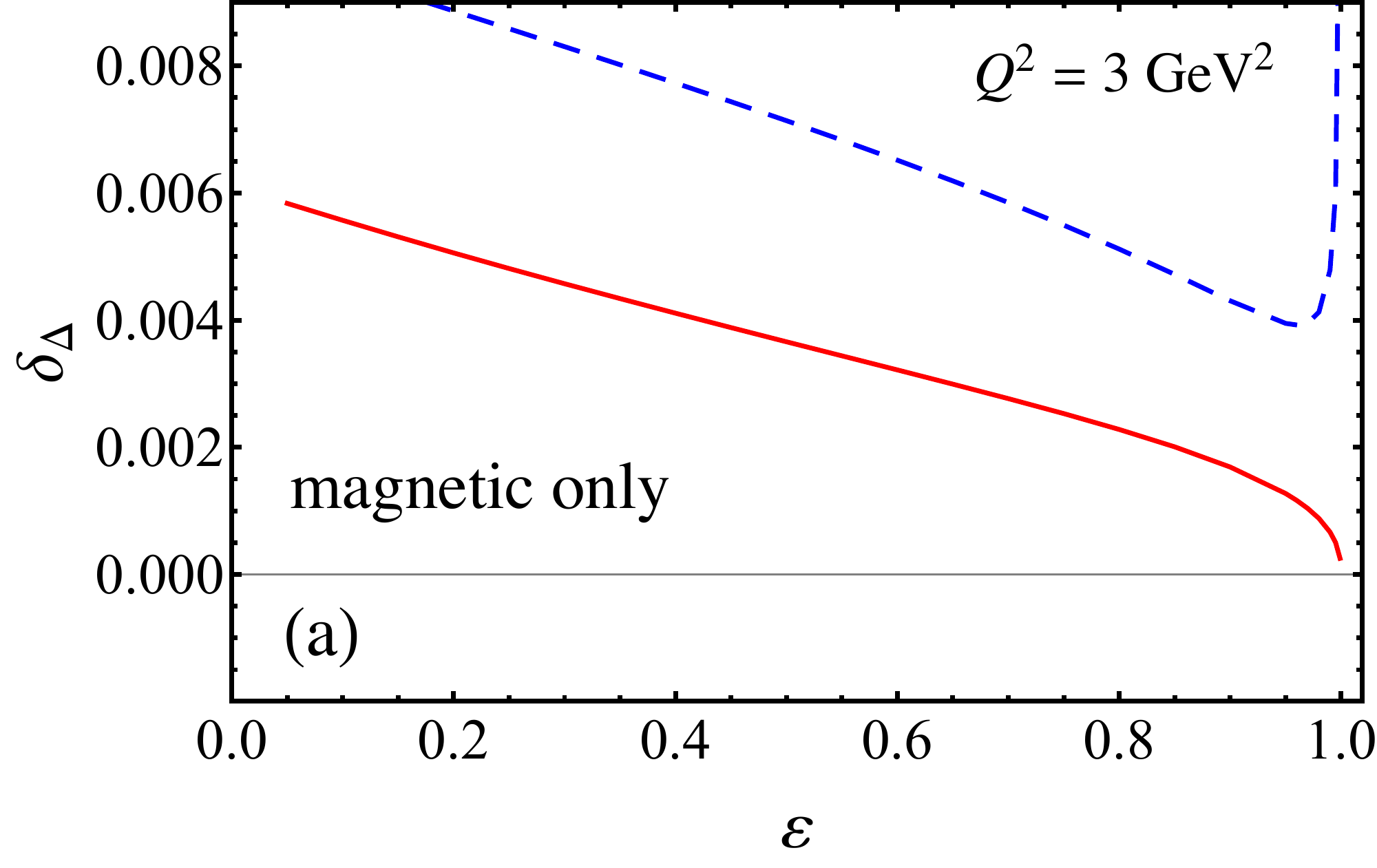}\hspace*{.2cm}
\includegraphics[width=0.48\textwidth]{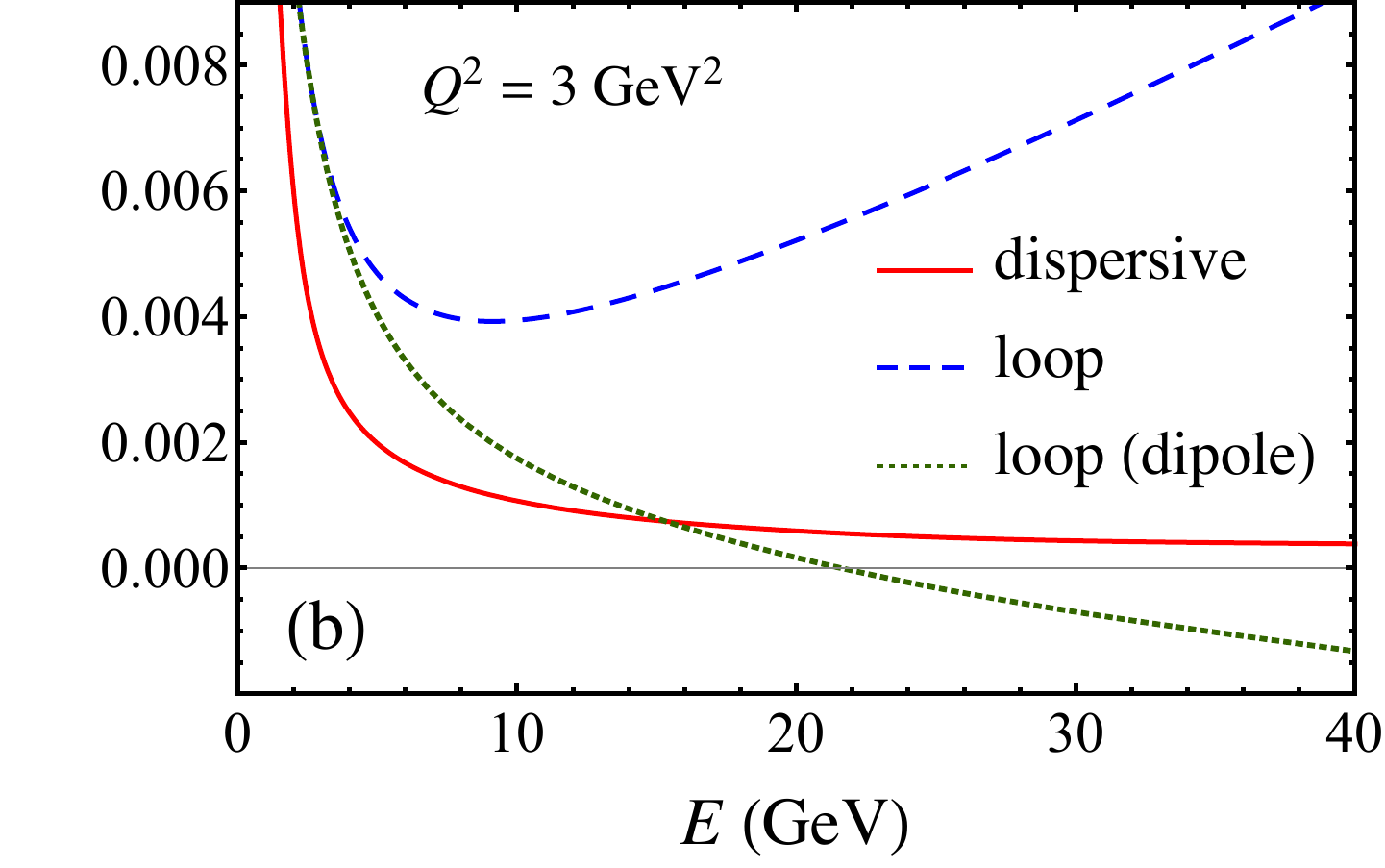}
\caption{Comparison of the $\Delta$ intermediate state contribution
	to the TPE cross section, $\delta_\Delta$, for the dispersive
	(solid red curves) and loop (dashed blue curves) calculations
	at fixed $Q^2 = 3$~GeV$^2$:
	(a) versus $\eps$, and
	(b) versus electron energy $E$. Note that $\eps=0.9$
	corresponds to $E=5.9$~GeV for this $Q^2$.
	In (b) the loop corrections using a dipole approximation
	to the $\gamma^* N \Delta$ form factor with mass 0.75~GeV
	is also shown (dotted green curve).
	For simplicity only the dominant magnetic contribution has
	been considered.}
\label{fig:deltaD_loop}
\end{figure}

As was observed in previous calculations of loop corrections
\cite{Kondratyuk:2005kk, Drell:1959zz, Campbell:1969uc,
Greenhut:1970aq}, the contribution from $\Delta$ intermediate states,
$\delta_\Delta$, is generally of opposite sign to the nucleon
contribution $\delta_N$ for $Q^2 \gtrsim 1$~GeV$^2$, and increases in
magnitude with increasing $Q^2$, as Fig.~\ref{fig:deltaD} illustrates. 
Unlike previous loop calculations \cite{Kondratyuk:2005kk,
Zhou:2014xka}, however, the correction $\delta_\Delta$ in the dispersive
approach is well-behaved for all $\eps$, vanishing in the $\eps \to 1$
limit. In contrast, the correction $\delta_\Delta$ from the loop
calculation with off-shell $\Delta$ states diverges as $\eps \to 1$, as
illustrated in Fig.~\ref{fig:deltaD_loop}(a).  Note that
the imaginary parts of the TPE amplitudes are identical for both
calculations --- only the real parts differ.

The nature of this divergence can be seen by plotting $\delta_\Delta$
versus electron energy $E$ instead of $\eps$, as in
Fig.~\ref{fig:deltaD_loop}(b).  The linear divergence in $E$ indicates a
violation of the Froissart bound~\cite{Froissart:1961ux}, and the
breakdown of unitarity.  This ``pathological'' behavior is not due to
the inapplicability of hadronic models when $E \to \infty$ and $\eps \to
1$, as suggested in Ref.~\cite{Zhou:2014xka}. Rather, it arises from an
unphysical behavior of the off-shell contributions at high energies for
interactions with derivative couplings.

As Fig.~\ref{fig:deltaD_loop} illustrates, the loop calculations at
large $\eps$ are actually very sensitive to the shape of the $\gamma^* N
\Delta$ form factors employed.  Whether one use a dipole approximation
or a more realistic parametrization, leads to significant differences
with the dispersive approach already for $E \gtrsim 3$--4~GeV.  (Here,
for simplicity only the magnetic contribution to $\delta_\Delta$ is
shown, but the effects are similar for the other $\gamma^* N \Delta$
form factors also.) In fact, the differences between the dispersive and
loop calculations are significant not just near $\eps \approx 1$, but
also at lower $\eps$ values.  Generally, the magnitude of the dispersive
$\Delta$ corrections is smaller than the loop results, resulting in less
cancellation with the intermediate state nucleon contribution.

% ......................................................................
\subsection{Verification of TPE effects}

Having detailed the calculation of the TPE corrections from the nucleon
and $\Delta$ intermediate states, we next compare the role of these
corrections in observables that are particularly sensitive to effects
beyond the Born approximation. These include the ratio of unpolarized
$e^+ p$ to $e^- p$ elastic scattering cross sections, and polarization
transfer cross sections for longitudinally and transversely polarized
electrons and protons.

% . . . . . . . . . . . . . . . . . . . . . . . . . . . . . . . . . . .
\subsubsection{$e^+ p$ to $e^- p$ ratio}
\label{sss.e+e-}

One of the observables that is most sensitive to the effects of TPE is
the ratio of $e^+ p$ to $e^- p$ elastic cross sections, which in the
one-photon exchange approximation is unity. Since the TPE terms enter
the $e^+ p$ cross section with opposite sign to that in the $e^- p$
reaction, the ratio
\bea
R_{2\gamma}
&=& \frac{\sigma^{e^+}}{\sigma^{e^-}}\
% \approx\ 1 - 2\, \left( \delta_N + \delta_\Delta \right),
\approx\ 1 - 2\, \delta_{\gamma\gamma},
\eea
where $\sigma^{e^\pm} \equiv d\sigma({e^\pm p\to e^\pm p})/d\Omega$,
provides a direct measure of effects beyond the Born approximation.
Earlier data from elastic $e^+ p$ and $e^- p$ experiments in the 1960s
from SLAC~\cite{Browman:1965zz, Mar:1968qd},
Cornell~\cite{Anderson:1966zzf}, DESY~\cite{Bartel:1967dsa} and
Orsay~\cite{Bouquet:1968yqa} gave some hints of a small enhancement of
$R_{2\gamma}$ at forward angles and low $Q^2$, but were in the region
(at large $\eps$) where TPE is relatively small and were consistent
within errors with $R_{2\gamma} = 1$.

\begin{figure}[t]
\includegraphics[width=0.49\textwidth]{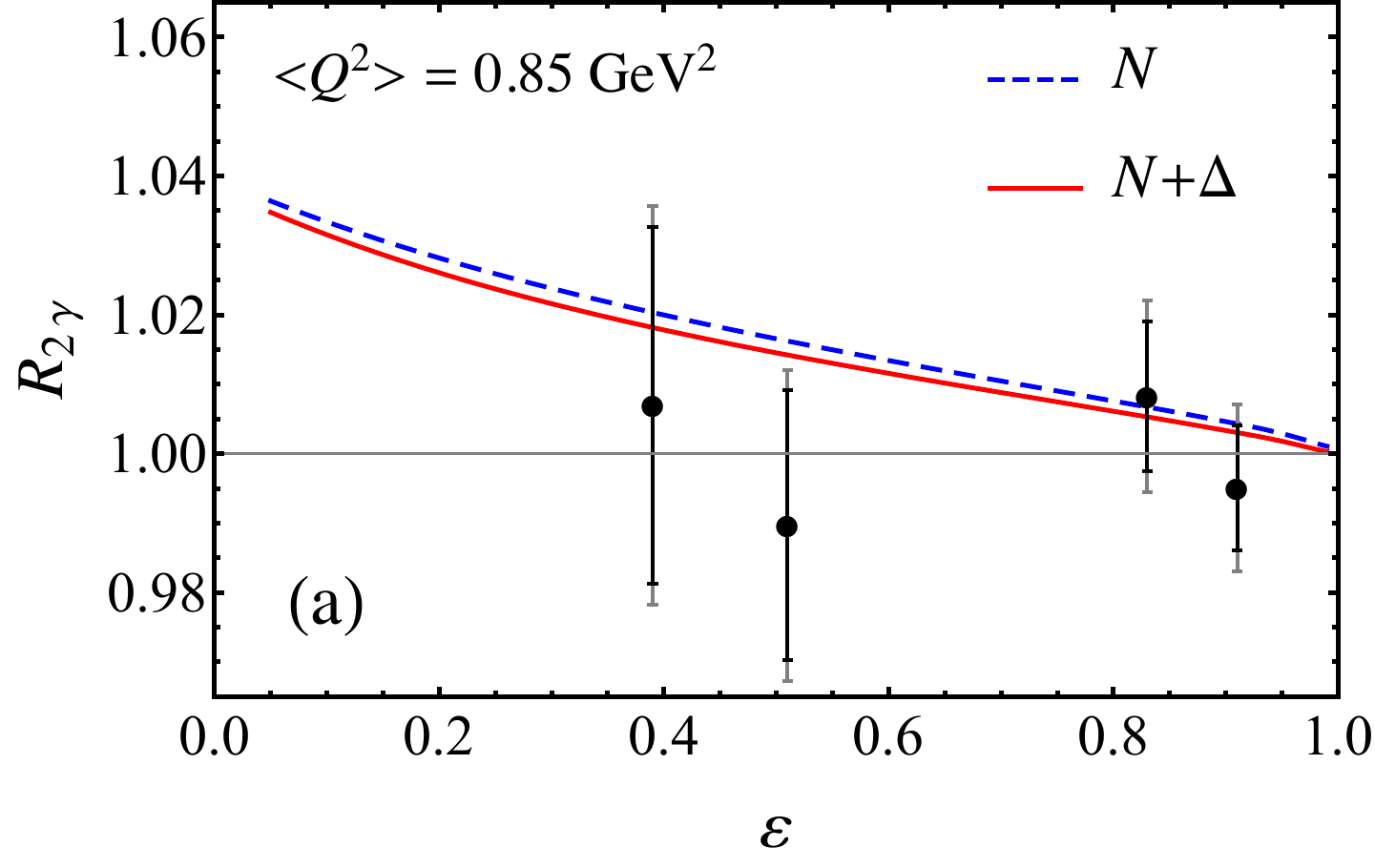}\hspace*{-0.11cm}
\includegraphics[width=0.49\textwidth]{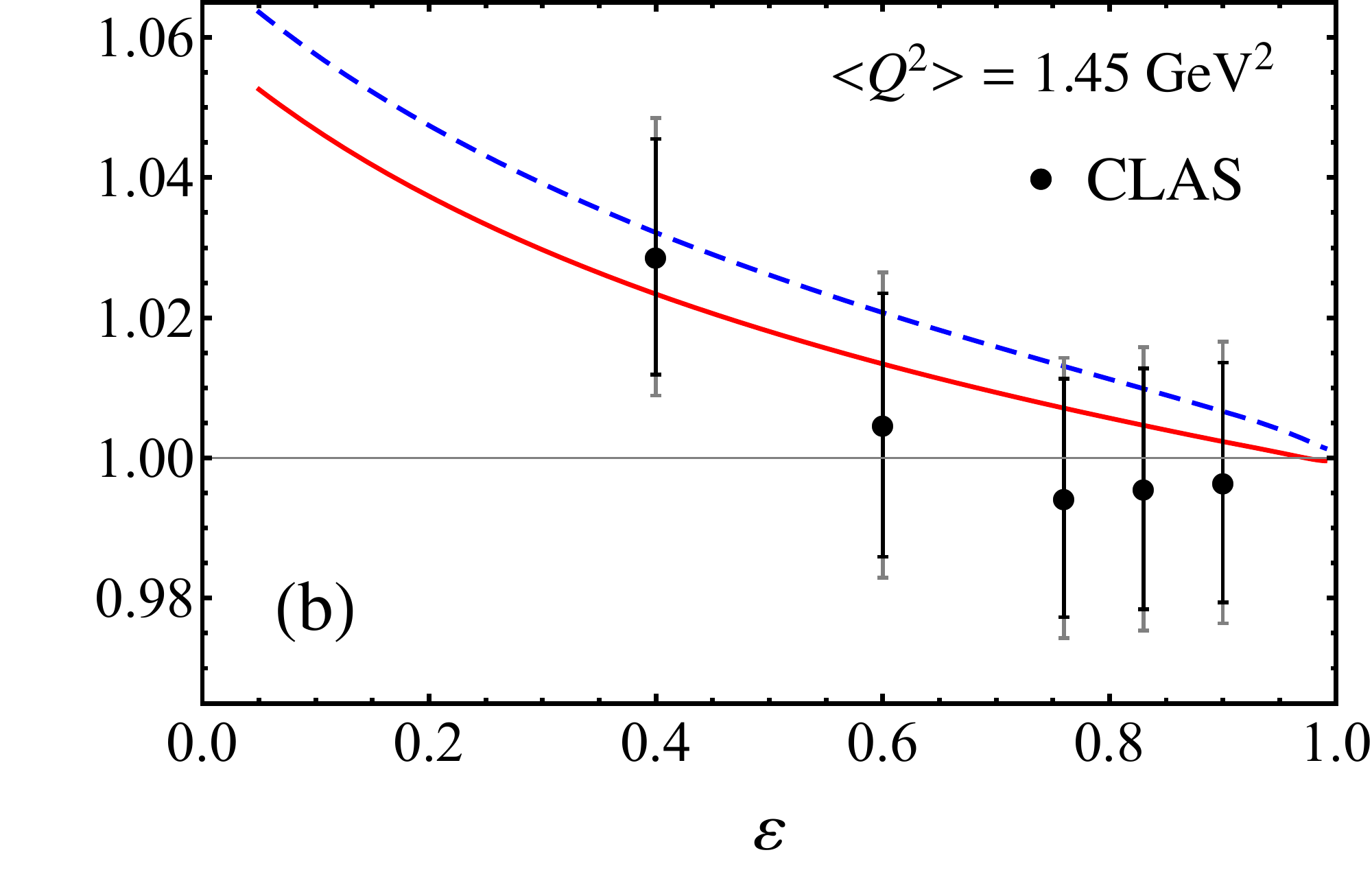}\\
\vspace*{0.3cm}\includegraphics[width=0.49\textwidth]{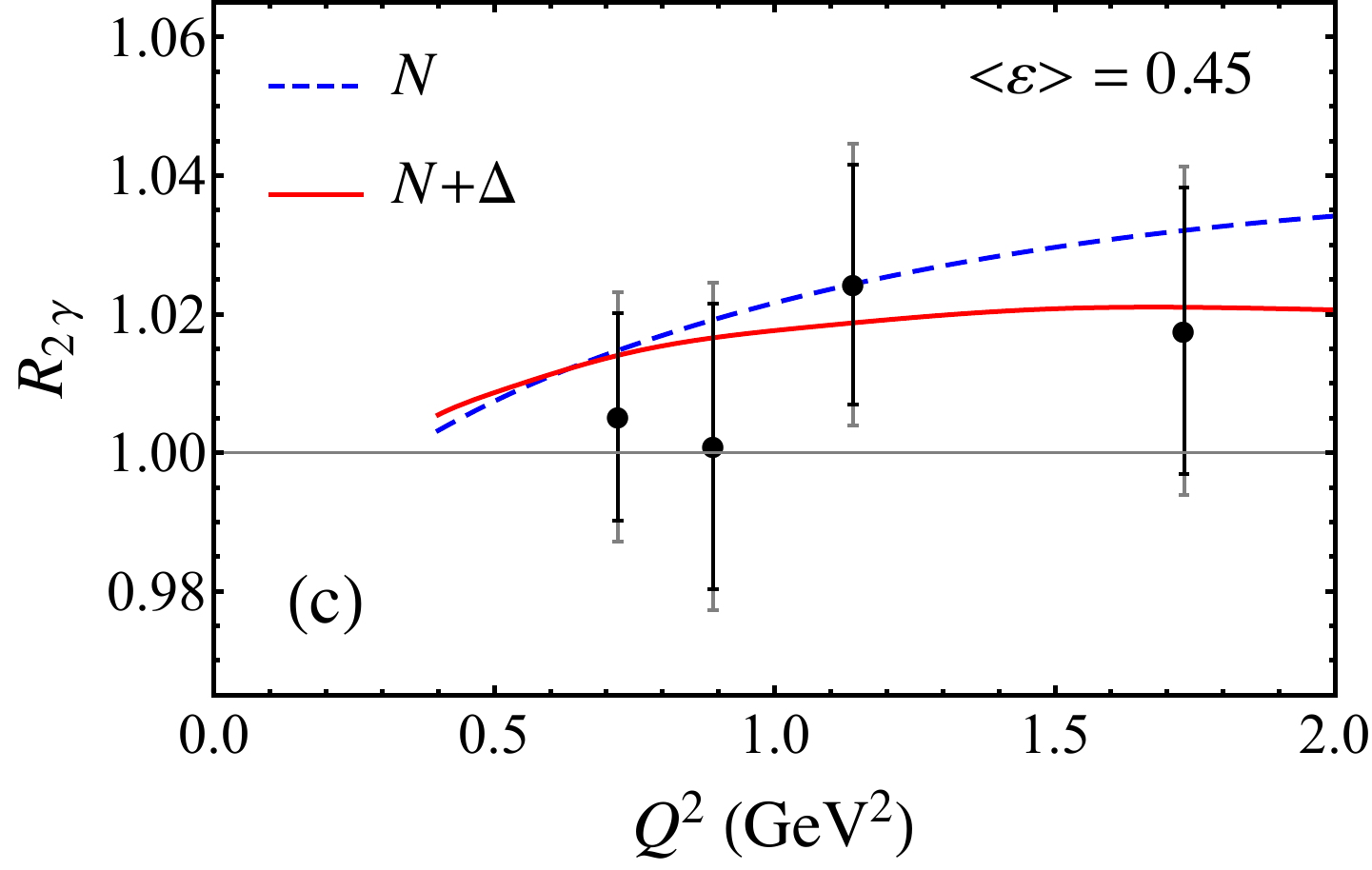}\hspace*{-0.11cm}
\includegraphics[width=0.49\textwidth]{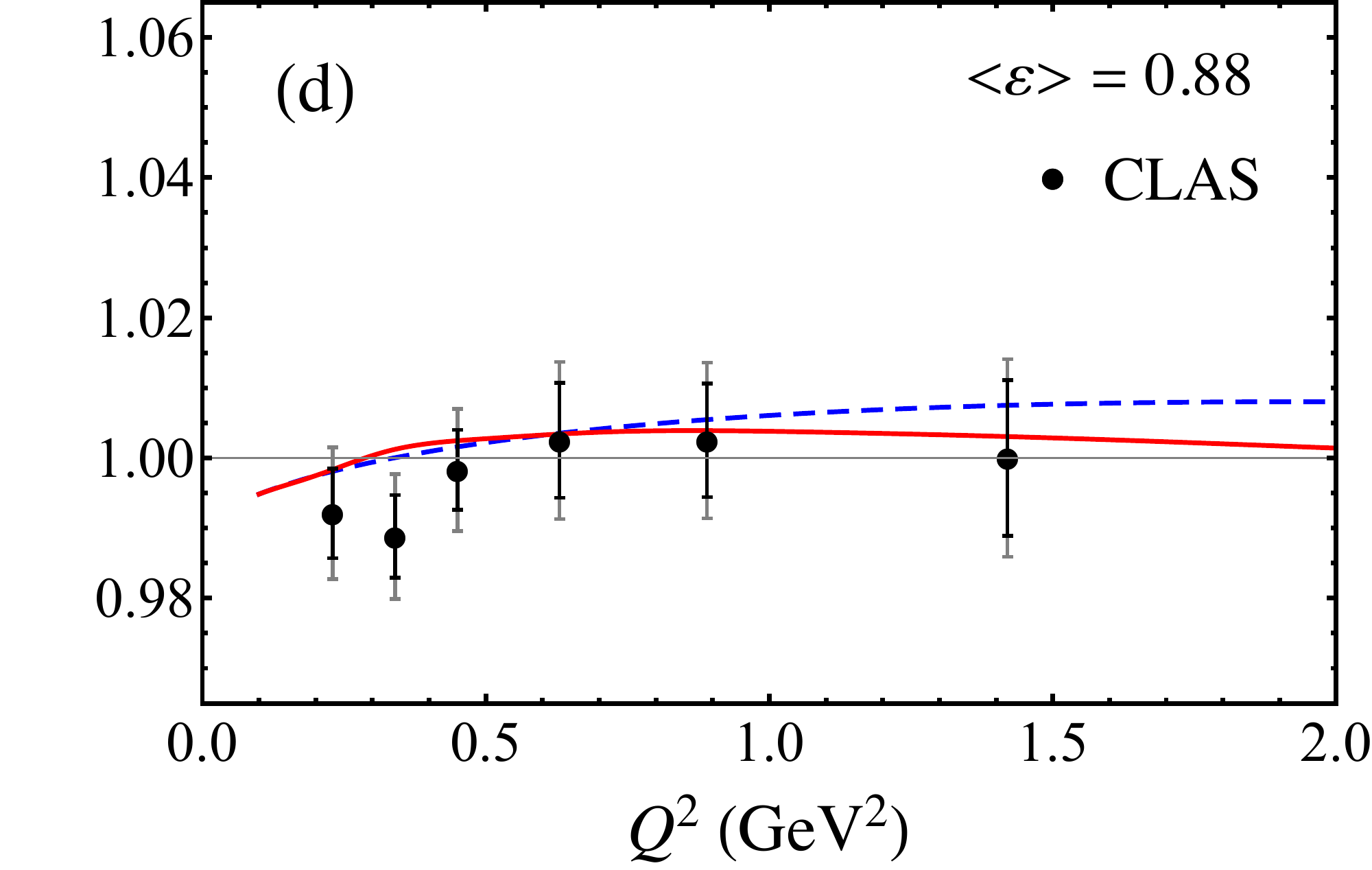}
\caption{Ratio $R_{2\gamma}$ of $e^+ p$ to $e^- p$ cross sections
	as a function of $\eps$ for
	(a) fixed $\langle Q^2 \rangle = 0.85$~GeV$^2$ and
	(b) fixed $\langle Q^2 \rangle = 1.45$~GeV$^2$, and
	as a function of $Q^2$ for
	(c) fixed $\langle \eps \rangle = 0.45$ and
	(d) fixed $\langle \eps \rangle = 0.88$.
	The contributions with nucleon only (dashed blue curves)
	and the sum of nucleon and $\Delta$ (solid red curves)
	intermediate states are compared with data from CLAS
	at Jefferson Lab (circles) \cite{Rimal:2016toz}, with the
	statistical and systematic uncertainties indicated by the
	(black) inner and (gray) outer error bars, respectively.}
\label{fig:eeCLAS}
\end{figure}

More recently, several dedicated $e^+ p$ to $e^- p$ ratio experiments
have been performed in CLAS at Jefferson Lab \cite{Rimal:2016toz},
VEPP-3 in Novosibirsk \cite{Rachek:2014fam, Nikolenko:2014uda}
and OLYMPUS at DESY \cite{Henderson:2016dea} aimed at providing
measurements of $R_{2\gamma}$ over a larger range of $\eps$ and
$Q^2$ with significantly reduced uncertainties.
In Fig.~\ref{fig:eeCLAS} the $R_{2\gamma}$ ratio from the CLAS
experiment is shown as a function of $\eps$ at averaged $Q^2$
values of $\langle Q^2 \rangle = 0.85$~GeV$^2$ and
$\langle Q^2 \rangle = 1.45$~GeV$^2$
[Figs.~\ref{fig:eeCLAS}(a) and (b), respectively],
and as a function of $Q^2$ at averaged $\eps$ values of
$\langle \eps \rangle = 0.45$ and
$\langle \eps \rangle = 0.88$~GeV$^2$
[Figs.~\ref{fig:eeCLAS}(c) and (d), respectively].
Most of the data at the larger $\eps$ values are consistent with unity
within the errors, but suggest a nonzero ratio, $\approx 2\%$ -- 4\%
greater than unity, at the lowest $\eps$ value for the higher-$Q^2$ set.
 The trend is consistent with the ratio calculated here, which shows a
rising $R_{2\gamma}$ with decreasing $\eps$.  At these kinematics the
calculated TPE correction is dominated by the nucleon elastic
intermediate state, with the $\Delta$ contribution reducing the ratio
slightly.
Note that both the data and the calculated TPE corrections here (and
elsewhere, unless otherwise stated) are shown relative to the Mo-Tsai
infrared result.

The same trend is seen when the $R_{2\gamma}$ data are viewed as a
function of $Q^2$ for fixed $\eps$.  At the larger average $\eps$ value,
$\langle \eps \rangle = 0.88$, the effects are consistent with zero as
well as with the small predicted TPE correction.  At the smaller value
$\langle \eps \rangle = 0.45$, on the other hand, the larger predicted
effect is consistent with the larger $R_{2\gamma}$ values with
increasing $Q^2$. Again the effects of the $\Delta$ intermediate state
are small at low $Q^2$ values, but become visible at larger $Q^2$, where
they improve the agreement between the theory and experiment.

\begin{figure}[t]
\includegraphics[width=0.49\textwidth]{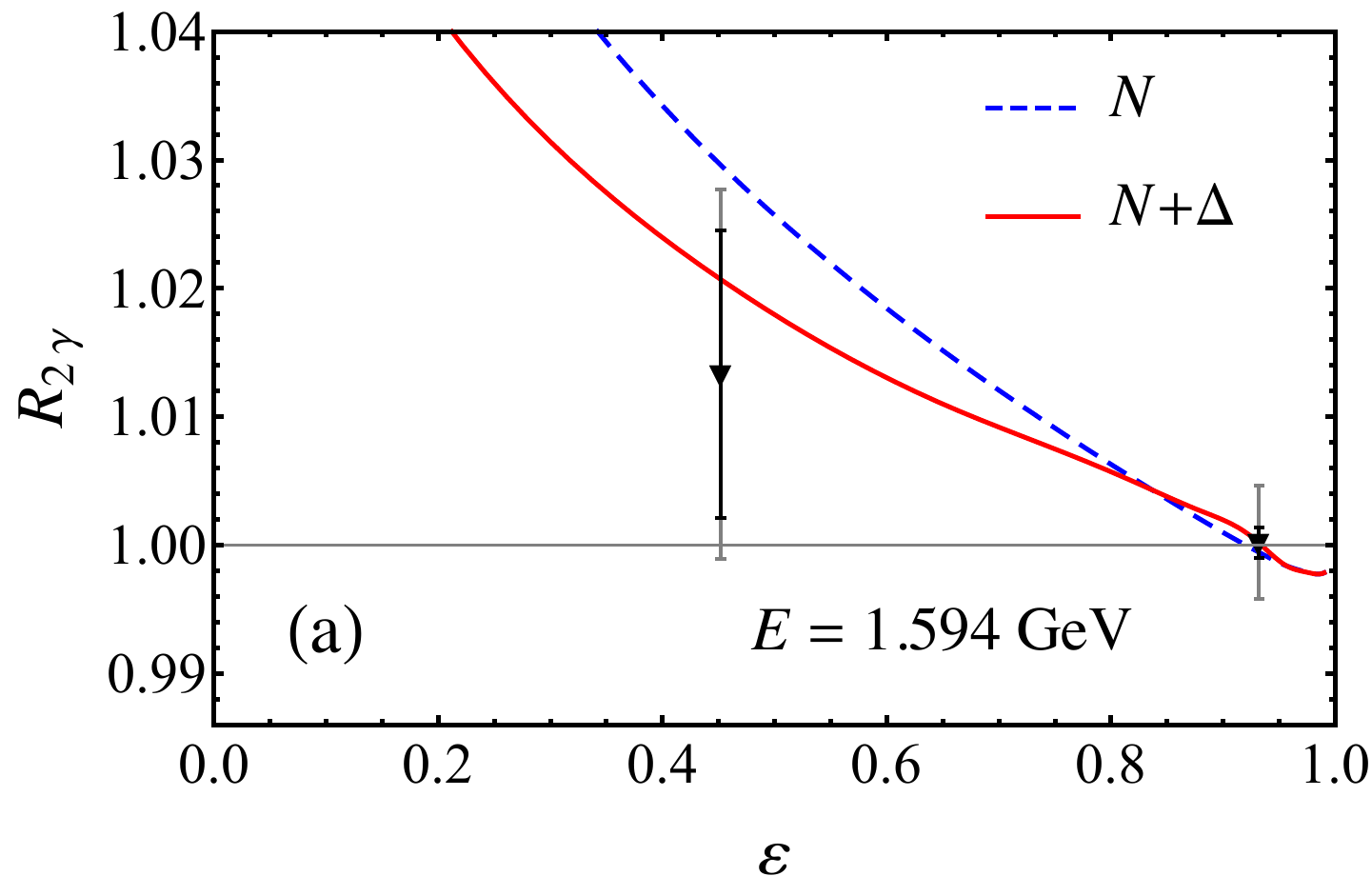}
\includegraphics[width=0.49\textwidth]{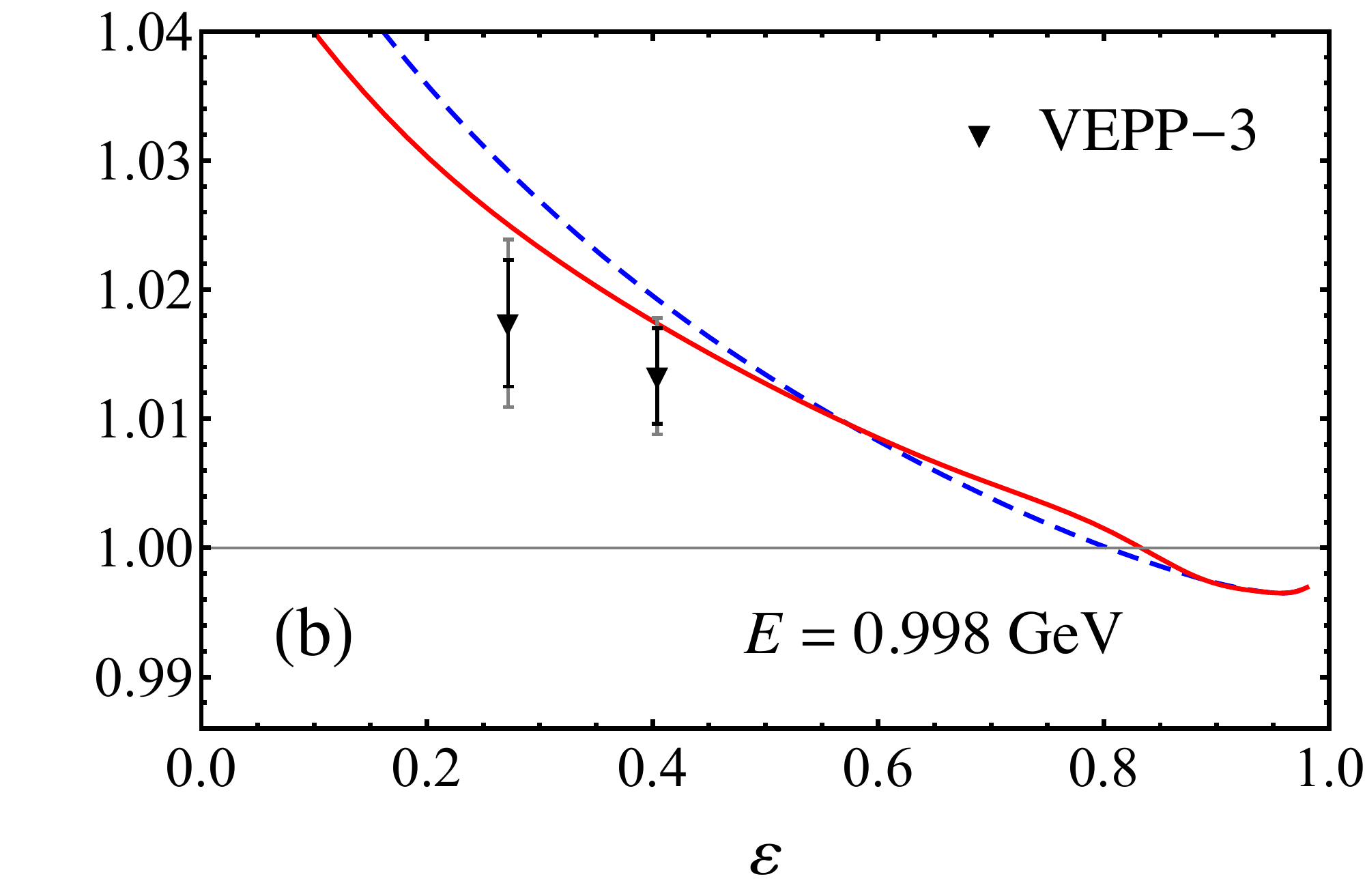}
\caption{Ratio $R_{2\gamma}$ of $e^+ p$ to $e^- p$ cross sections
        as a function of $\eps$ for fixed energy
	(a) $E = 1.594$~GeV and
	(b) $E = 0.998$~GeV.
        The contributions with nucleon only (dashed blue curves)
        and the sum of nucleon and $\Delta$ (solid red curves)
	intermediate states are compared with data from the
	\mbox{VEPP-3} experiment (triangles) \cite{Rachek:2014fam},
	with the statistical and systematic uncertainties indicated
	by the (black) inner and (gray) outer error bars, respectively.}
\label{fig:eeVEPP}
\end{figure}

Data from the VEPP-3 experiment at Novosibirsk \cite{Rachek:2014fam,
Nikolenko:2014uda}, taken at energies $E \approx 1$~GeV and 1.6~GeV, are
shown in Fig.~\ref{fig:eeVEPP} as a function of $\eps$. The data
correspond to a $Q^2$ range between $\approx 0.3$~GeV$^2$ and $\approx
1.5$~GeV$^2$, with $\eps$ down to $\approx 0.3$. The ratio at the low
$\eps$ values shows an effect of magnitude 1\% -- 2\%, slightly below
but still consistent with the calculated TPE result at the $\approx
1\sigma$ level.

\begin{figure}[t]
\includegraphics[width=0.5\textwidth]{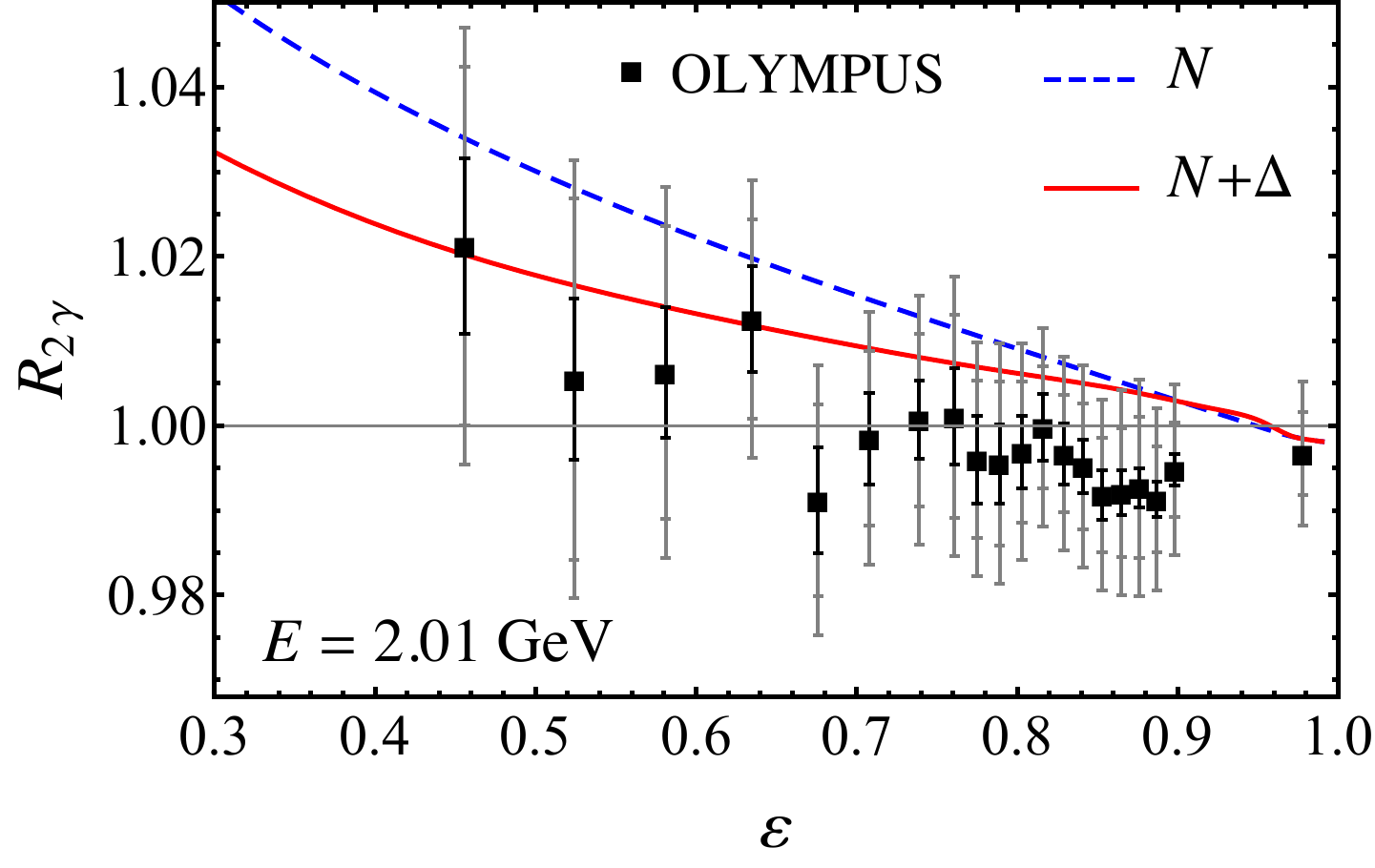}
\caption{Ratio $R_{2\gamma}$ of $e^+ p$ to $e^- p$ cross sections
	as a function of $\eps$ for fixed energy $E=2.01$~GeV.
	The contributions with nucleon only (dashed blue curves)
	and the sum of nucleon and $\Delta$ (solid red curves)
	intermediate states are compared with data from the
	OLYMPUS experiment \cite{Henderson:2016dea} (squares).
	The statistical and systematic uncertainties
	(correlated and uncorrelated) are indicated by the
	(black) inner and (gray) outer error bars, respectively.}
\label{fig:eeOLYM}
\end{figure}

Most recently, the OLYMPUS experiment at DESY \cite{Henderson:2016dea}
measured the ratio $R_{2\gamma}$ at an energy $E = 2.01$~GeV over a
large range of $\eps \sim 0.45 - 1$, corresponding to a $Q^2$ range from
$\approx 0.2$~GeV$^2$ to 2~GeV$^2$. The results for the ratio
$R_{2\gamma}$ are shown in Fig.~\ref{fig:eeOLYM}.  Interestingly, in
contrast to the results from the CLAS and VEPP-3 experiments in
Figs.~\ref{fig:eeCLAS} and \ref{fig:eeVEPP}, at large $\eps$ values the
trend in the data is towards values of the ratio slightly below unity,
whereas the calculated dispersive TPE corrections give a ratio that has
a small, $\lesssim 1\%$ enhancement above unity.  At the lower $\eps$
values, the trend is toward increasing values of $R_{2\gamma}$,
consistent with the TPE calculation.  Within the statistical and
systematic uncertainties, including the overall normalization
uncertainty of the OLYMPUS data, the theoretical result is consistent
with the data over the entire $\eps$ range. Note also that the $\lesssim
0.5\%$ correlated systematic (normalization) uncertainty quoted for the
OLYMPUS data \cite{Henderson:2016dea} is somewhat smaller than in the
other $e^+ p/e^- p$ experiments \cite{Rachek:2014fam, Nikolenko:2014uda}.

While the possibility of unexpected effects in the high-$\eps$ region
is intriguing, we should note that the ratio $R_{2\gamma}$ defined in
Ref.~\cite{Henderson:2016dea} is normalized by a ratio of $e^+$ to
$e^-$ events obtained from a Monte Carlo (MC) simulation of the
experiment designed to account for differences between electrons
and positrons,
$R_{2\gamma} \to
  (N_{\rm exp}^{e^+} / N_{\rm exp}^{e^-})
/ (N_{\rm MC}^{e^+}  / N_{\rm MC}^{e^-})$.
Here $N_{\rm exp}^{e^\pm}$ is the number of observed $e^\pm$ events, and
$N_{\rm MC}^{e^\pm}$ is the number of simulated $e^\pm$ counts, taking
into account radiative effects and various experimental settings. Of
course, in order to simulate the elastic scattering cross sections, some
input about the $e^\pm p$ interaction is needed for the MC, and it is
possible that this may introduce additional model dependence into the
procedure.  Indeed, simulations using radiative corrections computed to
order $\alpha^3$ versus those computed to all orders through
exponentiation show that the latter can give $R_{2\gamma}$ values as
much as 1\% higher at the lowest $\eps$ points \cite{Henderson:2016dea}.
In Fig.~\ref{fig:eeOLYM} the results shown correspond to the ratio
$R_{2\gamma}$ extracted with radiative corrections computed to all
orders in $\alpha$.

The relatively large overall uncertainties on all of the currently
available $R_{2\gamma}$ data unfortunately precludes any definitive
conclusions about TPE effects that can be reached, other than that
the effects are generally consistent with zero, as well as with the
signs and magnitudes expected from the dispersive TPE calculations.
This scenario calls for an urgent need for new measurements of
$e^+ p$ to $e^- p$ ratios at large $Q^2$, $Q^2 \gtrsim 2$~GeV$^2$,
and over a range of $\eps$ values below $\eps \sim 0.5$, where the
TPE effects are predicted to be large enough ($\sim 2\%$) to be
more clearly identified experimentally.
On the other hand, the negative values of the slope in $\eps$
predicted by the TPE calculations are generally consistent with the
data from each of the CLAS \cite{Rimal:2016toz}, VEPP-3
\cite{Rachek:2014fam, Nikolenko:2014uda} and OLYMPUS
\cite{Henderson:2016dea} $e^+ p/e^- p$ experiments.

% . . . . . . . . . . . . . . . . . . . . . . . . . . . . . . . . . . .
\subsubsection{Polarization observables}
\label{sss.PT}

A complementary set of observables that can provide information on TPE
effects involves polarization transfer in the elastic scattering of
longitudinally polarized electrons from (unpolarized) protons, with
measurement of the polarization of the final state proton, $\vec{e} p
\to e \vec{p}$. Defining $P_T$ and $P_L$ to be the polarizations of
recoil protons in the transverse and longitudinal directions relative to
the proton momentum in the scattering plane, one has
\cite{Guichon:2003qm}
\begin{subequations}
\bea
P_T
&=&-\frac{\sqrt{2\tau\eps (1-\eps)}}{\sigma_R}
    \Big[G_E G_M + G_M \Re \big(\fonep -\tau \ftwop \big)
        + G_E \Re \big( \fonep+\ftwop +\frac{\nu}{\tau} \gap \big)
    \Big],		\\
P_L
&=& \frac{\tau \sqrt{1-\eps^2}}{\sigma_R}
    \Big[G_M^2
     + 2 G_M \Re \big( \fonep+\ftwop + \frac{\nu}{\tau(1+\eps)} \gap
		       \big)
    \Big],
\eea
\end{subequations}
where $\sigma_R = \sigma_R^{\rm Born}(1+\delta_{\ggam})$, with
the reduced Born cross section $\sigma_R^{\rm Born}$ given in
Eq.~(\ref{eq:sigmaR}), and $\delta_{\ggam}$ in Eq.~(\ref{eq:delgg}).
Note that the IR subtractions we have made in $\fonep$ and
$\ftwop$ are such that terms in the numerator and denominator of
$P_L$ and $P_T$ cancel exactly, independent of regularization scheme.
Taking the ratio of the transverse to longitudinal polarizations,
we define
\bea
R_{TL}
&=& -\mu_p \sqrt{\frac{\tau(1+\eps)}{2\eps}}\ \frac{P_T}{P_L}\, ,
\eea
where $\mu_p$ is the proton's magnetic moment. In the Born
approximation, this reduces to a simple ratio of the electric to
magnetic form factors, $R_{TL} \to \mu_p G_E/G_M$, which is a function
only of $Q^2$ and is independent of $\eps$.

\begin{figure}[t]
\includegraphics[width=0.49\textwidth]{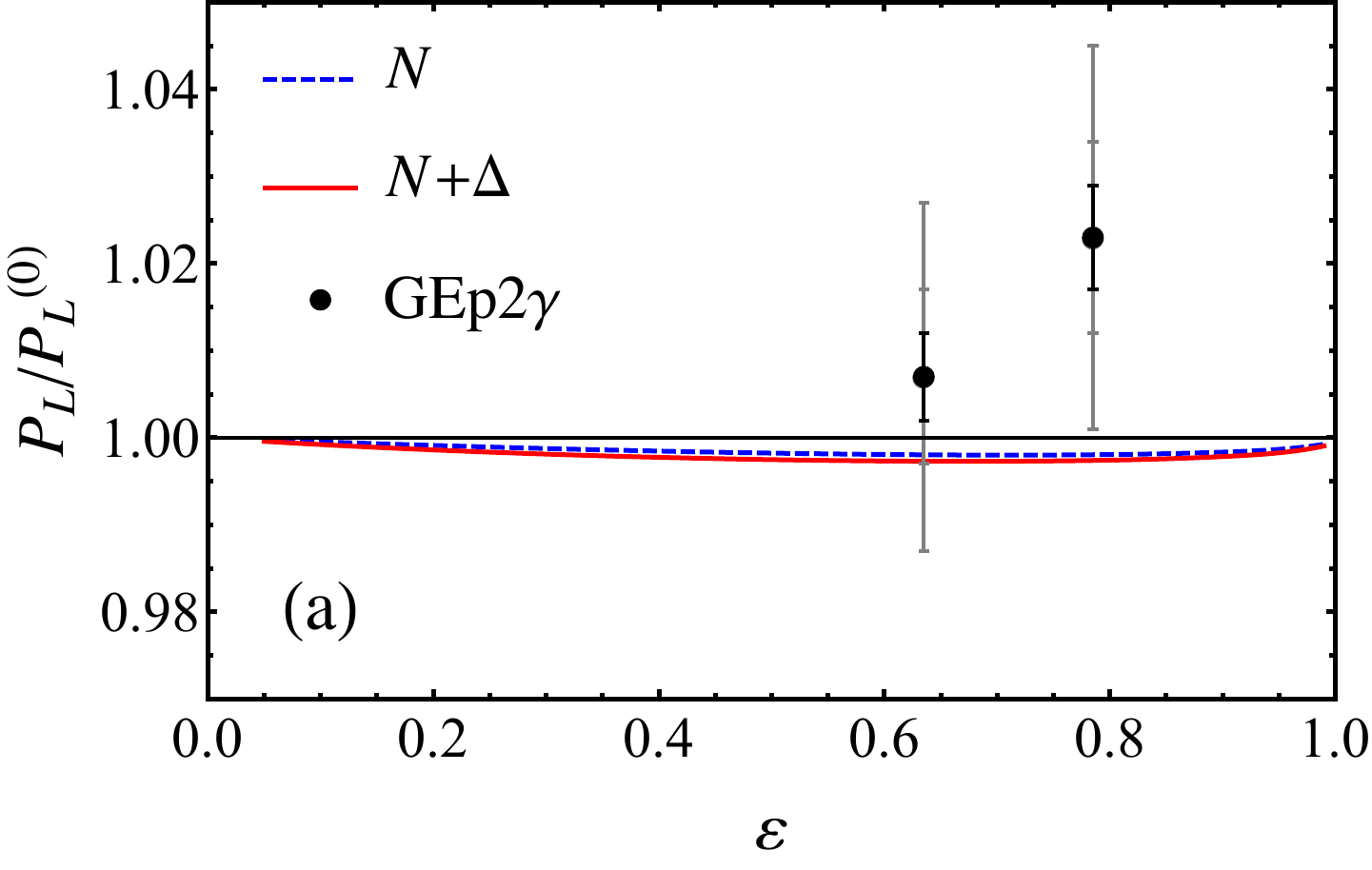}\hspace*{0.2cm}
\includegraphics[width=0.49\textwidth]{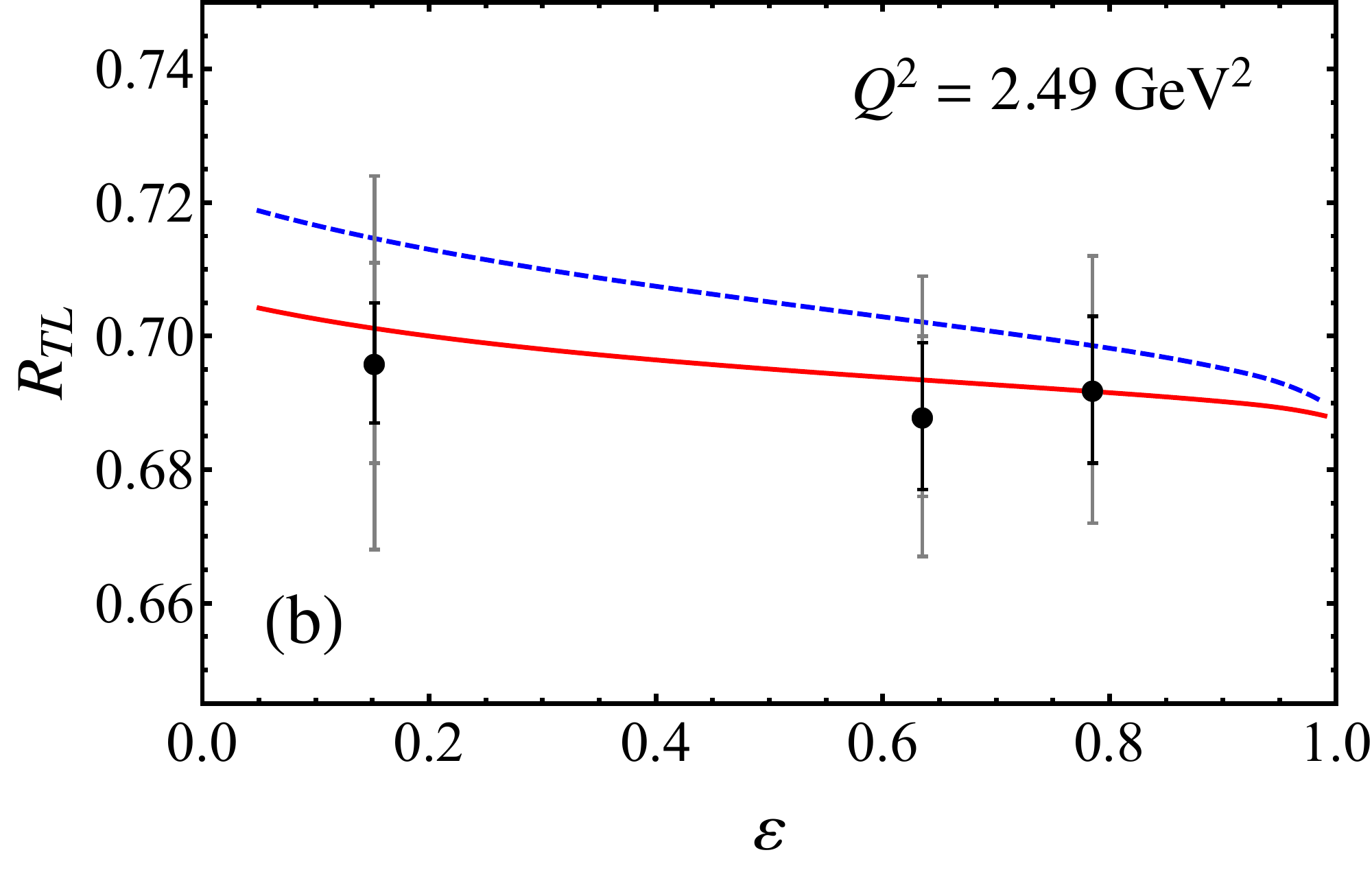}
\caption{(a) Ratio of the total longitudinal recoil proton polarization
	$P_L$ to the Born contribution $P_L^{(0)}$, and
	 (b) ratio $R_{TL}$ of the transverse to longitudinal
	polarizations relative to the Born contribution, as a
	function of $\eps$ for fixed $Q^2=2.49$~GeV$^2$.
	The dispersive calculation including nucleon only
	(dashed blue curves) and the sum of nucleon and $\Delta$
	(solid red curves) intermediate states is compared with
	the data (circles) from the GEp2$\gamma$ experiment at
	Jefferson Lab \cite{Meziane:2010xc}, with the statistical
	and systematic (correlated and uncorrelated) uncertainties
	indicated by the (black) inner and (gray) outer error bars,
	respectively.}
\label{fig:pol}
\end{figure}

The GEp2$\gamma$ experiment at Jefferson Lab \cite{Meziane:2010xc}
measured the ratios $R_{TL}$ and $P_L/P_L^{(0)}$, where $P_L^{(0)}$ is
the Born level longitudinal polarization, at several values of
$\eps$ for fixed $Q^2 = 2.49$~GeV$^2$. These are shown in
Fig.~\ref{fig:pol} as a function of $\eps$, compared with the
dispersive TPE calculations including nucleon and $\Delta$ intermediate
states. The TPE effect on the longitudinal polarization ratio is very
small, with the $P_L/P_L^{(0)}$ ratio only marginally below unity for
all $\eps$ values. Although the trend of the data suggests an
increasing effect at high $\eps$, the data are consistent with
the TPE calculation if (correlated and uncorrelated) systematic
uncertainties are taken into account.

The $\eps$ dependence of the $R_{TL}$ ratio in Fig.~\ref{fig:pol} is
also very weak, and in good agreement with the dispersive TPE
calculation, especially once the $\Delta$ intermediate states are
included. As for the $R_{2\gamma}$ ratio, higher-precision measurements
of the polarization observables at larger $Q^2$ and lower $\eps$ values
would be valuable in more definitively identifying effects beyond the
Born approximation.

%%%%%%%%%%%%%%%%%%%%%%%%%%%%%%%%%%%%%%%%%%%%%%%%%%%%%%%%%%%%%%%%%%%%%%%%
%%%%%%%%%%%%%%%%%%%%%%%%%%%%%%%%%%%%%%%%%%%%%%%%%%%%%%%%%%%%%%%%%%%%%%%%
\section{Outlook}
\label{sec.outlook}

In this paper we have presented a new approach to evaluating two-photon
exchange effects in elastic electron--proton scattering, based on a
dispersion relation analysis of the scattering amplitudes. We considered
two methods for evaluating the imaginary parts of the loop diagrams,
using analytic and numerical methods, and including the contributions
from nucleon and $\Delta$ resonance intermediate states.

In contrast to previous calculations based on the direct evaluation of
loop integrals, the dispersive analysis avoids uncertainties associated
with off-shell intermediate states, and guarantees the correct behavior
in the high energy limit.  This problem is particularly egregious for
the case of derivative interactions, such as for the $\Delta$ baryon,
where the TPE amplitude in the off-shell calculation diverges linearly
with energy in the forward limit.  The dispersive approach, on the other
hand, respects unitarity and is well-behaved at all energies.

The analytic dispersive method, which has been used recently in the
literature \cite{Borisyuk:2008es, Borisyuk:2012he, Borisyuk:2013hja,
Borisyuk:2015xma, Tomalak:2014sva}, has the advantage of allowing
closed analytic expressions for the amplitudes in terms of simple
logarithms, provided the vertex form factors can be parametrized
in terms of sums or products of monopole functions.  This has the
virtue of increased speed of computation, but is limited by the
accuracy of the monopole parametrization of the proton's electric
and magnetic form factors, which typically deteriorates markedly
for $Q^2 \gtrsim 5$--6~GeV$^2$.

The numerical contour method, in contrast, allows for a wide range
of form factor parametrizations, and is relatively straightforward
to implement.  We find that in practice a simple contour is valid
for any parametrization which has poles in the timelike region of
$Q^2$.  For more elaborate parametrizations that have with poles in
the spacelike region, a special choice of contour can be used up to
some maximum $Q^2$ that depends on the exact location of the poles.

To verify the utility of the dispersive approach, we have compared the
results of the numerical TPE calculations with the most recent data on
the ratio of $e^+ p$ to $e^- p$ elastic scattering cross sections, which
is directly sensitive to electromagnetic effects beyond the Born
approximation.  We find good agreement with the data from the CLAS
\cite{Rimal:2016toz} and \mbox{VEPP-3} \cite{Rachek:2014fam,
Nikolenko:2014uda} experiments. The results are also consistent, within
the experimental uncertainties, with the more recent OLYMPUS experiment
\cite{Henderson:2016dea}, which suggests a trend in the opposite
direction at near-forward angles compared with the TPE calculations and
the other data sets \cite{Rimal:2016toz, Rachek:2014fam,
Nikolenko:2014uda}.

For the future, it will be important to extend the present framework to
inelastic non-resonant intermediate states, including the $\pi N$
continuum, and allowing for widths of resonances.  Efforts in this
direction have been made recently in the literature
\cite{Borisyuk:2015xma, Tomalak:2016vbf}, and will be aided by better
knowledge of the empirical virtual Compton scattering amplitudes at
non-forward angles. Beyond this, a longer term challenge is to further
generalize the dispersive approach to higher $Q^2$ and intermediate
state masses, merging the phenomenological hadronic description with one
that expresses the TPE amplitudes explicitly in terms of partonic
degrees of freedom.

On the experimental front, new, higher precision data on $e^+ p$ and
$e^- p$ cross sections at larger $Q^2$ values, $Q^2 \gtrsim 3$~GeV$^2$,
and lower $\eps$ values, $\eps \ll 1$, are needed to unambiguously
identify TPE effects directly.  The possibility of achieving this with a
dedicated positron source at the 12~GeV Jefferson Lab facility remains
an exciting prospect \cite{Arrington:2009qd}. The technology described
here for the TPE calculations can also be readily applied to the
evaluation of $\gamma Z$ interference contributions in parity-violating
electron--proton scattering, in the extraction of the strange
electroweak form factors and the weak charge of the proton
\cite{Arrington:2011dn, Afanasev:2005ex, Gorchtein:2008px,
Gorchtein:2011mz, Rislow:2010vi, Tjon:2009hf, Sibirtsev:2010zg,
Hall:2013hta, Hall:2013loa, Hall:2015loa}.

\appendix
%%%%%%%%%%%%%%%%%%%%%%%%%%%%%%%%%%%%%%%%%%%%%%%%%%%%%%%%%%%%%%%%%%%%%%%%
%%%%%%%%%%%%%%%%%%%%%%%%%%%%%%%%%%%%%%%%%%%%%%%%%%%%%%%%%%%%%%%%%%%%%%%%
\section{Generalized form factors}
\label{app:genFF}

Beyond the Born approximation, the total amplitude for the elastic $e p
\to e p$ scattering process can, for a massless electron, be decomposed
into three independent amplitudes, or generalized form
factors~\cite{Guichon:2003qm}.  In this appendix we describe a practical
method for extracting these generalized form factors from the total
amplitude.

The objective is to map ${\cal M}_{\ggam}$ of Eq.~\eqref{eq:Mggbox} onto
the generalized matrix element $\widehat{\cal M}_{\rm
gen}(\fonep,\ftwop,\gap)$ given by Eq.~\eqref{eq:Mgen},
\bea
\widehat{\cal M}_{\rm gen}(\fonep,\ftwop,\gap)
&=& -\frac{e^2}{q^2}\,
\bar{u}_e(k') \gamma_\mu u_e(k)\ 
\bar{u}_N(p')\, \widehat{\Gamma}^\mu(\fonep,\ftwop,0;q)\, u_N(p) \nn\\
&&-\, \frac{e^2}{q^2}\, 
\bar{u}_e(k') \gamma_\mu \gamma_5 u_e(k)\ 
\bar{u}_N(p')\, \widehat{\Gamma}^\mu(0,0,\gap;q)\, u_N(p)\, ,
\eea
where the generalized current operator $\widehat\Gamma^\mu$ is
defined as
\be
\widehat{\Gamma}^\mu(\fonep,\ftwop,\gap;q)
= \fonep(Q^2,\nu)\,\gamma^\mu
+ \ftwop(Q^2,\nu)\, \frac{i \sigma^{\mu\nu} q_\nu}{2 M}
+ \gap(Q^2,\nu) \gamma^\mu \gamma_5\, ,
\ee
and the generalized form factors are functions of two variables,
taken to be $Q^2$ and $\nu$.

The basic idea behind this method is to project
$\widehat{\cal M}_{\rm gen}$ onto three linearly independent quantities
(pseudo-observables) using Dirac trace techniques. The pseudo-observable
projections are linear combinations of the amplitudes $\fonep$,
$\ftwop$, and $\gap$. The same projections are also made for
${\cal M}_{\ggam}$. One can then invert the transformation matrix to
obtain $\fonep$, $\ftwop$, and $\gap$ in terms of the pseudo-observable
projections of ${\cal M}_{\ggam}$.

While any three linearly independent projections will suffice,
it is convenient to use the same functional form given by
$\widehat{\cal M}_{\rm gen}$.  Consider the pseudo-observable
\bea
\sigma_{\rm gen} &=& \widehat{\cal M}_{\rm gen}^\dagger(A,B,C)\,
\widehat{\cal M}_{\rm gen}(\fonep,\ftwop,\gap)\nn\\
&=& A x + B y + C z\, ,
\label{eq:sigmagen}
\eea
where $x$, $y$, and $z$ are linear combinations of $\fonep$, $\ftwop$,
and $\gap$, and $A$, $B$, and $C$ are placeholder coefficients
representing the three independent projections.  The overall factor
$(-e^2/q^2)^2$ is irrelevant for this derivation, and can be absorbed
into the coefficients $\{A,B,C\}$.  The expression for $\sigma_{\rm gen}$
can be obtained from
\be
\sigma_{\rm gen} = L_{\alpha\mu} H^{\alpha\mu}\, ,
\ee
where the leptonic and hadronic tensors are given by
\begin{subequations}
\bea
L_{\alpha\mu}
&=& \frac{1}{2}
    \Tr \left[ \gamma_\alpha (\slashed{k}-\slashed{q})
               \gamma_\mu \slashed{k}
        \right]\, ,                             \\
H^{\alpha\mu}
&=& \frac{1}{2}
    \Tr \biggl[ \widehat{\Gamma}^\alpha(A,B,C;-q)
		(\slashed{p}+\slashed{q}+M)
		\widehat{\Gamma}^\mu(\fonep,\ftwop,\gap;q)
		(\slashed{p}+M)
	\biggr]\, ,
\eea
\end{subequations}
respectively. Note that rather than keeping the vector and axial-vector
terms separate, we have combined them into one compact expression. This
is possible because the parity-violating terms in the combined amplitude
vanish after taking the traces. Evaluating the traces, one finds
\be
\sigma_{\rm gen}
= (A,B,C) \cdot \left(\begin{array}{c}x\\y\\z\end{array}\right)
= (A,B,C) \cdot \mathbb{M} \cdot
  \left(\begin{array}{c} \fonep \\ \ftwop \\ \gap \end{array}\right)\, ,
\ee
where the transformation matrix $\mathbb{M}$ is given by
\be
\mathbb{M}= 16 M^4
\left(
\begin{array}{ccc}
 \nu ^2+\tau^2 - \tau  & 2 \tau ^2 & 2 \nu  \tau  \\
 2 \tau ^2 & \tau  \left(\nu ^2-\tau ^2+\tau \right) & 2 \nu  \tau  \\
 2 \nu  \tau  & 2 \nu  \tau  & \nu ^2+\tau ^2+\tau  \\
\end{array}
\right)\, .
\label{eq:Mtrans}
\ee
Inverting the matrix $\mathbb{M}$ will obtain the relationships of
interest,
\bea
\left(\begin{array}{c} \fonep\\ \ftwop \\ \gap \end{array}\right) &=&
\frac{1}{16 M^4 \left(\nu ^2-\tau  (\tau +1)\right)^2}\ \nn\\
&&
\times
\left(
\begin{array}{ccc}
 \nu ^2+\tau ^2-\tau  & 2 \tau  & -2 \nu  \tau  \\
 2 \tau  & (\nu ^2-\tau ^2+\tau)/\tau & -2 \nu  \\
 -2 \nu  \tau  & -2 \nu  & \nu ^2+\tau ^2+\tau  \\
\end{array}
\right)
\left(\begin{array}{c} x\\ y \\ z \end{array}\right).
\label{eq:Minv}
\eea
The same projections as in Eq.~\eqref{eq:sigmagen} can now be made
for ${\cal M}_{\ggam}$,
\bea
\sigma_{\rm gen}
&=& \widehat{\cal M}_{\rm gen}^\dagger(A,B,C)\, {\cal M}_{\ggam}\nn\\
&=& A x + B y + C z\, .
\label{eq:sigmagengg}
\eea
The set of projected functions $\left\{x,y,z\right\}$ are then combined
to give the TPE amplitudes using Eq.~\eqref{eq:Minv}.  There is an
apparent kinematic singularity in Eq.~\eqref{eq:Minv} at $\nu^2 =
\tau(\tau+1) = \nu_{\rm ph}^2$, which is the threshold between the
physical and unphysical regions. However, the full expressions for the
imaginary parts of $\fonep(Q^2,\nu)$, $\ftwop(Q^2,\nu)$, and
$\gap(Q^2,\nu)$ are continuous, smooth, and finite across this boundary.
Nevertheless, for numerical work we avoid directly using this kinematic
point.

From Eq.~(\ref{eq:Mggbox}), the pseudo-observable $\sigma_{\rm gen}$
has the form
\be
\sigma_{\rm gen}
= \frac{\alpha}{4\pi} Q^2 \frac{1}{i\pi^2}\int d^4 q_1\ 
  \frac{L_{\alpha\mu\nu} H^{\alpha\mu\nu}}
       {(q_1^2 -\lambda^2)(q_2^2-\lambda^2)}\, ,
\ee
where the leptonic and hadronic tensors of
Eq.~(\ref{eq:Lgg})--(\ref{eq:Hgg}) are
\bea
L_{\alpha\mu\nu}
&=& \frac{1}{2}
\Tr\left[ \gamma_\alpha (1+\gamma_5)
	  (\slashed{k}-\slashed{q}) \gamma_\mu
	  S_F(k-q_1) \gamma_\nu \slashed{k}
   \right]\, ,					\\
H^{\alpha\mu\nu}
&=& \frac{1}{2}
\Tr\biggl[ \widehat{\Gamma}^\alpha(A,B,C;-q)
	   (\slashed{p}+\slashed{q}+M)  	\nn\\
&& \hspace*{0.5cm}
   \times\ \Gamma_{R\to N\gamma}^{\mu\lambda}(p+q_1,-q_2)\,
 	   S_{\lambda\rho}(p+q_1,M_R)\,
	   \Gamma_{\gamma N\to R}^{\rho\nu}(p+q_1,q_1)
	   (\slashed{p}+M)
   \biggr]\, .
\eea
The decomposition of the total amplitude into a basis of generalized
form factors is not unique.  Another convention in the literature
\cite{Chen:2004tw, Tomalak:2014sva} is to use the generalized matrix
element
\bea
\widetilde{\cal M}_{\rm gen}(\fonem,\ftwom,\fthreem)
&=& -\frac{e^2}{q^2}\,
     \bar{u}_e(k') \gamma_\mu u_e(k)			\nn\\
& & \times\
    \bar{u}_N(p')
    \biggl[ \fonem(Q^2,\nu)\,\gamma^\mu
	  + \ftwom(Q^2,\nu)\, \frac{i \sigma^{\mu\nu} q_\nu}{2 M}
	  + \fthreem(Q^2,\nu) \frac{\slashed{k}\,p^\mu}{M^2}
    \biggr] u_N(p)\, ,					\nn\\
& &
\eea
The relationship between the set $\{\fonep,\ftwop,\gap\}$ and the set
$\{\fonem,\ftwom,\fthreem\}$ is
	$\fonep=\fonem+\nu \fthreem$,
	$\ftwop=\ftwom$, and
	$\gap = -\tau \fthreem$~\cite{Chen:2004tw}.
This relationship can be easily derived by contracting
$\widetilde{\cal M}_{\rm gen}$ with
$\widehat{\cal M}_{\rm gen}^\dagger$, in analogy with
Eq.~(\ref{eq:sigmagen}), and comparing the transformation matrices.

%%%%%%%%%%%%%%%%%%%%%%%%%%%%%%%%%%%%%%%%%%%%%%%%%%%%%%%%%%%%%%%%%%%%%%%%
%%%%%%%%%%%%%%%%%%%%%%%%%%%%%%%%%%%%%%%%%%%%%%%%%%%%%%%%%%%%%%%%%%%%%%%%
\section{Analytic expressions for imaginary parts of Passarino-Veltman
	\mbox{functions}}
\label{app:PV}

In the notation of the LoopTools package~\cite{Hahn:1998yk}, the full
dependence of the $s$-channel PV functions on kinematic variables is
\begin{subequations}
\bea
B_0(s)
&\equiv& B_0(s; m_e, W)\, ,			\\
C_0(s;\Lambda)
&\equiv& C_0(M^2,m_e^2,s; W, \Lambda, m_e)\, ,	\\
D_0(s;\Lambda_1,\Lambda_2)
&\equiv& D_0(M^2,M^2,m_e^2,m_e^2,t,s;\Lambda_1,W,\Lambda_2,m_e)\, .
\eea
\end{subequations}
Following the discussion in Sec.~\ref{ssec:analytic}, the full 
expressions for the imaginary parts of the these functions are
\begin{subequations}
\label{eq:b0c0d0}
\bea
b_0(s)
&=& \frac{\pi s_W}{s}\, \theta(s_W)\, ,		\\
c_0(s;\Lambda)
&=&-\frac{\pi}{s_M}
    \log\left(\frac{\Lambda^2 s + s_M s_W}{s \Lambda^2}\right)\,
    \theta(s_W)\, ,				\\
d_0(s;\Lambda_1,\Lambda_2)
&=& \frac{s_W}{4 s} J\,\theta(s_W)\, ,
\label{eq:d0}
\eea
\end{subequations}
where $s_M$ and $s_W$ are given by Eq.~(\ref{eq.sMsW}), the
quantity $J$ is given by
\bea
J &=& \frac{4\pi}{Y\sqrt{1-z^2}}
      \log\left(\frac{1+\sqrt{1-z^2}}{z}\right)\, ,
\eea
and we have introduced the dimensionless variable $z=Z/Y$, with
\bea
Z &=& \frac{\Lambda_1 \Lambda_2
            \sqrt{(\Lambda_1^2 s+s_M s_W) (\Lambda_2^2 s+s_M s_W)}}
	   {s}\, ,\\
Y &=& \frac{2 \Lambda_1^2 \Lambda_2^2 s
	    +s_M s_W \left(\Lambda_1^2+\Lambda_2^2\right)+Q^2 s_W^2}
	   {2 s}\, .
\eea
The expression for $d_0(s;\Lambda_1,\Lambda_2)$ in Eq.~\eqref{eq:d0} has
been rewritten in a form that is numerically stable for very small
values of $\Lambda_i$.  By comparison, the form given in
Eq.~\eqref{eq:Jdef}, with $X^2=Y^2-Z^2$, is susceptible to roundoff
error in this limit.  We also note that while the real part of $D_0$ has
a logarithmic dependence on $m_e$, the imaginary part does not, and is
therefore finite in the limit $m_e\to 0$, which we have used throughout
this paper.

The PV functions $c_0(s;\Lambda)$ and $d_0(s;\Lambda_1,\Lambda_2)$ are
IR-divergent in the limit $\Lambda^2\to 0$.  Replacing $\Lambda^2 \to
\lambda^2$, and keeping only logarithmic terms in $\lambda^2$, we have
the three IR-divergent combinations
\begin{subequations}
\label{eq:cdlam}
\bea
c_0(s;\lambda)
&=& -\frac{\pi}{s_M}
    \log\left(\frac{s_M s_W}{\lambda^2 s}\right)\, \theta(s_W)\, ,\\
d_0(s;\lambda,\Lambda)
&=& \frac{\pi}{\Lambda^2 s_M+Q^2 s_W}
    \log\left(
	\frac{s_W \left( \Lambda^2 s_M+Q^2 s_W\right)^2}
	     {\lambda^2 \Lambda^2 s_M
		\left(\Lambda^2 s+s_M s_W\right)}
        \right)\, \theta(s_W)\, ,				\\
d_0(s;\lambda,\lambda)
&=& \frac{2 \pi}{Q^2 s_W}
    \log\left( \frac{Q^2 s_W}{\lambda^2 s_M}\right)\, \theta(s_W)\, .
\eea
\end{subequations}
One can use these expressions to explicitly show that there is no
residual dependence on $\lambda$ in the total TPE amplitudes, after
subtracting the model-independent Maximon and Tjon IR-divergent terms. 
For numerical calculations, it is more convenient to use the full
expressions~\eqref{eq:b0c0d0}, with a value for $\lambda$ satisfying the
criterion $\lambda^2 \ll m_e^2$.

\begin{table}[thb]
\centering
\caption{\label{tab:Iij}
	Table of values for $I_{ij}$ in Eq.~(\ref{eq:Iij}), keeping
	only logarithmic terms in the infinitesimal regulator $\lambda$.
	The dependence on $s$ and other kinematic variables is
	suppressed for clarity.}
\begin{ruledtabular}
\begin{tabular}{cccc}
$N$	& $i$ & $j$ & $I_{ij}$ \\
\hline
0 & 0 & 0 & $d_0(\lambda,\lambda) - d_0(\Lambda_1,\lambda) -
				d_0(\lambda,\Lambda_2) + d_0(\Lambda_1,\Lambda_2)$\\
1 & 1 & 0 & $\Lambda_1^2 \left(d_0(\Lambda_1,\lambda) - d_0(\Lambda_1,\Lambda_2)\right)$\\
1 & 0 & 1 & $\Lambda_2^2 \left(d_0(\lambda,\Lambda_2) - d_0(\Lambda_1,\Lambda_2)\right)$\\
2 & 2 & 0 & $-\Lambda_1^2 \left(c_0(\lambda) - c_0(\Lambda_2) +
              \Lambda_1^2 \left(d_0(\Lambda_1,\lambda) - d_0(\Lambda_1,\Lambda_2)\right)\right)$\\
2 & 0 & 2 & $-\Lambda_2^2 \left(c_0(\lambda) - c_0(\Lambda_1) +
              \Lambda_2^2 \left(d_0(\lambda,\Lambda_2) - d_0(\Lambda_1,\Lambda_2)\right)\right)$\\
2 & 1 & 1 & $\Lambda_1^2 \Lambda_2^2 d_0(\Lambda_1,\Lambda_2)$\\
\end{tabular}
\end{ruledtabular}
\end{table}

For sums of monopole form factors, the integrals of interest are sums of
primitive integrals $I_{ij}$ given in Eq.~(\ref{eq:Iij}), with $N=i+j$. 
Table~\ref{tab:Iij} gives the values for $I_{ij}$ up to $N=2$, which
suffices for nucleon intermediate states. In general, for other states
like the $\Delta$ one needs up to $N=3$, which requires the form factor
to behave like $1/Q_i^4$ asymptotically, but the procedure follows
analogously.

%%%%%%%%%%%%%%%%%%%%%%%%%%%%%%%%%%%%%%%%%%%%%%%%%%%%%%%%%%%%%%%%%%%%%%%%
%%%%%%%%%%%%%%%%%%%%%%%%%%%%%%%%%%%%%%%%%%%%%%%%%%%%%%%%%%%%%%%%%%%%%%%%
\section{Form factor reparametrizations}
\label{app:param}

Here we present the reparametrizations of the nucleon and $\Delta$ vertex
form factors in terms of sums and/or products of monopoles, suitable for
use in the analytic expressions of Sec.~\ref{ssec:analytic}.  Fits are
over the range $0<Q^2<8$~GeV$^2$, with $F_1(Q^2)$ being a 5-parameter
fit, while all others are 4-parameter fits.

The nucleon form factors are fit to the parametrization of
Ref.~\cite{Venkat:2010by},
\bea
F_1(Q^2) &=& \frac{0.334}{1+Q^2/0.209}
           + \frac{1.228}{1+Q^2/0.805}
           - \frac{0.562}{1+Q^2/1.898}\, ,\\
F_2(Q^2) &=& \frac{\kappa}{1+Q^2/3.502}
             \left( \frac{1.165}{1+Q^2/0.364}
                  - \frac{0.165}{1+Q^2/2.675}
             \right)\, ,
\eea
with $Q^2$ in GeV$^2$. Defining
\be
g(Q^2) = \frac{(M_\Delta+M)^2}{Q_+^2}
       = \frac{1}{1+Q^2/(M_\Delta+M)^2}\, ,
\ee
the $\Delta$ transition form factors are fit to the parametrization
of Ref.~\cite{Aznauryan:2011qj,Aznauryan:2016pc},
\bea
g(Q^2)\, G_{M,E}^*(Q^2)
&=& G_{M,E}^*(0)\frac{1}{1+Q^2/3.177}
    \left( \frac{2.474}{1+Q^2/0.575} - \frac{1.474}{1+Q^2/1.000}
    \right)\, ,                         \\
g(Q^2)\, G_C^*(Q^2)
&=& G_C^*(0) \frac{1}{\left(1+Q^2/1.102\right)^2}
    \left( \frac{0.813}{1+Q^2/0.0684} + \frac{0.187}{1+Q^2/0.895}
    \right)\, .
\eea
The form factors $g_1(Q^2)$, $g_2(Q^2)$, and $g_3(Q^2)$ can be obtained
from simple combinations of these parametrizations, while still allowing
for implementation in analytic form.

The use of these reparametrizations in the numerical integration method
allows for a test of our codes against the analytic results. We were
routinely able to obtain a relative agreement at the level of five
significant digits.  Similarly, the validity of using these
reparametrizations in the analytic codes can be tested against the
numerical results using the original functional forms.  In general, we
found that the relative differences between the analytic and numerical
evaluations of the imaginary parts of the TPE amplitudes were comparable
to the relative differences between the original and reparametrized
forms over the relevant range of $Q^2$. As the original and
reparametrized vertex form factors given in this appendix agree at
roughly the 2\% level when averaged over the range $0<Q^2<8$~GeV$^2$, we
find a 2\% agreement in $\fonep(Q^2,\nu)$, $\ftwop(Q^2,\nu)$, and
$\gap(Q^2,\nu)$ up to $Q^2=4$~GeV$^2$. This suggests that the TPE
results are not very sensitive to pole structure of the vertex form
factor parametrizations in the complex plane.

%%%%%%%%%%%%%%%%%%%%%%%%%%%%%%%%%%%%%%%%%%%%%%%%%%%%%%%%%%%%%%%%%%%%%%%% 
\section*{Acknowledgements}

We thank J.~Bernauer, D.~Higinbotham and A.~Schmidt for helpful
discussions and communications, and I.~Aznauryan for providing a
parametrization of the $N \Delta$ transition form factors. PGB thanks
Jefferson Lab for support during a sabbatical leave, where part of
this work was completed. This work was supported by NSERC (Canada)
and DOE Contract No.~DE-AC05-06OR23177, under which Jefferson Science
Associates, LLC operates Jefferson Lab.

% ----------------------------------------------------------------------
%
%  Bibliography
%
% ----------------------------------------------------------------------
\raggedright

%\bibliography{TPEdisp}

\end{document}